# PROCESSES OF THE CORRELATION OF SPACE (LENGTHS) AND TIME (DURATIONS) IN HUMAN PERCEPTION

Lev I Soyfer

To study the processes and mechanisms of the correlation between space and time, particularly between lengths and durations in human perception, a special method (device and procedure) to conduct this experiment was designed and called LDR (Length Duration Relation) In the present study a pilot and three series of the primary experiment were conducted under conditions of different levels of uncertainty.
In all types of experiments, signals of a certain duration and modality were presented twice in random order to the subjects. Observers had to respond to time signals of different durations by choosing a corresponding space interval. The data which were obtained during the 1st and the 2nd time signal presentations were examined separately. The comparative data analysis of the experiment showed significant differences between the 1st and 2nd presentation of signals in the quantity of correct responses, the responses distribution along the scale of stimuli, the phenomena which occurred during the experiment. The higher level of uncertainty condition under which a certain type of the experiment was conducted, the more clearly this difference was manifested.

Based on results of the experiments comparative data analysis, one can suppose that the perceptive mechanism, named by us as an **innate mechanism of proportionality**, performed the correlation of these intervals into two stages: adaptation and activation. In the **adaptation stage** (which took place during the 1st time signals presentation) observers searched for a corresponding modulus, limits of the scale of presented stimuli, and created a temporary mental scale of possible stimuli. In the experiment conducted under the high level of uncertainty this scale was divided by an internal limit into two parts according to the principle of the golden section and detected locations of other conjugate pairs even though he or she did not recognize all of them. The purpose of this scale division into two parts was the creation of conditions which facilitated the activation of the remaining points on the scale. In the **activation or recognition stage** (which took place during the 2nd time signals presentation) the above were applied for correlation purposes. Based on the data analysis of the experiment one can suggest an existence of two interacting measurement systems that participated in the correlation processes: the **innate biological metrics** that underlie the organization of both human and animal physiological, behavioral and mental processes and the **acquired social metrics** which people learned from their society where they live and use it in their daily cognitive activity

It is clear that people use a relation between space and time in their behavior, For example, the driver of an automobile can quite precisely apply the brakes in a certain time (duration) in order to stop so that he does not hit the auto in front of him. Similarly, a baseball player swings his bat in just the right place and at the right time in order to send the ball flying out of the ballpark. These interactions between time intervals and space intervals are common, after a certain amount of learning.

A space - time relationship is the necessary component of physical and mental organization, not only of the coordination of different motions of man, but also his/her cognitive processes, including thinking and speech (Moroz V.N.,1962) Most of the psychological literature on time and space perception has concerned space or time. Some work has also concerned interactions between time and space perception, like those illustrated in the practical examples above. The earliest Gestalt-psychology observations were made with a toy tachistiscope, to show that the time interval between successive visual images determined whether they were seen as continuous motion or as successive. The work of Helson (1930), Helson and King (1931), Abe (1935) and others showed that changing the time interval between three successive touches on the skin or three visual brief stimuli could affect the apparent distances between pairs of them (the *tau* effect). And conversely, Cohen, Hansel, Sylvester (1955) showed that a change of the spatial separations could change the perceived temporal interval (the *kappa* – effect). Recent review of the literature related to these effects see in Hausmann thesis (1996) Cohen et al extended these early findings to auditory stimuli. Hirsh, Bilger & Deatherage (1956) presented an evidence for a strong relation between psychological time and level of auditory stimulation.

Is there a primordial "extensity" in perception, that applies to both spatial and temporal extent, such that observers could report an equivalence between a temporal and a spatial extent, a length-duration relation? Suppose we extend our interest to the durational and spatial aspects of experience.

The driver or the batter executes the movements too quickly to have allowed an introspective judgment of duration or of distance. But let us contrive an artificial or experimental situation in which we ask an observer to judge what length or spatial extent corresponds to a duration or temporal extent.. It is in answer to such a question that the following experiment was carried out.

## I. GENERAL METHOD

### Participants

The majority of observers participating in the experiment were university students, ages 18 to 25.To find out if the phenomenon of the correlation between space and time intervals was observed in the experiment with people of other ages we also drew into the experiment six children ages 6 through 10 and two elderly persons ages 80 to 82.

These observers gave the same estimation which agrees with the length - duration relations given by the experimenter and by the students of the university. All together 200 observers participated in different types of the experiment. In this work we did not attempt to make a comparative analysis of the data related to different ages.

## The experimental method[1]

### The Apparatus and Stimuli

**Apparatus**

[1] **The apparatus, procedure and data processing were computer simulated on 2009**

To study processes and a mechanism of correlation between space and time particularly between lengths and durations in human perception, a special method (device and procedure) to conduct the experiment was designed.
The LDR (Length – Duration Relation) installation was comprised of four blocks:  Block 1 was the panel at which an observer was tested.
On the screen of this panel was a set of different size blue plastic rods that were placed in random order.
Block 2 was the time-signal device that produced in the random order of the light, sound and light-sound time signals of different durations.
The same block registered the time of the observer's decision (time reaction).
Block 2 A was the push button timer device. By pressing the button of the timer, an observer could reproduce the duration of different signals.
Block 3 was the device for observer's response recording.

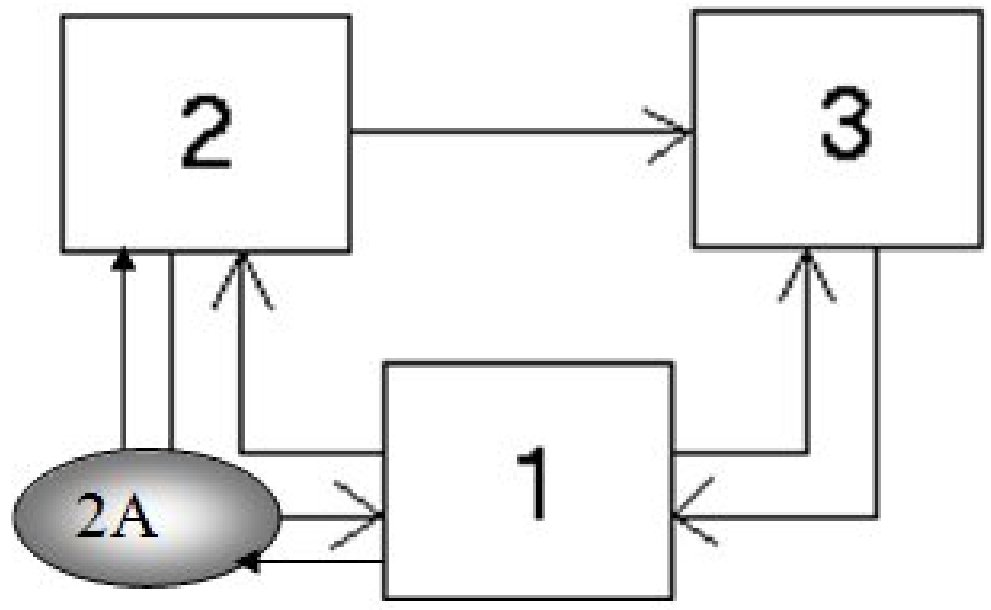

Flow diagram of the LDR – installation

Figure 1

In the present study, a pilot and a primary experiment were conducted. The primary experiment consisted of the 1$^{st}$ series, the 2$^{nd}$ A series, the 2$^{nd}$ B series, and subseries 1a. The differences between these types of experiments, their peculiarities, features, conditions of conducting, and results of the comparative analysis of their data will be considered in detail and discussed later.

**Stimuli presented to observers**

Observers were presented two sets of stimuli: the first set was of different duration time signals and the second set was blue plastic rods of different length placed on the screen in random order. Elements of each set were in a certain proportion to others according to a certain modulus.

These sets were called scales of temporal and spatial intervals.

The ratios between elements of each scale and their moduli were determined by the experimenter.

According to the ratio between these moduli, a coefficient of proportionality of the conjugate pairs of space and time intervals was built.

Thus, **a conjugate pair** was a combination of a certain space interval length and certain time interval duration based on a certain coefficient of proportionality.

The investigator gave each conjugate pair a relevant connection between a space interval of a definite length, and a time interval of a definite duration based on a certain coefficient of proportionality.

There were three principles on which these conjugate pairs could be constructed.

**The one – to – one direct proportionality principle** – occurs when the numerical value in seconds of a presented time signal was the same as the numerical value in centimeters of a space interval. For example, 1 second – 1 centimeter, 2 seconds – 2 centimeters,… 9 seconds – 9 centimeters and so on.

**The direct proportionality principle only** – occurs when the numerical values in seconds of temporal signals and the numerical values of centimeters of spatial intervals which when put together became a conjugate pair that is different although based on a certain coefficient of proportionality. In the case of this experiment, numerical values in seconds were considerably less than numerical values in centimeters. For example, 0.3 sec – 0.5 cm, 0.6 sec – 1 cm, … 4.5 sec – 7.5 cm and so on (The coefficient of proportionality was 1.66.)

**The absence of both the One – to One and the Direct proportionality Principles** – the experimental condition happens when tested observers were forced not to recognize, but to construct conjugate pairs from different size temporal and spatial signals given to them. This condition was not applied in the presented experiment but needs special investigation.

**The Structure of Conjugate Pairs of Space and Time Interval Scales**

A set of conjugate pairs of space and time intervals of different sizes based on the same coefficient of proportionality was considered as a scale of conjugate pairs of space / time intervals.

The next types of this scale were suggested.

**The scale of presented stimuli** was a set of conjugate pairs of space and time intervals of different sizes based on the same coefficient of proportionality physically presented to an observer. In the experiment there were two types of this scale: **complete and partly complete** scales.

**The complete scale** was a scale where conjugate pairs were placed uniformly step by step on every point of this scale from its minimum to maximum limits and a numerical value of such a step was equal to a numerical value of a certain modulus pair (see table # 1) This scale was only used in the pilot experiment.

**The assumed complete scale of possible stimuli** was a scale which contained information of the quantity of all possible conjugate pairs from its minimum to maximum limits including those which were not physically presented in the perceived set of stimuli but could be presented according to a certain coefficient of proportionality given in the primary experiment (see tables # 1, 2). The assumed scale of all possible stimuli was used for comparison purposes and the understanding of the results of the main experiment.

**The partly complete scale** was a scale where conjugate pairs were not placed uniformly everywhere on this scale. This scale was used in the primary experiment. In the primary experiment this scale consisted of two parts – **uniform** and **non-uniform** ones. In the uniform part – the first part of the partly complete scale, conjugate pairs were placed uniformly. In the non-uniform part - the second part of the partly complete scale conjugate pairs were placed non – uniformly where neighboring conjugate pairs could be on several steps from each other (see tables ## 2a, 2 b).

## II .THE STRUCTURE OF THE EXPERIMENT AND CHARACTERISTICS OF EACH IT'S SERIES

### Levels of Uncertainty in Conditions of the Experiment Performance

The term "uncertainty" has different meanings depending on the area of science in which it is used. From our point of view, uncertainty in the case of the given experiment should be considered as the absence – presence of information about a principle of the organization of the structure of relations between space and time intervals presented to an observer. Such information should contain: a) knowledge about limits of a presented stimuli set b) a modulus which underlies relationships between the different lengths and durations making this set c) thereby allowing an observer to predict correlations of different space and time intervals which can belong to such a set of signals but not yet presented to him or her.

Under the level of uncertainty in conditions of the experiment performance it was considered that the absence or presence of: a) a modulus pair within the set of signals, b) a special mention to an observer on this modulus presence within the set of signals, c) preliminary information of the limit pairs. Based on those considerations one can distinguish the next levels of such uncertainty.

**Lower uncertainty condition**– is where the conjugate modulus pair was within the set of signals and an observer was previously acquainted with its maximum and minimum limits.

**Low uncertainty condition**– is where the conjugate modulus pair was not within the set of signals, but an observer was previously acquainted with its maximum and minimum limits.

**Middle uncertainty condition** – is where the conjugate modulus pair was within the presented set of signals, but an observer was informed neither of maximum and minimum limits of this set nor of the presence of this modulus within presented stimuli.

**High uncertainty condition** – is where the conjugate modulus pair was not within the set of signals and the observer had no information about the limits of this set.

As mention above (see p..), in the present study the pilot and three series and one subseries of the primary experiment were conducted. All of these experiments consisted of the main part where an observer had to respond to time signals of different duration by choosing a corresponding space interval and auxiliary parts where the quantity in the pilot experiment and primary experiments were different.

## The pilot experiment

The pilot experiment consisted of two parts: the main part and one auxiliary part – which was a questionnaire that was presented to observers to determine their own mental operations which they performed while correlating space and time intervals, their personal peculiarities and their motivations.

In the pilot experiment, the complete scale of conjugate pairs was used. In this experiment there was no difference between assumed and presented conjugate pair scales.

Table 1

The complete scale of conjugate pairs in the pilot experiment.
(the "one – to – one" principle)

| | | Numbers of places on the scale | | | | | | | | | | | | |
|---|---|---|---|---|---|---|---|---|---|---|---|---|---|---|
| Stimulus categories | | 1 | 2 | 3 | 4 | 5 | 6 | 7 | 8 | 9 | 10 | 11 | 12 | 13 |
| Duration | Seconds | 1 | 2 | **3** | 4 | 5 | 6 | 7 | 8 | 9 | 10 | 11 | 12 | 13 |
| Length | Centimeters | 1 | **2** | **3** | **4** | 5 | 6 | 7 | 8 | 9 | 10 | 11 | 12 | 13 |

Numbers denoting the quantity of centimeters in
red are correct responses
blue are approximately correct responses.
(For example, numbers noted in red and blue colors relate only to one of the signals.)

The conjugate pairs were built on the one-to-one direct proportionality principle. It was assumed that such a structure of conjugate pairs could allow an observer to make a correlation of these pairs elements easier because he or she could compute these intervals by so called "social metrics" or a measuring system learned from society where he or she lives. Using this metrics he or she could easily determine the modulus pair 1 sec – 1 cm which was within a presented set of stimuli or recognize relations between elements of other conjugate pairs as well. For instance, when a three second time signal was received and correctly calculated, the quantity in seconds of the given signal chosen should be a length of three centimeters. The numerical value of these seconds he or she could apply for searching the corresponding length of three centimeters.

Because the modulus pair was within the presented set of stimuli, the condition of the pilot experiment performance should be referred to the middle level of uncertainty.

**The primary experiment**

The primary experiment consisted of the main part of the experiment and six auxiliary parts in which an observer participated immediately after conducting of the main part. The auxiliary parts of the primary experiment were performed as follows: 1. Observers were presented space intervals served before as stimuli in the main part to evaluate them by pressing the button of a timer, as if they had to "give back" these space interval times. 2. Observers had to nonverbally reproduce duration of time signals used in the main part of the experiment by pressing the button of a timer. 3. Observers had to evaluate verbally the same time intervals in seconds. 4. Observers had to nonverbally reproduce lengths used in the main part of the experiment by drawing them on paper. 5. Observers had to express verbally the same space intervals in centimeters. 6. Observers had to determine their own mental operations which they performed while correlating space and time intervals, their personal peculiarities and their motivation by the presentation of a special questionnaire. The goal of auxiliary parts was to find out how an observer correlated and established a connection between presented temporal and spatial intervals in the main part of the primary experiment and what mental processes could have taken place.

In the primary experiment, unlike the pilot experiment, the conjugate pairs were built on the direct proportionality principle only i.e. the one-to-one principle was broken. In this case as mentioned above, numerical values in seconds were considerably less than numerical values in centimeters. Unlike the pilot in the primary experiment, the presented scale was a partly complete one and because of that it was different from the assumed mental temporary scale of possible stimuli. The assumed scale of all possible stimuli was used for comparison purposes and the understanding of the results of the main experiment.

Table 2

The assumed complete scale of conjugate pairs in the main experiment

| | | Numbers of places on the scale | | | | | | | | | | | | |
|---|---|---|---|---|---|---|---|---|---|---|---|---|---|---|
| Stimulus categories | | 1 | 2 | 3 | 4 | 5 | 6 | 7 | 8 | 9 | 10 | 11 | 12 | 13 |
| Duration | Seconds | 0.3 | 0.6 | 0.9 | 1.2 | 1.5 | **1.8** | 2.1 | 2.4 | 2.7 | 3 | 3.3 | 3.6 | 3.9 |
| Length | Centimeters | 0.5 | 1 | 1.5 | 2 | **2.5** | **3** | **3.5** | 4 | 4.5 | 5 | 5.5 | 6 | 6.5 |

| Continuation | | Numbers of places on the scale | | | | | | | | |
|---|---|---|---|---|---|---|---|---|---|---|
| Stimulus categories | | 14 | 15 | 16 | 17 | 18 | 19 | 20 | 21 | |
| Duration | Seconds | 4.2 | 4.5 | 4.8 | 5.1 | 5.4 | 5.7 | 6 | 6.3 | |
| Length | Centimeters | 7 | **7.5** | 8 | 8.5 | 9 | 9.5 | **10** | 10.5 | |

Numbers denoting the quantity of centimeters in
red are correct responses
blue are approximately correct responses.
(For example, numbers noted in red and blue colors relate only to one of the signals.)

In the primary experiment, a partly complete scale was used in order to determine in particular, how the structure of a presented signal scale influences the success of conjugate pairs recognition. If this success depends on the uniformity of a presenting stimuli disposition on the scale, then the most quantity of correct responses should be received in response to all or most of the signals placed on the first half of the scale.

If the quantity of correct responses to signals placed on its second half of the scale are received equally, and even to a greater extent than to the signals placed on the first part of the scale, then in the success and dynamics of the given conjugate pairs recognition is not the uniformity of a presenting stimuli disposition, but other factors play a part. Specifically, this may be caused by the way in which the information of conjugate pairs moduli was obtained, and the limits of the scale.

The next reason for using the partly complete scale was to learn, when tested, if an observer felt a difference between the structures of the first and second parts of this scale. When this difference was discerned, the observer will work to restore the broken uniformity, rhythm of the second part of the scale and bring it into conformity with the first part to mentally set up the complete scale as a whole. Should this occur, the observer will use this complete mental scale for the correlation process and recognition of conjugate pairs of space and time intervals.

One can show **two types** of the partly complete scales. **The first** of them contained the uniform part which began with the first point of the assumed scale of all possible stimuli. The minimal limit of this scale was the conjugate pair 0.3 sec – 0.5 cm. This type was used in the 1st series and the subseries 1a of the primary experiment.(See Table 2 a)

## Table 2 a

### 1st series

### The partly complete scale of conjugate pairs presented to observers

| | | Numbers of places on the complete scale | | | | | | | | | | | | |
|---|---|---|---|---|---|---|---|---|---|---|---|---|---|---|
| Stimulus categories | | 1 | 2 | 3 | 4 | 5 | 6 | 7 | 8 | 9 | 10 | 11 | 12 | 13 |
| Duration | Seconds | 0.3 | 0.6 | 0.9 | 1.2 | 1.5 | **1.8** | 2.1 | 2.4 | | | 3.3 | 3.6 | |
| Length | Centimeters | 0.5 | 1 | 1.5 | 2 | **2.5** | **3** | **3.5** | 4 | | | 5.5 | 6 | |
| Numbers of places on the presented scale | | 1 | 2 | 3 | 4 | 5 | 6 | 7 | 8 | | | 9 | 10 | |

| Continuation | | Numbers of places on the complete scale | | | | | | | | |
|---|---|---|---|---|---|---|---|---|---|---|
| Stimulus categories | | 14 | 15 | 16 | 17 | 18 | 19 | 20 | 21 | |
| Duration | Seconds | | 4.5 | | | **5.4** | | 6 | 6.3 | |
| Length | Centimeters | | **7.5** | | | **9** | | **10** | 10.5 | |
| Numbers of places on the presented scale | | | 11 | | | 12 | | 13 | 14 | |

Numbers denoting the quantity of centimeters in
red are "correct" responses
blue are "approximately correct" responses.
(For example, numbers noted in red and blue colors relate only to one of the signals.)

**The second** one contained the uniform part which began with the third point of the assumed scale (the conjugate pair 0.9 sec – 1.5 cm) and served as the minimal limit of the presented scale. This type was used in the 2nd A and B series of the primary experiment.

Table 2 b
2nd series A, B
The partly complete scale of conjugate pairs presented to observers

| Stimulus categories | | Numbers of places on the complete scale | | | | | | | | | | | | |
|---|---|---|---|---|---|---|---|---|---|---|---|---|---|---|
| | | 1 | 2 | 3 | 4 | 5 | 6 | 7 | 8 | 9 | 10 | 11 | 12 | 13 |
| Duration | Seconds | | | 0.9 | 1.2 | 1.5 | **1.8** | 2.1 | 2.4 | 2.7 | | 3.3 | 3.6 | 3.9 |
| Length | Centimeters | | | 1.5 | 2 | **2.5** | **3** | **3.5** | 4 | 4.5 | | 5.5 | 6 | 6.5 |
| Numbers of places on the presented scale | | | | 1 | 2 | 3 | 4 | 5 | 6 | 7 | | 8 | 9 | 10 |

| Continuation | | Numbers of places on the complete scale | | | | | | | | |
|---|---|---|---|---|---|---|---|---|---|---|
| Stimulus categories | | 14 | 15 | 16 | 17 | 18 | 19 | 20 | 21 | |
| Duration | Seconds | | 4.5 | | | **5.4** | | 6 | 6.3 | |
| Length | Centimeters | | **7.5** | | | **9** | | **10** | 10.5 | |
| Numbers of places on the presented scale | | | 11 | | | 12 | | 13 | 14 | |

Numbers denoting the quantity of centimeters in
red are "correct" responses
blue are "approximately correct" responses.
(For example, numbers noted in red and blue colors relate only to one of the signals.)

## Series of the Primary Experiment

There were three series and one subseries of the primary experiment.

### The First Series

In the first series the modulus pair (0.3 sec / 0.5 cm) was given in the presented set of signals, but an observer was not informed of the maximum and minimum limits of the set, and not pointed specially to the modulus. This series was based on the partly complete scale the uniform part of which began with the first point of the assumed scale of all possible stimuli (the conjugate pair 0.3 sec – 0.5 cm. (see Table 2 a).

The modulus pair was within the presented set of stimuli, but an observer was neither acquainted with the maximum or minimum limits of this set nor pointed specially to the modulus pair. This condition of the first series of the primary experiment performance should be referred to as the middle level of uncertainty.

## Subseries 1a

Subseries 1a had the same peculiarities as the first series of the primary experiment except for the uncertainty condition. In this subseries an observer was preliminarily acquainted with maximum and minimum limits of a presented set of stimuli where the minimum limit was the same as the modulus pair. Such a condition of subseries 1a of the primary experiment performance should be referred to as the lower level of uncertainty.[2]

## The Second Series (A and B)

In the 2nd A and B series the modulus pair (0.3 sec / 0.5 cm) was not given in the presented set of signals. These series were based on the partly complete scale uniform part of which began with the third point of the assumed scale of possible stimuli (the conjugate pair 0.9 sec – 1.5 cm, see Table 2 b)

## The 2nd A series

In the 2nd A series, a modulus pair (0.3 sec – 0.5 cm) was not within the presented set of signals, but an observer was previously acquainted with its maximum and minimum limits. This condition of the 2nd A series of the primary experiment performance should be referred to as the low level of uncertainty.

## The 2nd B series

In the 2nd B series modulus pair (0.3 sec – 0.5 cm) was not within the presented set of signals and an observer was not previously acquainted with its maximum and minimum limits. This condition of the 2nd B series of the primary experiment performance should be referred to as the high level of uncertainty.

[2] The l subseries 1a was conducted on a small group of observers (six persons) for comparison purposes.

Table 3

The pilot and series of the primary experiment and their characteristics.

(Space and time intervals used in the main and auxiliary parts of the primary experiment were the same**)**

| ## | Conditions of the experiment | **The pilot Experiment** | The Primary experiment | | | |
|---|---|---|---|---|---|---|
| | | | **1st series** | **subseries 1 a** | **2nd A series** | **2nd B series** |
| 1 | The one-to-one directly proportional principle | + | - | - | - | - |
| 2 | The only directly proportional principle | - | + | + | + | + |
| 3 | The complete scale of conjugate pairs | + | - | - | - | - |
| 4 | The partly complete scale of conjugate pairs | - | + | + | + | + |
| 5 | The lower uncertainty condition | - | - | + | - | - |
| 6 | The low uncertainty condition | - | - | - | + | - |
| 7 | The middle uncertainty condition | + | + | - | - | - |
| 8 | The high uncertainty condition | - | - | - | - | + |
| Auxiliary parts and their numbers | | 1 | 6 | 6 | 6 | 6 |
| 1 | Evaluation space intervals by time intervals by pressing the button of a timer | - | + | + | + | + |
| 2 | Reproducing nonverbally, time intervals by pressing the button of a timer | - | + | + | + | + |
| 3 | Evaluation verbally, time intervals in seconds | - | + | + | + | + |
| 4 | Reproducing nonverbally, space intervals by drawing them on the paper | - | + | + | + | + |
| 5 | Evaluation verbally, space intervals in centimeters | - | + | + | + | + |
| 6 | Answering to a special questionnaire to determine mental operations which an observer performed while correlating space and time intervals. | | | | | |

# III. EXPERIMENT PERFORMANCE

## Specifications of stimuli presented to observers

In all types experiment observers were presented time signals of different duration which were of three modalities: light, sound, and light - sound together. Signals of a certain duration and modality were presented twice in random order during all types of conducted experiments. Observers were asked to choose among different rod-lengths the one that seemed to represent the signal duration. Any feedback was eliminated where possible. Observers and in particular those who participated in the 2nd B series of experiments had no preliminary information about the nature of the connections between the elements of presented conjugate pairs. In the process of the experiment they did not know if their responses were correct or not. Observers could not possibly be aware of the fact that a certain signal was presented to them the first time or the second.

One of the principal measuring characteristics of the experiment was the reaction time – the interval between the termination of the signal and the observer's response (the selection of certain lengths)

The number of presented intervals in the pilot experiment was 13. Therefore the total number of presentations in the pilot experiment was 13 x 2 x 3 = 78 (2 presentations and 3 types of durations: light, sound, and light-sounds).

The number of presented intervals in the primary experiment was 14. There were six presentations of the same duration signal (three times in the first presentation and three times in the second one). Therefore, the total number of presentations was 14 x 2 x 3 = 84.

In the course of the primary experiment, two length-panels with different sets of rods were presented:

Each length - panel was presented twice, i.e. total of the number of the length-panel presentation was 2 x 2 = 4.

The number of time interval presentations that related to the one length-panel presentation was equal to 84/4 = 21.This set of time interval presentations was termed as a "**signal presentation subset"** per one length-panel, or **"subset".**

The time for all 84 presentations was about 40 to 50 minutes without any breaks. (The reverse procedure, namely in which the length was chosen by the experimenter first and would respond with particular signal duration, will be reported in a subsequent paper).

The experimental data permitted the analysis of (1) all the presentations (without dividing them into the subsets), and (2) the separate subsets of presentations providing information about the recognition dynamics.

This information makes it possible to find out:

a) The dependence of correct responses on the type of presentation, whether it was the first presentation or the second one, whether these presentations belong to the first subset or a different one and so on.

b) The dependence of correct responses to the given presentation on all the previous presentations.

Such kind of the dependence can be expressed first of all, in the changing number of correct responses from subset to subset.

In the pilot experiment the set of presentations was not divided into such subsets.

**Data Analysis**

As mentioned above, signals of a certain duration and modality were presented twice in random order during both the pilot and the primary experiments. (see the section "Specifications of stimuli presentation to observers ”). Hence, there were two presentations of the same time signal: the **first presentation and the second presentation.** In this work for analysis purposes, the data of all types of conducted experiments were divided into two parts: data which were obtained during the 1st time signal presentations and the data which were obtained during 2nd time signal presentations. Each part of these data was examined separately. As will be shown below (See Tables 6 (a, b), 7 (a, b), during the 1st time signal presentation the quantity of correct responses and the level of conjugate pairs recognition were considerably less than those observed during the 2nd time signal presentation where this quantity and the level of conjugate pairs recognition were more than doubled. It was observed in all experiments with the exception of the subseries 1a of the primary experiment which was conducted under lower uncertainty conditions. But it may seem obvious and not worthy of discussion since it tacitly suggests that the 1st presentation of the signals is the learning or training phase of any experiment and that this phase can be neglected. However, during the first time signal presentation, conditions for the successful recognition of conjugate pairs during the 2nd time signal presentation were founded.[3] Because of that, the study of the perceptual processes and phenomena taking place during the first signal presentation was very significant for understanding (comprehending) a mechanism of correlation between perceived lengths and durations.

---

[3]Some can assert that equivalence between a temporal and spatial extent, a length duration reported by observers is results of two factors: first, the sets of stimuli were preselected, and second, the "correct" correspondence was prejudged. In connection with this one should remind that any feedback in these experiments was excluded. Observers and in particular those who participated in the 2nd B series of experiments had no any preliminary information about the nature of the connections between the elements of presented conjugate pairs. In the process of experiment they did not know were their responses correct or not. During the 1st signals presentation the absolute majority of responses of these observers were incorrect. Therefore an assertion that the observers gave correct responses, since the sets of stimuli were preselected and correct correspondence was prejudge is groundless.

In connection with this, the next questions can arise:

1) What function did the first presentation of signals carry out in processes of the correlation of perceived space and time intervals in a so-called learning phase in this case?

2) What actions did the perceptive system of observers perform during the first signals presentation as a result of which observers acquired a possibility to give correct responses to these signals during the second their presentation?

Answering to this question one can make at least two suggestions. 1. This phase is a base for a simple repetition of presented signals. 2. This phase is the process of acquisition of specific information from presented stimuli and creation of temporary mental structures based on this information whereby an observer can find a proper correlation between space and time intervals designated in the experiment and gives correct responses. The second suggestion was supported by the comparative analysis of experimental facts and phenomena which will be stated below.

**Kinds and Nature of Observers Responses**

Responses of observers, or recognition of conjugate pairs, were divided into three groups: **Correct response** - when an observer responded to a time signal and chose a space interval which when, put together, corresponded to a specific conjugate pair. Thus, it was as if he struck the center of a target scoring a “bull’s eye”.

For instance, when the observer responded to the time signal of 1.8 seconds and chose the length of 3 centimeters as shown in tables 2, 2 a and 2 b.

**An approximate correct response** - when an observer responded to a specific time signal but did not choose the correct space interval. However, when taken together, the results were a conjugate pair but this pair was a neighboring conjugate pair. This was as if he did not strike the center of a target, but at its closest circle.

Here, one can also mark two kinds of approximate correct responses.

The first was when an observer's choice of an interval was placed **immediately** on the next point of the scale before or after a point where a correct interval was located. This concerned those parts of a partly complete scale where conjugate pairs were placed **uniformly** step by step.

For instance, when responding to the time signal of 1.8 seconds, an observer chose a length of 2.5 or 3.5 centimeters instead of the correct length of 3.0 centimeters (see tables 2, 2 a, 2 b).

The second kind of approximately correct response was when an observer's choice of an interval was placed on the scale at a range of one or two steps before or after the correct interval location.

This concerned those parts of the partly completed scale, where conjugate pairs were placed **non – uniformly** between the closest neighboring presented stimuli. They could be part of several "unfilled" positions of the scale by presenting stimuli points on the assumed complete scale of possible stimuli of the primary experiment.

For instance, in responding to the time signal of 5.4 seconds, an observer chose a length of 7.5 to 10 centimeters instead of the correct length of 9 centimeters. (See and compare tables 2, 2 a, 2 b).

**Incorrect responses** - when an observer responded to a specific time signal and did not choose an even close to correct space interval. This signal was not a conjugate pair on any of the neighboring scales. For instance, in responding to the time signal of 1.8 seconds, if an observer chose neither the correct length of 3 centimeters nor the approximately correct lengths of 2.5 or 3.5 centimeters but lengths of 2 or less or 4 or more centimeters (see and compare tables 2, 2 a, 2 b). It was as if the target was missed completely.

Incorrect responses can be divided into two groups

1. **Overestimated incorrect responses** – when an observer responding to a specific time signal chose a longer interval than the correct one. For instance, an observer responding to the time signal of 1.8 sec chose the space interval of 4 centimeter instead of the interval of 3 cm, which when put together with this time signal made up the conjugate pair 1.8 sec – 3 cm.
2. **Underestimated incorrect responses** – when an observer responding to a specific time signal chose a shorter interval than correct one. For instance, an observer responding to the time signal of 1.8 sec chose the space interval of 2 centimeters instead of an interval of 3 cm, which when put together with this time signal made up the conjugate pair 1.8 sec – 3 cm.

The overestimated incorrect responses were located in the area adjacent to the maximum limit of the range, and showed the observer's tendency to approach this limit. In turn, the underestimated incorrect responses were located in the area adjacent to the minimum limit of the range and showed the observer's tendency to approach this limit.

One of the principal measured characteristics of the experiment was reaction time – the interval between the termination of the signal and the observer's response (the selection of certain lengths). This characteristic was divided into:

**Correct decision time** was a reaction time (latency) of the correct response.

**Incorrect decision time** was a reaction time (latency) of the incorrect response.

**Average decision time** was a reaction time including reaction time of both correct and incorrect responses to a specific time signal.

**Recognizability Levels and Strengths of Conjugate Pairs**

Different conjugate pairs placed on different points of the scale were recognized differently by observers in the first and the second presentations. One can distinguish three forms or three levels of this recognition.

**Absolute correct recognition -** were among the responses given to the presentation of a specific signal. Correct responses amounted to upwards of fifty per cent and most of the remaining responses were the approximately correct ones.

**Relatively correct recognition -** were among the responses given to the presentation of a specific signal. Correct and approximately correct responses together amounted to upwards of fifty per cent.

**Absence of recognition -** were among the responses given to the presentation of a certain signal. Correct and approximately correct responses together were less than fifty per cent.

As it appears in the above text, the level of pairs recognition depended on the quantity of correct responses to the presented time signals of different durations.

This quantity could be considered as an index of **the conjugate pair recognition strength**. The more correct responses to a given signal, the more recognition strength a certain conjugate pair had.

Based on the percentage of correct responses among all responses to a certain signal, one can classify conjugate pairs into the following:

**Strong pairs -** (SP) more than 50% of correct responses

**Relatively strong pairs -** (RSP) more than 40 % of correct responses close to the level of 50 %, and put together with approximately correct responses, more than 50 percent and higher.

**Weak pairs** - (WP) lower than 40% of correct responses and put together with approximately correct responses were 50% and higher.

**Weakest pairs** - (WtP) lower than 30 % of correct responses and put together with approximately correct responses, were less than 50 percent.

Strong, relative strong and weak pairs one can consider as recognized pairs in different degrees. Weakest pairs one could be considered as pairs which were not recognized.

Unlike the classification mentioned where both correct and approximately correct responses were considered, in the present classification only correct responses related to a certain conjugate pair scoring a "bull's eye" should be considered highly useful, especially for the analysis of the dynamics of conjugate pairs recognition.

## IV. RESULTS OF THE EXPERIMENT

### The results of the main part of the pilot and all series of the primary experiment

#### The quantity of correct and approximately correct responses among responses to all signal presentations as a whole in the pilot and primary experiments

**Table 4**

The comparison of the pilot experiment and three series and one subseries of the primary experiment by the quantity of correct and approximately correct responses (% of the total number of all responses during the 1st and 2nd time signal presentations)

| Experimental series | Correct responses | **Correct + approxima tely correct responses** |
|---|---|---|
| Pilot experiment | **30.8** | 61 |
| 1st series | **47.6** | 73.6 |
| The subseries 1st a | **55.8** | 80.2 |
| 2nd A series | **52** | 76.2 |
| 2nd B series | **43.9** | 66.2 |

An analysis of the data showed that a majority of observers who participated in the main parts of both the pilot and primary experiments gave correct or approximately correct responses to most of the presented time signals.

This phenomenon evidences a capability of the human psyche to extract from stimuli of any modality, their spatial and temporal characteristics, and correlate values of these characteristics with each other in a certain proportional way.

But there were differences between the data of the pilot experiment, the primary experiment as a whole, and between data of the three series of the primary experiments themselves. As one can see in Table 4, the largest quantity of correct and approximately correct responses can be observed in subseries 1 A where the experiment was performed under lower uncertainty condition and the smallest quantity in the pilot experiment.

It should be noted that in the pilot experiment which was performed under middle uncertainty condition the quantity of correct responses was considerably less than in the 2nd B series of the primary experiment performed under high uncertainty conditions. This and other differences between the data of the pilot experiment and the entire series of the primary experiment became more apparent upon examination of an observer's response dynamics.

### The observer responses dynamics

**The difference in the quantity of correct and approximately correct responses to the 1st and 2nd time signal presentations in the pilot and in all series of the primary experiment**

**Table 5**

The comparison of the pilot experiment and the series of the primary experiment by the quantity of correct and approximately correct responses. (% of the number of responses to signals of their first and second presentations)

| Experimental Series | Correct responses | | **Correct + approximately correct responses** | |
|---|---|---|---|---|
| | **1st presentation** | **2nd presentation** | **1st presentation** | **2nd presentation** |
| Pilot experiment | **26.3** | **35.7** | 56.7 | 66 |
| Primary 1st series | **42.4** | **52.8** | 67 | 79.9 |
| Primary 1st A subseries | **52** | **59.5** | 78.7 | 81.7 |
| Primary 2nd A series | **46.4** | **57.6** | 69.9 | 82.6 |
| Primary 2nd B series | **38.7** | **52.4** | 58.1 | 76.6 |

As can be seen from Table 5, the quantity of correct and approximately correct responses in the first presentation was the highest in the subseries 1st a of the primary experiment and the least in the pilot experiment. Moreover, the quantity of correct responses only in the subseries 1sta was already predominant in the first presentation in comparison with rest of the series of the primary experiment and especially of the pilot experiment.

The quantity of correct and approximately correct responses increased in the second presentation, both in the pilot and three series of the primary experiment. However, in all series of primary experiments the quantity of correct responses become predominant, more than 50% from all observer responses, whereas in the pilot experiment this quantity was at a much lower level.

**Levels of pairs recognizability**

As previously mentioned, different conjugate pairs were recognized differently by the observers. There were three levels of these pairs of recognition: absolute correct, relatively correct, and the absence of any recognition. (See in detail above)

Table 6

The comparison of the pilot and all series of primary experiment by levels of the conjugate pair recognition. (% of the number of conjugate pairs placed on the scale of presented stimuli)

Table 6 a

1st presentation

| **Level of recognition** | **Pilot Experiment** | Primary experiment | | | |
|---|---|---|---|---|---|
| | | **1st series** | Subseries 1 a | **2nd A series** | **2nd B series** |
| **Absolute recognition** | 0 | 28.6 | 57.1 | 42.8 | 21.4 |
| **Relative recognition** | 61.5 | 57.1 | 42.8 | 50 | 42.8 |
| **Absence of recognition** | 38.5 | 14.3 | 0 | 7.1 | 35.7 |

Table 6 b

2nd presentation

| Level of recognition | Pilot Experiment | Primary experiment | | | |
|---|---|---|---|---|---|
| | | **1st series** | Subseries 1 a | **2nd A series** | **2nd B series** |
| **Absolute recognition** | 15.4 | 57.1 | 64.3 | 85.7 | 50 |
| **Relative recognition** | 69.2 | 42.8 | 35.7 | 14.3 | 50 |
| **Absence of recognition** | 15.4 | 0 | 0 | 0 | 0 |

As can be seen from Tables 6 a and 6 b, the ratio between these levels was different and changed differently from the first time signal presentation to the second one under different experimental conditions.

In the first signals presentation, with exception of the subseries 1 a, there were cases with absent recognition in all series of the experiment and especially in the pilot and in the 2 B series of the primary experiment.

As to cases with absolute correct recognitions, their different quantities could be observed in all series of the primary experiment, most of them in the subseries 1a where they became predominant, and the 2nd A series as well, but there were none in the pilot experiment. In the second signal presentation, there were no cases of absent recognitions in the entire series of the primary experiments, but absent recognitions continued to remain in the pilot experiment.

As to cases with absolute correct recognition, in the second presentation, they were sharply increased and became nearly predominant in the entire series of the primary experiments, especially in the 2 A series.

Whereas in the pilot experiments, absolute correct recognition cases rose slightly and referred only to the first two on the scale of conjugate pairs i.e., 1 sec/cm and 2 sec/cm.

**Dynamics of the recognizability of different conjugate pairs which were placed on the scales of presented stimuli in different experimental conditions during two time signals presentations**

One can ask, what underlies such a difference in the recognizability of different conjugate pairs placed on the same scale of presented stimuli? The following are four possible reasons:

1. Levels of uncertainty in conditions of the experiment performance (lower, low, middle and high uncertainty conditions, see Table 3)
2. Peculiarities of the ratio of certain conjugate pair elements (the one – to – one direct proportionality principle and the direct proportionality principle only, see above*)
3. Structure of the scale of presented stimuli (complete and partly complete scales, see above*)
4. A conjugate pair location on this scale.

The first and the second of these problems will be considered in this section. The third and fourth problems will be considered in another section.
For the analysis of this problem one can use the classification of conjugate pairs by their recognizable strength. As mentioned above, conjugate pairs were divided into strong (SP), relatively strong (RSP), weak (WP), and weakest (WtP) pairs (see above in detail*).

Table 7

The changing of the conjugate pair recognition strength during time signals presentations (% of the number of conjugate pairs placed on the scale of presented stimuli).

Table 7 a

1st presentation

| Experimental Conditions | Levels of a conjugate pair recognition strength | | | |
|---|---|---|---|---|
| | Strong pairs | Relatively strong pairs | Weak pairs | Weakest pairs |
| Pilot experiment | 0 | 15.4 | 46.1 | 38.5 |
| 1st series of primary exp | 28.6 | 7.1 | 50 | 14.3 |
| Subseries 1a of primary exp | 57.1 | 21.4 | 21.4 | 0 |
| 2nd A series of primary exp | 42.9 | 7.1 | 42.9 | 7.1 |
| 2nd B series of primary exp | 21.4 | 21.4 | 21.4 | 35.7 |

Table 7 b

2nd presentation

| Experimental Conditions | Levels of a conjugate pair recognition strength | | | |
|---|---|---|---|---|
| | Strong pairs | Relatively strong pairs | Weak pairs | Weakest pairs |
| Pilot experiment | 15.4 | 30.7 | 38.5 | 15.4 |
| 1st series of primary exp | 57.1 | 35.7 | 7.1 | 0 |
| Subseries 1 a of primary exp | 64.3 | 21.4 | 15.4 | 0 |
| 2nd A series of primary exp | 85.7 | 0 | 14.3 | 0 |
| 2nd B series of primary exp | 50 | 35.7 | 14.3 | 0 |

As can be seen from Tables 7 a and 7 b, the pilot experiment and all series of the primary experiment differed sharply from each other by the quantity of strong, relatively strong, weak, and weakest pairs during both the first and the second time signal presentations. During the first time signal presentation in the primary experiment, the quantity of strong and relatively strong conjugate pairs strictly corresponded to the level of uncertainty conditions of each series of this experiment. The lower the level of uncertainty the higher the quantity of strong conjugate pairs and vice versa. The greatest quantity of strong pairs was observed and became predominant at once in the subseries 1a of the primary experiment which was conducted under lower uncertainty conditions. In this subseries there were no weakest pairs unlike another series of the primary experiment where these pairs were more or less observed especially in the 2nd B series which was conducted under high uncertainty conditions.

The quantity of strong conjugate pairs observed during the first presentation of time signals in the 2nd A series of the primary experiment which was conducted under the low uncertainty condition, ranked below such quantity observed in the subseries 1a but was not predominant.

It should be noted that during the first presentation of time signals the strong conjugate pairs were observed in all series of the primary experiment even the 2nd B series which was conducted under the high uncertainty condition.

As to the pilot experiment, unlike any series of the primary experiment, there were no strong conjugate pairs, but the highest quantity of weakest pairs were observed during the first presentation of time signals.

The difference between the pilot experiment and all series of the primary experiment in the level of conjugate pair recognition strength became more noticeable and significant during the second time signals presentation.

As can be seen from the table 7 b, this strength increased and turned from the lower level into the higher level differently depending on the experimental conditions. One can observe the next differences in the level of conjugate pair recognition strength between the pilot and all series of the primary experiments during the second time signals presentation:

1. The quantity of strong conjugate pairs became predominant or thereabout in all series of the primary experiment whereas in the pilot experiment this quantity was very low.
2. In all series of the primary experiment no weakest pairs were observed, whereas in the pilot experiment a certain quantity of these pairs remained.

The following is noteworthy. During the first time signal presentation the quantity of strong conjugate pairs observed in the subseries 1a was considerably higher than in the 2nd A series. During the second time signals presentation this quantity was just the opposite in the 2nd A series and was considerably higher than in the subseries 1a which was conducted under a lower uncertainty condition. This phenomenon of the conjugate pair recognition dynamics can be useful in studying how a specific temporary scale of possible stimuli was formed, and how the recognition of conjugate pairs takes place.

**The dependence of the conjugate pair recognition on its position on the scale of presented stimuli**

In numerous studies on memorization since H. Ebbinghaus, it was shown that ordinarily items toward the beginning and end of the series are easier to learn then those in the center. (e.g., Robinson, E.S., and M. A. Brown (1924) Later, this phenomenon was observed in absolute judgment or absolute identification processes in different perception conditions (e.g., Berliner, J.E.., Durlach, N.I., & Braida, L. D. (1977), Lacouture , Y. (1997). In these studies it was called the "edge" or "bow" effect. According to these studies, the "bow" effect consisted of performance deterioration as the stimulus set size increased with stimuli located towards the ends of the stimulus range. These were identified with greater accuracy than those located towards the middle of the range (Lacouture , Y. (1997) .

In accordance with this "bow" effect description, the disposition of conjugate pairs along the scale of presented stimuli by their level of strength should be as follows: 1. Strong pairs (**SP**) should occupy the first and the last places on the scale, i.e. its minimal and maximal limits (1st,2nd, and 14th 13th places). Relatively strong pairs (**RSP**) – 3rd , 4th and 12th , 14th places, Weak pairs (**WP**) – 5th , 6th , 9th and 10th places, and Weakest pairs (WtP) should occupy the middle of the scale, in this case 7th and 8th places (See Table 8).

Table 8

| Numbers of places on the presented scale | | | | | | | | | | | | | |
|---|---|---|---|---|---|---|---|---|---|---|---|---|---|
| 1 | 2 | 3 | 4 | 5 | 6 | 7 | 8 | 9 | 10 | 11 | 12 | 13 | 14 |
| **SP** | **SP** | **RSP** | **RSP** | **WP** | **WP** | **WtP** | **WtP** | **WP** | **WP** | **RSP** | **RSP** | **SP** | **SP** |

Legends:

**Strong pairs** (more than 50% of "correct" responses) - **SP**
**Relatively strong pairs** (more than 40 % close to the level at 50 % of "correct" responses and putting together with "approximately correct "responses more than 50 and higher percents) – **RSP**
**Weak pairs** (lower than 40% of "correct" responses and putting together with "approximately correct" responses more than 50 and higher percents as well) – **WP**
**Weakest pairs** (lower than 30 % of "correct" responses and not putting together with "approximately correct "responses less than 50 percents) – **WtP**

In this connection the question arises, if the "bow" effect is general for all perception conditions then the structure of correct responses distribution depends only on the stimulus set size. To answer to this question a comparative analysis was conducted which showed conjugate pairs distribution by their strength along the scale of presented stimuli of the pilot and all series of the primary experiments.

The goal of this analysis was to find out if there was a difference in such a distribution, as described in Table 8, between the pilot experiment and all series of the primary experiment which differed from each other by the structure of the presented stimuli scale and the level of uncertainty there in.

The data of conjugate pairs distribution by their strength along the scale are represented symbolically by letters in Tables 8 a, b, ba, c, d and were compared with an "ideal sample" of "bow" effect represented by Table 8.

Table 8 a

Pilot experiment

The strength of presented conjugate pairs along the complete scale

| | | Numbers of places on the complete scale | | | | | | | | | | | | |
|---|---|---|---|---|---|---|---|---|---|---|---|---|---|---|
| Stimulus categories | | 1 | 2 | 3 | 4 | 5 | 6 | 7 | 8 | 9 | 10 | 11 | 12 | 13 |
| Duration | Seconds | 1 | 2 | 3 | 4 | 5 | 6 | 7 | 8 | 9 | 10 | 11 | 12 | 13 |
| Length | Centimeters | 1 | 2 | 3 | 4 | 5 | 6 | 7 | 8 | 9 | 10 | 11 | 12 | 13 |
| 1st presentation | | **RSP** | **RSP** | **WP** | **WtP** | **WtP** | **WP** | **WtP** | **WtP** | **WtP** | **WP** | **WP** | **WP** | **WP** |
| 2nd presentation | | **SP** | **SP** | **RSP** | **WP** | **WP** | **WP** | **WtP** | **WP** | **WtP** | **WP** | **RSP** | **RSP** | **RSP** |

In the pilot experiment (see Table 8 a), during the 1st time signal presentation, no strong pairs were placed at any point of the scale even at its limits. Moreover, the maximum limit of this scale was occupied by a weak pair as a pair placed in the middle of the presented 13 stimuli scale. The following is noteworthy because this weak pair occupied 6th place on the scale, and was between the two weakest pairs. It was as if observers detected three points on the scale and then recognized a conjugate pair only in the middle of them. It should be recalled that weak pairs were connected with approximately correct responses, whereas weakest pairs connected with incorrect ones, i.e. they are not related to any conjugate pair fields. During the 2nd time signal presentation, only two strong pairs appeared on the 1st and 2nd points of the scale and four relatively strong pairs which appeared mainly on its last three points. Two weakest pairs were on 7th and 9th points. As stated above, the scale of the pilot experiment was complete and built on the one – to – one principle. Its size consisted of 13 points (see page 8 and Table 1 for details).

Unlike the pilot experiment, scales of presented stimuli of all series of the primary experiment were partly complete and were not built on the one –

to – one principle, and consisted of 14 points. As mentioned above, there were two types of partly completed scales. The first type was used in the first series and in the subseries 1st a, and the second type was used in the 2nd A and B series of the primary experiment (see pp.11 and Tables 2 a, 2 b, and 3 for details).

Unlike the pilot experiment, in all series of the primary experiment, the location of any conjugate pair can be considered and determined on two scales: 1. on its partly complete scale of presented stimuli, and 2. the assumed complete scale of possible stimuli. Such a consideration of the conjugate pairs distribution along a real partly complete scale and an assumed complete scale in our opinion, can be useful for both the analysis of experimental data and the understanding of the mechanism of correlation between perceiving lengths and durations.

Data from Tables 8 b, 8 b (a), 8 c, and 8 d showed the next peculiarities and differences between the series of the primary experiment in the distribution of conjugate pairs by their strength along scales during the first and the second time signal presentations. During the 1st time signals presentation in the 1st, 2nd A series, and especially in the 2nd B series of the primary experiment, the quantity of strong and relatively strong pairs was observed more in the region of the non-uniform part of the scale adjoined to its maximum limit than in the region of the uniform part of the scale. Unlike these series in subseries 1a, the quantity of strong and relatively strong pairs was observed more in the region of the uniform part of the scale which was adjoined to its minimum limit than in the region of the non-uniform part of scale adjoining the maximum limit. (see table 8 e)

**Table 8 e**

Quantity of strong and relative
strong conjugate pairs during the 1st presentation

| Series of the primary experiment | Parts of the scale | | The total strong and relative strong pairs |
|---|---|---|---|
| | Uniform | Non-uniform | |
| 1st series | **3** | **4** | **7** |
| 2nd A series | **3** | **4** | **7** |
| 2nd B series | **2** | **4** | **6** |
| Subseries 1 a | **6** | **5** | **11** |

In the subseries 1a, during the first time signal presentation, the maximum limit conjugate pair was relatively strong. The quantity of correct responses among responses given to the presentation of the longest time signal (maximum limit) was less than 50% whereas (see table 8 b(a) in 1st, 2nd A, and especially 2nd B series these pairs were strong. The quantity of correct responses to this signal was more than 50%. (see tables 8 b, 8 c, and 8 d). Unlike the subseries1a, in the 1st series, and 2nd A series, and especially in the 2nd B series during the 1st time signals presentation, most of the strong and relatively strong conjugate pairs were observed in the region of the non-uniform part of the presented stimuli scale. The quantity of correct responses to the longest time signal (maximum limit) in the 2nd B series was considerably higher not only in the subseries 1a, but in the 1st series and 2nd A series as well (see table 9, on pp ).

The comparison of minimum and maximum limit pairs by their strength and quantity of correct responses to presented stimuli in different series of primary experiment during the 1st presentation is shown as follows:

Table 8 b

Subseries $1^{st}$ a

The strength of presented conjugate pairs along the partly complete scale

| | | Numbers of places on the complete scale of possible stimuli | | | | | | | | | | | | |
|---|---|---|---|---|---|---|---|---|---|---|---|---|---|---|
| Stimulus categories | | 1 | 2 | 3 | 4 | 5 | 6 | 7 | 8 | 9 | 10 | 11 | 12 | 13 |
| Duration | Seconds | 0.3 | 0.6 | 0.9 | 1.2 | 1.5 | **1.8** | 2.1 | 2.4 | | | 3.3 | 3.6 | |
| Length | Centimeters | 0.5 | 1 | 1.5 | 2 | **2.5** | **3** | **3.5** | 4 | | | 5.5 | 6 | |
| **Numbers of places on the presented scale** | | 1 | 2 | 3 | 4 | 5 | 6 | 7 | 8 | | | 9 | 10 | |
| $1^{st}$ presentation | | **SP** | **SP** | **SP** | **SP** | **WP** | **RSP** | **SP** | **SP** | | | **WP** | **WP** | |
| $2^{nd}$ presentation | | **SP** | **SP** | **SP** | **SP** | **RSP** | **SP** | **WP** | **SP** | | | **SP** | **WP** | |

| Continuation | | Numbers of places on the complete scale | | | | | | | |
|---|---|---|---|---|---|---|---|---|---|
| Stimulus categories | | 14 | 15 | 16 | 17 | 18 | 19 | 20 | 21 |
| Duration | Seconds | | 4.5 | | | **5.4** | | 6 | 6.3 |
| Length | Centimeters | | **7.5** | | | **9** | | **10** | 10.5 |
| **Numbers of places on the presented scale** | | | 11 | | | 12 | | 13 | 14 |
| $1^{st}$ presentation | | | **RSP** | | | **SP** | | **SP** | **RSP** |
| $2^{nd}$ presentation | | | **RSP** | | | **RSP** | | **SP** | **SP** |

In the subseries 1 a during the first time signal presentation the maximum limit conjugate pair was relatively strong, i.e. the quantity of correct responses among responses given to the presentation of the longest time signal (maximum limit) was less than 50% whereas conjugate pairs located on the middle of both assumed and presented stimuli scales (places 7 and 8 ) were strong. The quantity of correct responses to time signals related to these pairs was more than 50 %. But according to the “ideal bow effect” distribution the maximum limit pair is strong and conjugate pairs located on the middle of this scale were weakest, i.e. the quantity of correct and approximately correct responses together were less than 50%. The minimum limit pair was stronger and the quantity of correct responses to the presented corresponding time signal was considerably higher than the maximum limit pair, (see tables 8 b (a), 9, and the table 2 (1) from the appendix 1).
The quantity of differences between strengths of the same conjugate pairs located in the same order along the "ideal bow effect" distribution scale and the scale of presented stimuli observed in the subseries 1a was 7 during both the 1st and the 2nd presentations (See table 1 from the appendix 2).
The main peculiarity of mentioned differences was that the all conjugate pairs excepting maximum limit pair during the 1st time signals presentation in subseries 1 a were considerably stronger than the same conjugate pairs placed in the same order on the "ideal bow effect" distribution scale, especially on the middle part of this scale. At the same time the maximum limit pair from subseries 1 a was less strong than on the "ideal bow effect" distribution scale in spite of the fact that observers were preliminarily acquainted with both ends of the subseries 1 a scale. The maximum limit pair along with two conjugate pairs placed inside the scale in subseries 1a became stronger only during the 2nd time signals presentation, whereas two other inside conjugate pairs became weaker. (See table 8 b (a) and table 1 from appendix 2).
Thus, the "ideal bow effect" distribution of correct responses did not take place in the subseries 1a of the primary experiment which was conducted under lower uncertainty condition.

Table 8 c

The 1st series

The strength of presented conjugate pairs along the partly complete scale

| | | Numbers of places on the complete scale of possible stimuli | | | | | | | | | | | | |
|---|---|---|---|---|---|---|---|---|---|---|---|---|---|---|
| Stimulus categories | | 1 | 2 | 3 | 4 | 5 | 6 | 7 | 8 | 9 | 10 | 11 | 12 | 13 |
| Duration | Seconds | 0.3 | 0.6 | 0.9 | 1.2 | 1.5 | **1.8** | 2.1 | 2.4 | | | 3.3 | 3.6 | |
| Length | Centimeters | 0.5 | 1 | 1.5 | 2 | **2.5** | **3** | **3.5** | 4 | | | 5.5 | 6 | |
| **Numbers of places on the presented scale** | | 1 | 2 | 3 | 4 | 5 | 6 | 7 | 8 | | | 9 | 10 | |
| 1st presentation | | **SP** | **SP** | **WP** | **RSP** | **WtP** | **WP** | **WP** | **WP** | | | **WtP** | **WP** | |
| 2nd presentation | | **SP** | **SP** | **SP** | **SP** | **RSP** | **WP** | **RSP** | **RSP** | | | **RSP** | **RSP** | |

| Continuation | | Numbers of places on the complete scale | | | | | | | | |
|---|---|---|---|---|---|---|---|---|---|---|
| Stimulus categories | | 14 | 15 | 16 | 17 | 18 | 19 | 20 | 21 | |
| Duration | Seconds | | 4.5 | | | **5.4** | | 6 | 6.3 | |
| Length | Centimeters | | **7.5** | | | **9** | | **10** | 10.5 | |
| **Numbers of places on the presented scale** | | | 11 | | | 12 | | 13 | 14 | |
| 1st presentation | | | **RSP** | | | **RSP** | | **SP** | **SP** | |
| 2nd presentation | | | **RSP** | | | **SP** | | **SP** | **SP** | |

Unlike subseries 1a , the distribution of conjugate pairs along the scale of presented stimuli in the 1st series of the primary experiment resembles the "ideal bow effect" distribution of correct responses but does not coincide completely. In the 1st series during the 1st time signal presentation, both minimum and maximum limit pairs were strong and there were no significant differences between them in the quantity of correct responses to presented shortest and longest time signals ($t = 1.6$, $p > .5$). See tables 8 b, 9, and the table 3 from the appendix 1 and table 2 from appendix 2. But there was a difference in the structure of this distribution between 1st series of the primary experiment and the “ideal bow effect” distribution of correct responses. On the middle of the scale on the 7th and 8th places, where according to the “ideal bow effect” distribution, weakest pairs should have occurred, however more weak pairs were observed in the 1st series of the primary experiment.

Weakest pairs were observed in this series on the 5th and 9th positions of the scale of presented stimuli whereas according to the “ideal bow effect” weak pairs have to be placed on these positions. Thus, in the 1st series of the primary experiment during the 1st time signals presentation on the middle part of the scale, a region of conjugate pairs relative correct recognition was formed on both sides which there were points where conjugate pairs were totally unrecognizable. Inside this region the recognizability of each conjugate pair was different. The most correct responses were given to the time signal of 2.4 seconds (the conjugate pair 2.4 sec – 4 cm). This conjugate pair located on the 8th place of both the scale of presented stimuli consisted of 14 points, and the assumed scale of possible stimuli consisted of 21 points. The quantity of correct responses to the time signal of 2.4 seconds exceeded the quantity of such responses to both neighboring time signals of 2.1 sec ($t = 2$, $p<.5$) and in 3.3 sec ($t = 2.7$, $p<.1$). (see Table 3 from appendix 1). On this scale the time signal of 2.1 seconds was the next step towards the minimum limit pair and belonged to the uniform part of the scale. The time signal of 3.3 seconds joined to the time signal of 2.4 seconds was indirectly two steps toward the maximum limit pair and belonged to the non-uniform part of the scale. Between the time signals of 2.4 seconds and 3.3 seconds the time signals of 2.7 and 3.0 seconds could have been placed. In this connection, two questions arise:

1. Did the absence of these signals among presented stimuli have an influence on neighboring pairs recognition? 2. If an observer during time signal presentations could detect, perhaps unconsciously, the interruption of a sequence of presented stimuli, could observers organize his or her strategy of corresponding conjugate pairs search and recognition? (it should be mentioned that all time signals were presented in random order).

These issues will be discussed below in further analysis of the experimental data. One subject of discussion will be the conjugate pair 0.9 sec – 1.5 cm. This pair occupied the 3rd space on the scale of presented stimuli in the 1st series as on the assumed scale of possible stimuli and the 1st space on the scales of presented stimuli in 2nd A and B series where it served as a minimum limit pair. As one can see from Table 2 (appendix 2) on the scale of the "ideal bow effect" distribution the pair 0.9 sec – 1.5 cm was relatively strong whereas on the scale of presented stimuli in the 1st series, it was weak. In the 1st series during the 2nd time signal presentation, the quantity of correct responses and the corresponding conjugate pairs strength were considerably increased. In the 1st series during the 2nd time signal presentation, as compared with the 1st time signal presentation, the quantity of correct responses was sharply decreased to the presented maximum limit time signal ($t = 2.8$, $p < .01$), and sharply increased to the presented minimum limit pair ($t = 2.9$, $p < .01$). As a result, the level of recognizability of the minimum limit pair became significantly higher ($t = 3.4$, $p < .001$) than the level of recognizability of the maximum limit as shown on Table 9.

In addition, two peculiarities of the conjugate pairs recognition dynamics in the 1st series should be mentioned. 1). The strength of the conjugate pair 1.8 sec – 3 cm located in the 6th place on the scale of the 1st series which was weak during the $1^{st}$ time signal presentation and remained weak during the $2^{nd}$ presentation as the strength of the same pair which was located in the same place on the scale of “ideal bow effect” distribution (see Table 2 from appendix 2). 2). The aforementioned conjugate pair 2.4 sec – 4 cm which was dominated by the quantity of correct responses of the neighboring conjugate pairs placed on the middle part of the scale, lost its dominants during the $2^{nd}$ time signal presentation, although the quantity of correct responses to this signal along with neighboring ones sharply increased. (See Table 3 from appendix 1). We will return to these phenomena in the analysis of experimental data in the sequel. Turning to Table 5, one can see that the quantity of correct responses given during the $2^{nd}$ time signal presentation in the $1^{st}$ series, 52.8% became practically equal to the quantity of such responses given during the $1^{st}$ time signal presentation 52% in the subseries 1 a that was conducted under lower uncertainty condition.

Table 8 d

The $2^{nd}$ A series

The strength of presented conjugate pairs along the partly complete scale

| | | Numbers of places on the complete scale of possible stimuli | | | | | | | | | | | | |
|---|---|---|---|---|---|---|---|---|---|---|---|---|---|---|
| Stimulus categories | | 1 | 2 | 3 | 4 | 5 | 6 | 7 | 8 | 9 | 10 | 11 | 12 | 13 |
| Duration | Seconds | | | 0.9 | 1.2 | 1.5 | **1.8** | 2.1 | 2.4 | 2.7 | | 3.3 | 3.6 | 3.9 |
| Length | Centimeters | | | 1.5 | 2 | 2.5 | **3** | 3.5 | 4 | 4.5 | | 5.5 | 6 | 6.5 |
| **Numbers of places on the presented scale** | | | | 1 | 2 | 3 | 4 | 5 | 6 | 7 | | 8 | 9 | 10 |
| $1^{st}$ presentation | | | | **SP** | **SP** | **RSP** | **WP** | **WP** | **WP** | **WP** | | **WtP** | **WP** | **WP** |
| $2^{nd}$ presentation | | | | **SP** | **SP** | **SP** | **WP** | **SP** | **SP** | **WP** | | **SP** | **SP** | **SP** |

The 2nd A series

| Continuation | | Numbers of places on the complete scale | | | | | | | | |
|---|---|---|---|---|---|---|---|---|---|---|
| Stimulus categories | | 14 | 15 | 16 | 17 | 18 | 19 | 20 | 21 | |
| Duration | Seconds | | 4.5 | | | 5.4 | | 6 | 6.3 | |
| Length | Centimeters | | 7.5 | | | 9 | | 10 | 10.5 | |
| **Numbers of places on the presented scale** | | | 11 | | | 12 | | 13 | 14 | |
| 1st presentation | | | **SP** | | | **SP** | | **SP** | **SP** | |
| 2nd presentation | | | **SP** | | | **SP** | | **SP** | **SP** | |

As mentioned above, the uniform portion of the partly complete scale in the 2nd A and B series of the primary experiment began with the third point of the assumed scale - the conjugate pair 0.9 sec – 1.5 cm. which served in these series as the minimum limit of their scale. Comparing the distribution of recognized conjugate pairs along the scale in the 2nd A series with the "ideal bow effect" distribution, one finds the following. One can see from Table 3 in appendix 2, that during the 1st time signal presentation the strength of most conjugate pairs (ten of fourteen) which were placed on the scale of presented stimuli in the 2nd A series, completely coincided with the strength of the same pairs located in the same places on the ideal" "bow effect" distribution scale. As in the ideal "bow effect" distribution, in the 2nd A series, both minimum and maximum limit pairs were strong and the conjugate pair which was located in the 8th place of the middle of the scale of presented stimuli was the weakest. Unlike the 1st series, where the quantity of correct responses to the time signal 2.4 seconds exceeded the quantity of such responses to both neighboring time signals in the 2nd A series, the quantity of correct responses to this time signal became slightly dominant over only one neighboring time signal of 2.7 sec. This was the next step towards the maximum limit pair and belonged to the uniform part of the scale. (See Table 4(1) from appendix 1). But the following is noteworthy. As mentioned above in the 2 A series, an observer was preliminarily shown a connection between the minimum (0.9 sec – 1.5 cm) and maximum (6.3 sec – 10.5 cm) limits of the presented temporal and spatial signals sets. However, the quantity of correct responses to the time signal connected with maximum limit pair exceeded the quantity of correct responses to the time signal connected with the minimum limit pair ($t = 2.1$, $p < .05$) (see tables 8 c, 9, and the table 4 (1) from appendix 1). The quantity of correct responses to time signals which were located on half of the scale, belonging to the maximum limit pair was considerably higher than the quantity of correct responses to time signals which were located on the other half of the scale belonging to the minimum limit pair. (See Table 4 (3) from appendix 1)

During the 2nd time signal presentation the complete resemblance between structures of the conjugate pairs distribution along scales of the ideal "bow effect" and the 2nd A series completely disappeared as did the "bow effect" itself. As one can see from Table 9, during this presentation the quantity of correct responses to the maximum limit pair sharply decreased, and to the minimum limit pair sharply increased and became predominant over all other conjugate pairs placed on the scale of presented stimuli (see Table 4 (1) from appendix 1). The quantity of correct responses to time signals which were located on half of the scale belonged to maximum limit pair became equal to the quantity of correct responses to time signals which were located on half of the scale belonging to the minimum limit pair. (See Table 4 (3) from appendix 1). It is as if the correct responses distribution "bowed" at the beginning during the 1st presentation to the maximum limit pair and at the end during the 2nd presentation to the minimum limit pair. It is interesting that the quantity of correct responses to the time signal of 2.4 seconds which was located on the 6th place of presented stimuli scale in the 2nd A and B series, and on the 8th place of the assumed scale of possible stimuli, during the 2nd time signal presentation sharply increased and became predominant not only the quantity of such responses to time signals placed on the middle of the scale but also to the maximum limit signal (see Table 4(1) from appendix 1). These phenomena observed during the 1st and the 2nd time signal presentations will be discussed later. The data seem to hint at certain processes of correlation between perceiving lengths and durations.

Table 8 e

The 2nd B series

The strength of presented conjugate pairs along the partly complete scale

| | | Numbers of places on the complete scale of possible stimuli | | | | | | | | | | | | |
|---|---|---|---|---|---|---|---|---|---|---|---|---|---|---|
| Stimulus categories | | 1 | 2 | 3 | 4 | 5 | 6 | 7 | 8 | 9 | 10 | 11 | 12 | 13 |
| Duration | Seconds | | | 0.9 | 1.2 | 1.5 | **1.8** | 2.1 | 2.4 | 2.7 | | 3.3 | 3.6 | 3.9 |
| Length | Centimeters | | | 1.5 | 2 | 2.5 | **3** | 3.5 | 4 | 4.5 | | 5.5 | 6 | 6.5 |
| **Numbers of places on the presented scale** | | | | 1 | 2 | 3 | 4 | 5 | 6 | 7 | | 8 | 9 | 10 |
| 1st presentation | | | | **WP** | **RSP** | **WtP** | **WtP** | **WtP** | **RSP** | **WtP** | | **WtP** | **WP** | **WP** |
| 2nd presentation | | | | **SP** | **SP** | **SP** | **WP** | **WP** | **SP** | **RSP** | | **RSP** | **RSP** | **RSP** |

The 2nd B series

| Continuation | | Numbers of places on the complete scale | | | | | | | | |
|---|---|---|---|---|---|---|---|---|---|---|
| Stimulus categories | | 14 | 15 | 16 | 17 | 18 | 19 | 20 | 21 | |
| Duration | Seconds | | 4.5 | | | 5.4 | | 6 | 6.3 | |
| Length | Centimeters | | 7.5 | | | 9 | | 10 | 10.5 | |
| **Numbers of places on the presented scale** | | | 11 | | | 12 | | 13 | 14 | |
| 1st presentation | | | **RSP** | | | **SP** | | **SP** | **SP** | |
| 2nd presentation | | | **SP** | | | **RSP** | | **SP** | **SP** | |

Legends:

**Strong pairs** (more than 50% of “correct” responses) - **SP**

**Relatively strong pairs** (more than 40 % close to the level at 50 % of “correct” responses and putting together with “approximately correct “responses more than 50 and higher percents) – **RSP**

**Weak pairs** (lower than 40% of “correct” responses and putting together with “approximately correct“ responses more than 50 and higher percents as well) – **WP**

**Weakest pairs** (lower than 30 % of “correct” responses and not putting together with “approximately correct “responses less than 50 percents) – **WtP**

During the 1st time signals presentation, the structure of the conjugate pairs distribution by their strength along the scale of presented stimuli especially on the left half of this scale belonging to the minimum limit pair in the 2nd B series did not coincide with the structures of ”the ideal” “bow effect” and the 2nd A series of the primary experiment. This can be seen from (Tables 8 (d), 9, and Table 4 from appendix 2). According to ”the ideal bow effect” distribution, both maximum and minimum conjugate pairs must be strong, i.e. the absolute majority of responses to longest and shortest signals from the set signals must be correct. As can be seen from Tables 8 (d), 9, and Table 5(1) from appendix 1 and Table 4 from appendix 2, the minimum limit pair in the 2nd B series during the 1st time signal presentation was weak whereas the maximum limit pair was very strong. The quantity of correct responses to the longest time signal was more than double to that of the shortest signal. During the 1st time signal presentation, most of strong and relatively strong pairs (4 of 6) were observed on half of the scale which belonged to the maximum limit pair, whereas the absolute majority of weakest pairs (5 of 6) were observed on half of the scale which belonged to the minimum limit one. The absolute majority of correct responses (59.6%) was given to time signals which were located on this part of the scale as well (see Table 5 (3) from appendix 1).Thus, on half of the scale belonging to the minimum limit in the 2 B series of the primary experiment, which was conducted under a higher uncertainty condition, during the 1$^{st}$ time signal presentation the “bow” effect did not take place.

But the following is noteworthy. As can be seen from Table 8 (d) and from Table 4 from appendix 2, among the weakest pairs which were an absolute majority on this half of the scale, two relatively strong pairs were observed. One of them, the conjugate pair 1.2 sec – 2 cm was placed on the 2$^{nd}$ point of the scale of presented stimuli (the 4$^{th}$ place of the scale of possible stimuli) between the minimum limit pair 0.9 sec – 1.5 cm which was a weak pair and the weakest or not recognized pair 1.5 sec – 2.5 cm. The second relatively strong pair 2.4 sec – 4 cm was placed on the 6$^{th}$ point of the scale of presented stimuli (the 8$^{th}$ place of the scale of possible stimuli) between two weakest (not recognized) pairs 2.1 sec – 3.5 cm and 2.7 sec – 4.5 cm (see Table 8 (d), fig.2, and table 5(1) from appendix 1). As one can see, a gradient junction between conjugate pairs of different strength was not observed during the 1$^{st}$ time signals presentation in spite of the fact that these pairs were inside the uniform part of the scale. Both mentioned pairs, which were within three points of each other on the scale, where unrecognized weakest pairs were placed. The relatively strong pairs 2.4 sec – 4 cm was inside a region of weakest pairs dividing them into two parts belonged to the minimum and maximum limits of the scale. This phenomenon indicated that during the 1$^{st}$ time signals presentation, an observer made a limit inside of the scale, and detected locations of other conjugate pairs even though he or she did not recognize all of them. We called this phenomenon as **an outburst of recognition activity in the middle part of the scale - (ORAMPS phenomenon)**

SERIES 2B

1st Presentation

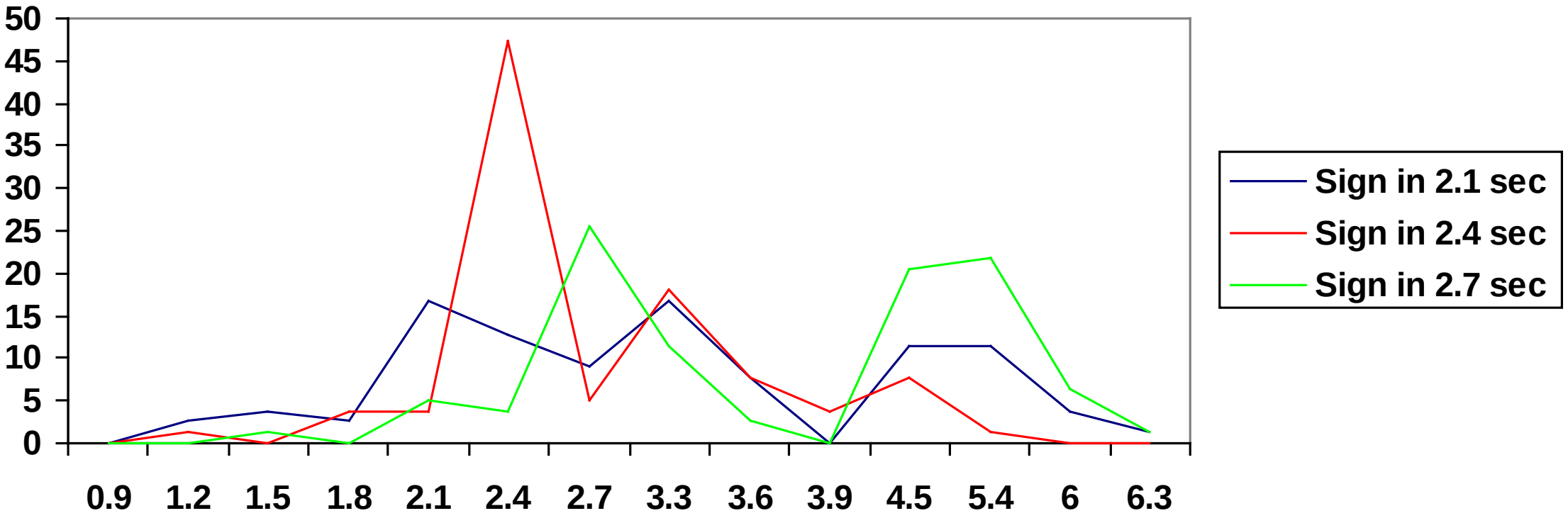

The Distribution of Responses to Signals in 2.1, 2.4, 2.7 seconds

Notice Only ones of the members of conjugate pairs – time intervals are shown on the abscissa

Space intervals are not shown for convenience Figure 2

In this connection a question arises – *what kind of scale does an observer divide into two parts? Is it the scale of presented stimuli or the mental full scale of possible stimuli, which he or she could construct while receiving time signals?* These questions will be considered below inasmuch as it is related to problems of correlation between perceiving lengths and durations. During the 2nd time signals presentation the level of conjugate pairs recognition sharply increased. As can be seen from Table 8 (6) and Table 4 from appendix 2, no weakest, or unrecognized pairs, remained and most of them became strong and relative strong pairs. During this presentation, as in the 1st series and 2ndA series, in the 2nd B series a redistribution of correct responses to maximum and minimum limit signals took place. The quantity of correct responses to the maximum limit signal decreased and to the minimum limit sharply increased and became equal to each other (See Table 9). Moreover, the quantities of correct responses to the signals which were placed on both halves of the scale belonging to the minimum and maximum limits became equal as well (see Table 5 (3) from appendix 1).

Table 9

Dynamics of the Response Distribution

Difference in quantity of correct responses to presented minimum and maximum limit signals (average % of responses quantity)

| Series of the Experiment | Pilot experiment | | Primary experiment | | | | | | | |
|---|---|---|---|---|---|---|---|---|---|---|
| | | | 1st series | | Subseries 1a | | 2nd A series | | 2nd B series | |
| Presentations | **Minim Limit** | **Maxim limit** | **Minim limit** | **Maxim limit** | **Minim limit** | **Maxim Limit** | **Minim limit** | **Maxim limit** | **Minim Limit** | **Maxim Limit** |
| **1st presentation** | **45.7** | **38** | **71.5** | **68.3** | **77.8** | **44.4** | **65.2** | **75.8** | **37.2** | **79.5** |
| **2nd presentation** | **69.5** | **46.6** | **88.2** | **50.8** | **88.9** | **66.6** | **84.8** | **54.5** | **65.4** | **65.4** |
| Difference | **23.8 +** | **8.6 +** | **16.7 +** | **17.5 -** | **11.7+** | **22.2 +** | **19.6 +** | **21.3 -** | **28.2 +** | **14.1 -** |
| Difference | T=3.2 P< .1 | T=1 p>.5 | T= 2.8 P< .5 | T=3.5 P< .01 | T=.2.4 P< .5 | T=3.7 P< .1 | T=2.8 P< .1 | T=3 P< .1 | T=3.6 P< .1 | T= 2.9 P< .1 |

Legends

**+ increase**

**- decrease**

**A comparison of the data analysis results of the primary experiment series conducted under different uncertainty conditions**

One can find both similarities and differences between the three main series of the primary experiment in tendencies and peculiarities of the correct and incorrect responses distribution along the scale of presented stimuli during the 1st and the 2nd time signal presentations. Data of the pilot experiment represented in Table 10 will be discussed later.

**Similarities**

During the 1st time signal presentation

1. In the 1st, 2nd A, and 2nd B series, most correct responses were given to the time signals which were located on half of the scale belonging to the maximum limit (see Tables 3(3), 4(3), 5(3) from appendix 1).

2. In the 2nd A and especially in the 2nd B series, the quantity of correct responses to the maximum limit was considerably higher than the quantity of correct responses to the minimum limit. In the 1st series these quantities were equal (see Table 9).

3. In all three series of the primary experiment especially in the 2nd B series, the quantity of overestimated incorrect responses was considerably higher than the quantity of the underestimated ones (see Table 10).

4. Despite the difference in the structure of presented stimuli scales (see Tables 2 a and 2 b), one can find a specific similarity between the 1st and 2nd B series in the structure of correct responses distribution along the scale. During the 1st time signals presentation, the structure of such a distribution along the first three points on the scale of the 2nd B series (conjugate pairs 0.9 sec – 1.5 cm, 1.2 sec – 2 cm, 1.5 sec – 2.5 cm) coincided with such a structure along the part of the points of the scale of the 1st series where these conjugate pairs were placed. See Tables 8(4), 8 (6), and Tables 4 from appendix 2, and Tables 3 (1), 5 (1) from appendix 1. The strength of the conjugate pair 0.9 sec – 1.5 cm and the percentage of correct responses to the time signal of 0.9 sec which was the minimum limit in the 2nd B series, were the same as in the 1st series where this conjugate pair was not a minimum limit pair. In both cases this pair was weak and the percentage of correct responses to the time signal of 0.9 sec was 37.2% in the 2nd B series and 36.6% in the 1st series.

Table 10

**Dynamics in shifting tendencies to both limits of the scale**

(Average % of incorrect responses quantity)

| **Series of the Experiment** | Pilot Experiment | | Primary experiment | | | | | | | |
|---|---|---|---|---|---|---|---|---|---|---|
| | | | Subseries 1a | | 1st series | | 2nd A series | | 2nd B series | |
| **Distribution tendencies** | + | − | + | − | + | − | + | − | + | − |
| 1st presentation | 55 | 24.8 | 30.3 | 23.2 | 48.4 | 14.7 | 37.8 | 19.8 | 52.3 | 13.3 |
| 2nd presentation | 40.2 | 29.4 | 22.6 | 22.7 | 34 | 18.2 | 16 | 26.3 | 25.6 | 25.7 |

Legends:

+ Overestimated incorrect responses distribution towards a maximum limit pair

− Underestimated incorrect responses distribution towards a minimum limit pair

During the 2nd time signal presentation

1. In all three series and especially in the 2nd B series of the primary experiment, the quantity of correct responses to the time signals which were located on half of the scale which belonged to the minimum limit increased, and became equal to the quantity of such responses to time signals which were located on half of the scale belonging to the maximum limit (see Tables 3(3), 4(3), 5(3) from appendix 1).

2. In all three series of the primary experiment, the quantity of correct responses sharply decreased to the maximum limit signal, and sharply increased to the minimum limit. In the 1st and 2nd A series the quantity of correct responses to the minimum limit signal became considerably higher than to the maximum limit. In the 2nd B series the quantities of correct responses to both limit signals became equal (see Table 9).

3. In all three series of the primary experiment, the quantity of overestimated incorrect responses was considerably decreased, whereas the quantity of underestimated incorrect responses were increased (see Table 10).

**Differences**

During the 1st time signal presentation

1 The quantity of correct responses and strong and relatively strong conjugate and weakest pairs corresponded strictly to the level of uncertainty conditions of each series of this experiment. The lower the level of uncertainty the higher the quantity of correct responses, and therefore also with strong and relatively strong conjugate pairs and vice versa. Correspondingly, the higher the level of uncertainty, the higher the quantity of weakest unrecognizable conjugate pairs was shown and vice versa. Most weakest pairs were observed in the 2nd B series of the primary experiment which was conducted under high uncertainty conditions and no weakest pairs were observed in the subseries 1 a conducted under lower uncertainty conditions (See table 5, 8(b), 8(c), 8(d), 8 (e).

2. Levels of the minimum and maximum limit pairs recognizability, and the ratio between them depended on the level of uncertainty conditions under which series of the primary experiment were conducted. During the 1st time signal presentation among all series of the primary experiment, the greatest quantity of correct responses to the maximum limit signal and the least quantity of correct responses to the minimum limit were observed in the 2nd B series, which was conducted under a high uncertainty condition. As mentioned above, in this series the quantity of correct responses to the maximum limit signal was exceeded by twice the quantity of such responses to the minimum limit. On the other hand, the greatest quantity of correct responses to the minimum limit signal, and the least quantity of correct responses to the maximum limit signal were observed in the subseries 1 a, which was conducted under a lower uncertainty condition (see Table 9).

3. There was a difference between series of the primary experiment in quantities of overestimated and underestimated incorrect responses distribution. The lower the uncertainty condition the less overestimated and the more underestimated incorrect responses were observed and vice versa. Most overestimated and less underestimated incorrect responses were observed in the 2nd B series and in the 1st series as well. Less overestimated and more underestimated incorrect responses were observed in the subseries 1a and in the 2nd A series as well. (See Table 10)

4. One can show a difference between the series of the primary experiment in the peculiarities of the correct response distribution along the middle of the scale. As mentioned above (see p. 44), during the 1st time signal presentation in the 2nd B series inside the uniform part of the scale, the relatively strong pair of 2.4 second – 4 cm was observed between weakest pairs which contained nothing recognizable.

In other words, an observer gave the correct responses only to the signal of 2.4 seconds placed on the 6th point of the scale of presented stimuli or on the 8th point of the scale of possible stimuli. Whereas his or her responses to signals placed before and after this point were incorrect. Along the scale there was no gradient junction between this relatively strong pair and the weakest ones. This phenomenon became more or less apparent in other series of the primary experiment. One can show an important difference between the 2nd B series and other series of the primary experiment with the appearance of this phenomenon. As mentioned above, a large quantity of correct responses were given to the time signal of 2.4 seconds. But among the incorrect responses to its neighboring time signals, it was as if suddenly there was no gradient junction between them whereas in the rest of series this junction was clearly defined. However as mentioned above, the quantity of correct responses to the time signal of 2.4 seconds exceeded the quantity of such responses to both neighboring time signals with significant difference (see in detail p 35). In the 2nd A series where an observer was preliminarily shown the minimum (0.9 sec – 1.5 cm) and maximum (6.3 sec – 10.5 cm) limits of the presented temporal and spatial signals sets during the 1st time signals presentation this phenomenon was hardly noticeable. The quantity of correct responses to the time signal of 2.4 sec became slightly predominant over only one neighboring time signal of 2.7 sec with no significant difference (See Table 4(1) from appendix 1).

In the subseries 1a which was conducted under a lower uncertainty condition during the 1st time signal presentation, the conjugate pair 2.4 seconds – 4 cm was as strong as the conjugate pair 2.1 seconds – 3.5 cm, its neighbor on the scale. There was no significant difference between the quantities of correct responses to time signals of 2.1 and 2.4 seconds. However the quantity of correct responses to the time signal of 2.4 seconds exceeded with a significant difference, the quantities of such responses not only to the neighboring time signal of 3.3 seconds but to all signals located on the scale after this signal which included the maximum limit pair (See Table 2 (1) from appendix 1).

<u>During the 2nd time signal presentation</u>

1. During the 2nd time signal presentation the conjugate pairs strength and the quantity of correct responses in different series of the primary experiment increased also differently. Most of all, such an increase was observed in the 2nd A and especially in the 2nd B series which was conducted under a high uncertainty condition, and least of all in the subseries 1a which was conducted under a lower uncertainty condition (see Table 11).

Table 11

The differences between the series of the primary experiment showed increasing conjugate pairs strength, and the quantity of correct responses during the 2nd time signals presentation (% of all observers responses).

| Series of the primary experiment | Time signal presentation numbers | The quantity of different strength conjugate pairs | | | | The quantity of correct responses |
|---|---|---|---|---|---|---|
| | | SP | RSP | WP | WtP | |
| The subseries 1a | The 1st presentation | 8 | 3 | 3 | 0 | 52 |
| | The 2nd presentation | 9 | 3 | 2 | 0 | 59.5 |
| | Difference | +1 | 0 | - 1 | 0 | 7.5 |
| The 1st series | The 1st presentation | 4 | 3 | 5 | 2 | 42.4 |
| | The 2nd presentation | 7 | 6 | 1 | 0 | 52.8 |
| | Difference | +3 | +3 | - 4 | - 2 | 10.4 |
| The 2nd A series | The 1st presentation | 6 | 1 | 6 | 1 | 46.4 |
| | The 2nd presentation | 12 | 0 | 2 | 0 | 57.5 |
| | Difference | +6 | -1 | - 4 | - 1 | 11.1 |
| The 2nd B series | The 1st presentation | 3 | 3 | 3 | 5 | 38.7 |
| | The 2nd presentation | 7 | 5 | 2 | 0 | 52.3 |
| | Difference | +4 | +2 | -1 | - 5 | 13.6 |

Legends:

**SP - Strong Pairs** (more than 50% of correct responses)

**RSP - Relatively Strong Pairs** (more than 40 % close to the level at 50 % of correct responses and putting together with approximately correct responses more than 50 and higher percentage).

**WP - Weak Pairs** (lower than 40% of correct responses and putting together with approximately correct responses more than 50 and higher percentage as well).

**WtP - Weakest Pairs** (lower than 30 % of correct responses and not putting together with approximately correct responses less than 50 percent). Weakest pairs were considered as unrecognized pairs.

+ Increase of the quantity of a certain conjugate pair

- Decrease of the quantity of a certain conjugate pair

In Table 11, one can find a similarity between the data of the 1st series and the 2nd B series during the 2nd time signal presentation despite the difference in the structure of their presented stimuli scales. Unlike the other series of the primary experiment, the 1st and 2nd B series had the same quantities of strong pairs and correct responses. As shown above, the 1st series was based on the partly complete scale, the uniform part of which began with the first point of the assumed scale of all possible stimuli (the conjugate pair 0.3 sec – 0.5 cm., see Table 2 a). Whereas the 2nd B series was based on the partly complete scale, the uniform part of which began with the third point of the assumed scale of possible stimuli (the conjugate pair 0.9 sec – 1.5 cm, see Table 2 b).

2. As stated above, during the 2nd time signal presentation in the 1st, 2nd A, and 2nd B series of the primary experiment, the quantity of correct responses decreased sharply in regard to the maximum limit signal, and sharply increased to the minimum limit. In the subseries 1a however, this phenomenon did not take place. During the 2nd time signal presentation, the quantity of correct responses increased to both minimum and maximum limit signals, and the quantity of such responses to the minimum limit remained higher than to the maximum one.

In the 1st, 2nd A, and 2nd B series, this phenomenon appeared differently. In the 1st and 2nd A series the quantity of correct responses to the minimum limit signal became considerably higher than to the maximum limit. However, in the 2nd B series the quantities of correct responses to both limit signals became equal (See Table 9).

One can see a difference between the series of the primary experiment in the changing peculiarities of the quantity of correct responses to the minimum and maximum limit signals found during the 2nd time signal presentation.

The greatest increase in the quantity of correct responses to the minimum limit signal among all series was observed in the 2nd B series. The increase in this series exceeded a decrease of the quantity of correct responses to the maximum limit signal by twice the amount. The largest decrease in the quantity of such responses to the maximum limit signal was observed in the 2nd A series (See Table 9). In the 1st series, the quantity of correct responses to the minimum limit signal was increased by the same amount, and thus decreased the quantity of those responding to the maximum limit signal.

In the subseries 1a during the 2nd time signal presentation, the quantity of correct responses to the minimum and maximum limit signals increased. The quantity of correct responses to the maximum limit signal increased twice as much as the minimum limit signal. As a result, during the 2nd time signal presentation the maximum limit pair became strong. The pilot experiment data which is represented in Table 9 will not be considered at this point, but will be discussed later.

3. As mentioned above, during the 2nd time signal presentation in the 1st, 2nd A and 2nd B series of the primary experiment, the increase of underestimated and decrease of overestimated incorrect responses were a general rule. But a significant difference in the ratio between underestimated and overestimated responses in different series was observed. As can be seen from Table 10, during the 2nd time signal presentation in the 2nd A series, the quantity of underestimated incorrect responses prevailed greatly over the quantity of overestimated responses. In the 2nd B series both of these quantities became equal. In the 1st series, only overestimated responses decreased significantly whereas underestimated responses increased to a small degree, and the quantity of overestimated responses exceeded the quantity of underestimated responses. In the subseries 1a which was conducted under a lower uncertainty condition, only the quantity of overestimated responses was decreased but the quantity of underestimated responses was not changed. Both of these quantities became equal. Thus in the 1st, 2nd A, and especially in the 2nd B series of the primary experiment during the 1st time signal presentation, the observer's cognitive activity (the conjugate pairs recognition and the responses distribution curve) was directed towards the maximum limit pair whereas during the 2nd time signal presentation, this activity was directed toward the minimum limit. This caused a transformation of the conjugate pairs recognizability from a lower to a higher level. For instance, all of the weakest pairs in all three series of the primary experiment during the 2nd time signal presentation became either strong or relatively strong, or weak i.e. became recognizable in different degrees. Such a pendulous response distribution swings, and their results asked the next three questions:

(1). Would these pendulous response distribution swings continue if not two but three or more time signal presentations had been given to a observer?

(2). Why did unrecognizable weakest pairs during the 1st time signal presentation become recognizable in different degrees during the 2nd presentation? Does this mean that at the beginning, an observer did not notice the points of the scale where these pairs were placed as "holes" which were deliberately made by the experimenter in the non-uniform part of the partly completed scale?

(3). Towards what minimum limit pair was an observer's cognitive activity directed during the 2nd time signal presentation in the 2nd A and B series? Was it the minimum limit of the scale of presented stimuli (the conjugate pair 0.9 sec – 1.5 cm) or the minimum limit of the scale of possible stimuli (the conjugate moduli 0.3 sec – 0.5 cm)?

4. As shown above, during the 1st time signals presentation, a quantity of correct responses to a time signal of 2.4 seconds which was located in the middle part of the scale was greater than the quantity of such responses to the signals which bordered it. Clearly, this phenomenon was observed in the 2nd B series of the primary experiment, which was conducted under a high uncertainty condition (See fig 2). During the 2nd time signal presentation in several different series of the experiment, the quantity of correct responses to the time signal of 2.4 seconds changed differently. In the subseries 1a, which was conducted under a lower uncertainty condition, this quantity became lower ($t = 2.4$, $P > .5$). However, the quantity of correct responses to the neighboring signal of 2.1 seconds located toward the minimum limit signal sharply decreased and became significantly lower than the quantity of such responses to the time signal of 2.4 seconds ($t = 3.7$, $p < .1$).

The quantity of correct responses to the neighboring time signal of 3.3 seconds located toward the maximum limit signal increased significantly and became equal to the quantity of such responses to the time signal of 2.4 seconds. As shown above, during the 1st time signal presentation, correct responses to the time signal of 2.4 seconds exceeded with a significant difference the quantity of such responses to the time signal of 3.3 seconds. See figure 1, Table 8 (3), Table 2 (1) from appendix 1, Table 1 from appendix 2. In the 1st series during the 2nd time signal presentation, a quantity of correct responses to time signals of 2.1 and 3.3 seconds, which were neighbors to the time signal of 2.4 seconds, increased sharply ($t = 2.6$, $P < .1$ and $t = 3.4$, $P < .01$ respectively). The quantity of correct responses to the time signal of 2.4 seconds increased very little ($t = 1.9$, $P > .5$), and did not prevail over quantities of correct responses to neighboring signals as it was observed during the 1st time signal presentation (See figure 2, Table 2 from appendix 3, and Table 3 (1) from appendix 1).

In the 1st series during the 2nd time signal presentation, the quantity of correct responses sharply decreased to the maximum limit time signal. In the 2nd A series, unlike the 1st time signal presentation, this phenomenon was hardly noticed, but during the 2nd time signal presentation the quantity of correct responses to the time signal of 2.4 seconds increased sharply ($t = 3.8$, $p < .01$). And it became predominant not only to the quantity of such responses to time signals placed on the middle of the scale, but also to the maximum limit signal. The quantities of correct responses to neighboring signals of 2.1 and 2.7 seconds increased to a lesser degree ($t = 2.8$, $p < .1$) and $t = 2.1$, $p < .5$ respectively). At the same time, quantities of correct responses to the time signals of 6.0 and especially of 6.3 seconds sharply decreased (see Table 3 and figure 3 from appendix 3 and Table 4(1) from appendix 1). In the 2nd B series as in the 2nd A series during the 2nd time signal presentation, the quantity of correct responses to the signal of 2.4 seconds increased ($t = 2.1$, $p < .5$), and reached the same level. However, this increase in the 2nd B series was considerably less than in the 2nd A series (10.3% and 22.8% respectively) although both quantities reached the same level (see Tables 3, 4 and figures 3, 4 from appendix 3 and Tables 4(1), 5(1) from appendix 1). Nevertheless, the increase in the quantity of correct responses to the neighboring signals of 2.1 and 2.7 seconds was considerably larger than to the signal of 2.4 seconds (23% $t = 2.8$, $p < .1$ and 23.1% $t = 2.8$, $p < .1$ respectively). The corresponding conjugate pairs of 2.1 sec – 3.5 cm and 2.7 sec – 4.5 cm during the 2nd time signal presentation were transformed from the weakest unrecognizable pairs into recognizable pairs of a different degree. At the same time, quantities of correct responses to three time signals that were placed on half the scale belonging to the maximum limit of 5.4, 6.0, and especially 6.3 seconds were all decreased.

**Similarities and Dissimilarities between the series of the primary experiment to the latency of responses to time signal presentations**

Similarities and differences between series of the primary experiment were observed in a latency of responses and particularly in a ratio between latencies of correct and incorrect responses during both the 1st and the 2nd time signal presentations.

Table 12

Latency of responses to time signal presentations in all series of the primary experiment (in seconds)

| Responses kinds | Correct responses | | | | Incorrect responses | | | | Average | | | |
|---|---|---|---|---|---|---|---|---|---|---|---|---|
| **Series of experiments** | **Subseries 1a** | **1st series** | **2nd A series** | **2nd B series** | **Subseries 1a** | **1st series** | **2nd A series** | **2nd B series** | **Subseries 1a** | **1st series** | **2nd A series** | **2nd B series** |
| **1st presentation** | 3.24 | 3.27 | 4.6 | 4.65 | 4.13 | 3.7 | 4.83 | 4.54 | 3.56 | 3.52 | 4.74 | 4.58 |
| **2nd presentation** | 2.57 | 3.07 | 3.78 | 4.06 | 2.83 | 3.32 | 3.99 | 4.25 | 2.67 | 3.2 | 3.91 | 4.15 |
| **Average** | 2.9 | 3.16 | 4.19 | 4.36 | 3.48 | 3.53 | 4.41 | 4.4 | 3.11 | 3.35 | 4.33 | 4.37 |

As can be seen from Table 12, during the 1st time signal presentation the latency of correct responses in the subseries 1a coincides with the latency of correct responses in the 1st series, and the latency of correct responses in the 2nd A series coincides with the latency of correct responses in the 2nd B series. At the same time, latencies of correct responses in the subseries 1a, and the 1st series, where a modulus within presented stimuli, was considerably shorter than in the 2nd A ($t = 2.8$, $P < .1$) and B series ($t = 2.7$, P

< .1) where this modulus was absent within presented stimuli. One can notice a difference between latencies of correct and incorrect responses in different series of the primary experiment. A latency of correct responses was lower than a latency of incorrect responses in the subseries 1 a (t = 2.4, P < .05), the 1st series (t = 2, p <.5). But in the 2nd A series there was no significant discrepancy (t = 1.6, p > .5). These numerical results were higher than the latency of incorrect responses in the 2nd B series where there was no significant discrepancy either (t = 1.4, p > .5). A latency of all responses as a whole, both correct and incorrect ones, in the series 1 a and the 1st series was shorter ( t = 2.7, P < .1) than in the 2 A and the 2 B series.

During the 2nd time signal presentation the latency of both correct and incorrect responses decreased in all series of the primary experiment. However, a ratio between subseries 1 a and the 1st series on the one hand, and the 2nd A and the 2nd B series on the other hand, the latencies of responses as a whole and a ratio between latencies of correct and incorrect responses as well remained the same as it was during the 1st time signal presentation.

## Discussion

### Phenomena which were observed during the first and the second time signal presentations

As mentioned above, a time signal of a certain duration and modality was presented twice to an observer during the main part of the primary experiment. Time signals were presented and repeated in a random order and an observer did not know if a certain signal was presented for the first time or the second time. The data analysis showed a difference between the 1st and 2nd time signal presentations in the quantity of correct responses, the levels of conjugate pair recognition, the peculiarities of a responses distribution along the scale, and time reactions under different uncertainty conditions. Therefore the perceptual phenomena which were observed during the first and second time signals presentation in the primary experiment will be considered separately.

## Phenomena which were observed during the first time signal presentation

1. The quantity of correct responses and strong and relatively strong conjugate pairs strictly corresponded to the level of uncertainty conditions in each series of this experiment. The lower the level of uncertainty the higher the quantity of strong conjugate pairs. The higher this level the more unrecognizable weakest pairs were observed.

The entire series of the primary experiment can be arranged by the quantity of correct responses and level of the recognition of conjugate pairs as follows: In the first place the subseries 1a is where an observer was first acquainted with the maximum and minimum limits of a presented set of stimuli and where the minimum limit was the same as the modulus pair (the lower level of uncertainty). The absolute majority of responses (52%) were correct. Correct and approximately correct responses together comprised a total of 78.7 %. There were no unrecognizable weakest pairs.

The second place was occupied by the 2$^{nd}$ A series where a modulus pair (0.3 sec – 0.5 cm) was not within the presented set of signals, but an observer was previously acquainted with its maximum and minimum limits (the low level of uncertainty). Correct responses composed 46.4% of total number of all responses, and put together with approximately correct responses comprised 69.9%. Only one unrecognizable weakest pair was observed.

In third place was the 1st series where the modulus pair (0.3 sec - 0.5 cm) was within the presented set of stimuli, but a observer was neither acquainted with the maximum or minimum limits of this set nor pointed specially to the modulus pair (the middle level of uncertainty). Correct responses composed 42.4% of the total number of all responses, and put together with approximately correct responses the results were 67%. Two unrecognizable weakest pairs were observed.

The fourth place was occupied by the 2nd B series where the modulus pair (0.3 sec – 0.5 cm) was not within the presented set of signals, and a observer was not previously acquainted with its maximum and minimum limits (the high level of uncertainty). Correct responses were 38.7% of the total number of all responses, and put together with approximately correct responses the results were 58%. Five unrecognizable weakest pairs were observed.

2. The nature, the tendency, and the peculiarity of a response distribution along the scale depended on the structure of the scale of presented stimuli, and the level of uncertainty conditions under which the experiment was conducted. In all series of the primary experiment and in the pilot experiment as well, during the 1st time signal presentation one can observe the same tendency. The quantity of overestimated incorrect responses was higher than the quantity of the underestimated ones. This tendency is shown in different degrees in different series of the primary experiment and it depends strictly on the level of uncertainty. The higher this level, the greater the degree to which a quantity of overestimated responses was dominant over the quantity of underestimated responses. Among all the series of the primary experiment, this tendency was observed most in the 2nd B series, which was conducted under the high level of an uncertainty condition. In the 2nd B series, the quantity of overestimated responses exceeded the quantity of underestimated responses almost four to one. In the 1st series, which was conducted under the middle level of uncertainty condition, a quantity of overestimated responses exceeded a quantity of underestimated responses by a ratio a little more than 3 to 1. In the 2nd A series, which was conducted under a low level of uncertainty condition, a quantity of overestimated responses exceeded a quantity of underestimated responses by almost 2 to 1. In the subseries 1a, which was conducted under the lower level uncertainty condition, the quantity of overestimated responses exceeded the quantity of underestimated responses by only 1.3 times. Data in the pilot experiment will be discussed later.

The above mentioned tendency influenced the correct response distribution along the scale of presented stimuli differently in different series of the primary experiment. This influence depended more on the structure of the scale than on the uncertainty condition.

During the first time signal presentation the absolute majority of correct responses were given to the time signals located on the half scale belonging to the maximum limit in the 2nd A, and especially the 2nd B series (see tables 4(3), 5(3) from appendix 1). As mentioned above, in these series, scales of presenting stimuli began not with the first point of the assume scale of possible stimuli which was the conjugate pair 0.3 sec – 0.5 cm but with the third point of this scale which was the conjugate pair 0.9 sec – 1.5 cm and served as the minimum limit of the presented scale. In the first series and the subseries 1a the scale of presented stimuli began with the first point of the assumed scale of possible stimuli (the conjugate pair 0.3 sec – 0.5 cm) that was also the minimum limit of the scale. In the first case, the quantity of correct responses to time signals placed on both halves of the scale was distributed almost equally (in the first series). In the second case the quantity of correct responses to signals located on the half scale, which belonged to the minimum limit, was greater than quantity of correct responses given to signals located on the half scale belonging to the maximum limit (in the subseries 1a).(See tables 2(3), 3(3) from appendix 1) The same tendency of correct response distribution was observed regarding the ratio between the level of recognition, and the quantities of correct responses to the maximum and minimum limits of presented stimuli scales. In the 2nd A series, a quantity of correct responses to the maximum limit exceeded with significant difference the quantity of correct responses to the minimum limit despite the fact that observers were preliminarily informed of both limits (see p.41). In the 2nd B series where observers were not informed of both limits of the scale, a quantity of correct responses to the maximum limit exceeded the quantity of correct responses to the minimum limit by more than twice (see Table 9). In the 1st series and the subseries 1a the quantity of correct responses to both limits were either equal (the 1st series) or a quantity of correct responses to the minimum limit exceeded with significant difference a quantity of correct responses to the maximum limit (the subseries 1a). It was observed that the percentage of correct responses to the time signal of 0.9 seconds in the 1st series and the 2nd B series was the same despite the fact that in the 1st series this signal was located on the third point of the scale of presented stimuli, whereas in the 2nd B series it was located on the 1st point of the scale and served as its minimum limit (see p. 48)

3. An outburst of recognition activity in the middle part of the scale - (ORAMPS phenomenon) - the recognition of the conjugate pair 2.4 sec – 4 cm.

In the 2nd B series the quantity of correct responses to the time signal of 2.4 seconds, which was located on the 6th place of the scale of presented stimuli or in 8th place of the scale of possible stimuli (the middle part of the range of stimuli) was significantly higher than the quantity of such responses to the minimum limit (the time signal of 0.9 seconds), and especially to the neighboring time signals of 2.1 and 2.7 seconds. (see p.45 and Table 8 (d), fig.2, and table 5(1) from appendix 1). These signals belonged to conjugate pairs (2.1 sec – 3.5 cm and 2.7 sec – 4.5 cm respectively) were not recognized during the 1st time signal presentation. There was no gradient junction (smooth transition) between levels of recognition of neighboring conjugate pairs and the pair 2.4 sec – 4 cm. In other words, observers gave a large quantity of correct responses only to the signal of 2.4 seconds. Whereas their responses to signals placed on the scale before and after this signal were incorrect. Unlike other series of primary experiment, only in the 2nd B series did this phenomenon appear in its pure form during the 1st time signal presentation. The main difference between the 2nd B series and other series of the primary experiment was that in the 2nd B series there was not gradient junction (smooth transition) between levels of recognition of neighboring conjugate pairs and the pair 2.4 sec – 4 cm. However, in all other series such a junction (smooth transition) existed. In the 2nd A series, which was conducted under a low uncertainty condition during the 1st time signals presentation, this phenomenon was hardly noticeable. In the 1st series of the primary experiment during the 1st time signals presentation, the conjugate pair 2.4 seconds – 4 cm and its neighbors on the scale (conjugate pairs 2.1 sec – 3.5 cm and 3.3 sec – 5.5 cm) were weak, i.e. their levels of recognition were the same. But the quantity of correct responses to the time signal of 2.4 seconds exceeded the quantity of such responses to both neighboring time signals of 2.1 sec and of 3.3 sec with a significant difference.

In the subseries 1a, which was conducted under a lower uncertainty condition during the 1st time signal presentation, there was no significant difference between the quantities of correct responses to the time signal of 2.4 seconds and to its neighbor on the scale time signals of 2.1 seconds. However, the quantity of correct responses to the time signal of 2.4 seconds exceeded with a significant difference the quantities of such responses not only to the neighboring time signal of 3.3 seconds but to all signals located on the scale after this signal which included the maximum limit pair.

4. A latency of responses as a whole, and especially correct responses in the 1st series and the subseries 1 a, where a modulus was within presented stimuli, was considerably shorter than in both the 2nd A and 2nd B series where this modulus within presented stimuli was absent. In the 1st series and the subseries 1 a, latency of correct responses was lower than a latency of incorrect responses. Whereas in both the 2nd A and 2nd B series between these latencies there was no significant difference.

**Phenomena which were observed during the second time signal presentation**

1. In all three series and especially in the 2nd B series of the primary experiment, the quantity of correct responses to the time signals which were located on half of the scale which belonged to the minimum limit increased, and became equal to the quantity of such responses to time signals which were located on half of the scale belonging to the maximum limit.

During this presentation, a transformation of the conjugate pairs recognizability from a lower to a higher level took place. All of the weakest or unrecognized pairs in all three series of the primary experiment became recognizable in different degrees such as strong or relatively strong, or weak pairs. The conjugate pairs strength and the quantity of correct responses in different series of the primary experiment increased differently as well. Most of all, such an increase was observed in the 2nd A and especially in the 2nd B series which was conducted under a high uncertainty condition, and least of all in the subseries 1a which was conducted under a lower uncertainty condition. The quantity of correct responses given during the 2nd time signal presentation in the first series was 52.8%, and became practically equal as a whole to the quantity of such responses given during the 1st time signal presentation of 52% in subseries 1 a . A similarity between the data of the 1st series and the 2nd B series during the 2nd time signal presentation arose despite the difference in the structure of their presented stimuli scales. Unlike the other series of the primary experiment in the 1st and 2nd B series, the quantities of strong pairs and correct responses became the same.

2. During the 2nd time signal presentation the nature, tendency, and peculiarity of a response distribution along the scale were changed as follows: In the 1st, 2nd A and 2nd B series of the primary experiment, the increase of underestimated and the decrease of overestimated incorrect responses were a general rule. But a significant difference in the ratio between underestimated and overestimated responses in different series was observed. In the 2nd A series, the quantity of underestimated incorrect responses prevailed greatly over the quantity of overestimated responses. In the 2nd B series, both of these quantities became equal. In the 1st series, only overestimated responses decreased significantly whereas underestimated responses increased to a small degree, and the quantity of overestimated responses exceeded the quantity of underestimated responses. In the subseries 1a which was conducted under a lower uncertainty condition, only the quantity of overestimated responses was decreased but the quantity of underestimated responses was not changed. Both of these quantities became equal. This phenomenon reflected the general tendency of the response distribution of the scale. In all series and especially in the 2nd B series of the primary experiment, the quantity of correct responses decreased sharply in regard to the maximum limit signal, and sharply increased to the minimum limit. But under different experimental conditions, this phenomenon appeared differently. In the 1st and 2nd A series the quantity of correct responses to the minimum limit signal became considerably higher than to the maximum limit signal. However, in the 2nd B series the quantities of correct responses to both limit signals became equal. The greatest increase in the quantity of correct responses to the minimum limit signal among all series was observed in the 2nd B series. The increase in this series exceeded a decrease in the quantity of correct responses to the maximum limit signal by twice the amount. The largest decrease in the quantity of such responses to the maximum limit signal was observed in the 2nd A series. In the 1st series, the quantity of correct responses to the minimum limit signal was increased by the same amount, and thus decreased the quantity of those responding to the maximum limit signal.

In the subseries 1 a however, this phenomenon did not take place. In this subseries, the quantity of correct responses increased to both minimum and maximum limit signals, and the quantity of such responses to the minimum limit remained higher than to the maximum one. The quantity of correct responses to the maximum limit signal increased by twice as much as the minimum limit signal. As a result, during the 2nd time signal presentation the maximum limit pair became strong. Thus in the 1st, 2nd A, and especially in the 2nd B series of the primary experiment during the 1st time signal presentation, the observer's cognitive activity (the conjugate pairs recognition and the responses distribution curve) was directed towards the maximum limit pair whereas during the 2nd time signal presentation, this activity was directed toward the minimum limit. This is one of the reasons for the mentioned above transformation of the lowest level of conjugate pair recognition into the higher level of recognition.

3. During the 2nd time signal presentation, an Outburst of Recognition Activity in the Middle part of the Scale Phenomenon – ORAMPSP (the recognition of the conjugate pair 2.4 sec – 4 cm) changed its appearance differently in different series of the primary experiment.

In the 1st series this phenomenon disappeared. The quantity of correct responses to the time signals of 2.1 and 3.3 seconds, which were neighbors of the time signal of 2.4 seconds, increased sharply whereas the quantity of correct responses to the time signal of 2.4 seconds increased insignificantly and stopped prevailing over quantities of correct responses to neighboring signals as it did during the first time signal presentation.

In the subseries 1 a, the structure of distribution of the correct responses along the middle part of the scale during the second time signal presentation became the opposite of that which was observed during the first time signal presentation. The quantity of correct responses to the signal of 2.4 seconds became lower, but the conjugate pair 2.4 sec – 4 cm remained strong. However, the quantity of correct responses to the neighboring signal of 2.1 seconds located toward the minimum limit signal decreased sharply and became significantly lower and the conjugate pair 2.1 sec – 3.5 cm was transformed from a strong pair into a weak pair. The quantity of correct responses to the neighboring time signal of 3.3 seconds located toward the maximum limit signal increased significantly and the conjugate pair 2.1 sec – 3.5 cm was transformed from a weak pair into a strong pair.

In the 2nd A series unlike the 1st time signal presentation where this phenomenon was only slightly noticed, during the 2nd time signal presentation the quantity to the time signal of 2.4 seconds increased sharply and became predominant not only in the quantity of such responses to time signals placed on the middle of the scale, but even to the maximum limit signal. The quantities of correct responses to neighboring signals of 2.1 and 2.7 seconds increased to a lesser degree. At the same time, quantities of correct responses to time signals of 6.0 and especially to 6.3 seconds (the maximum limit) decreased sharply.

In the 2nd B series, the quantity of correct responses to the time signal of 2.4 seconds just adjacent to time signals of 2.1 and 2.7 seconds increased. Nevertheless, the increase in the quantity of correct responses to neighboring signals of 2.1 and 2.7 seconds was considerably larger than to the signal of 2.4 seconds. A sharp difference in the levels of recognition of the conjugate pair 2.4 seconds – 4 cm and conjugate pairs adjacent to it, were observed during the 1st time signal presentation disappeared and was replaced by a smooth transition from one level to another. The conjugate pairs 2.1 sec – 3.5 cm and 2.7 sec -4.5 cm during the 2nd time signal presentation were transformed from weakest unrecognizable pairs into recognizable pairs of a different degree. At the same time, quantities of correct responses to three time signals, which were placed on half of the scale belonging to the maximum limit of 5.4, 6.0, and especially 6.3 seconds, were decreased.

4. During the 2nd time signal presentation, the latency of both correct and incorrect responses decreased in all series of the primary experiment. However, a ratio between the subseries 1a and the 1st series on the one hand, and the 2nd A and the 2nd B series on the other, it was found that the latencies as a whole, and a ratio between latencies of correct and incorrect responses as well remained the same as it was during the 1st time signal presentation.

**A Comparison of the Primary and the Pilot Experiments results**

Some similarities and significant differences were noticed between the pilot experiment and the different series of the primary experiment in quantities of correct responses, in the levels of the recognition of conjugate pairs, in the dynamics and tendencies of the responses distribution along the scale of presented stimuli during the first, and especially the second presentation of time signals.

Similarities

1. One can find a similarity between the pilot experiment and the different series of the primary experiment in the tendency of overestimated and underestimated responses distribution. In all the series of the primary experiment and in the pilot experiment as well, during the 1st time signal presentation the quantity of overestimated incorrect responses were higher than the quantity of the underestimated ones. This tendency was shown to different degrees in different series of the primary experiment, and it depended strictly on the level of uncertainty. The higher this level, the greater the degree to which a quantity of overestimated responses was dominant over the quantity underestimated responses. The smallest difference between overestimated and underestimated responses (only 7.1%) was observed in the subseries 1a which was conducted under a lower uncertainty condition. The largest difference between the overestimated and underestimated responses (38.9%) was observed in the 2nd B series that was conducted under a high uncertainty condition.

In the pilot experiment and in the 1st series of the primary experiment, this difference was practically the same - 30% and 33.7% respectively.

During the 2nd time signal presentations in all series of the primary and also in the pilot experiments, the increase of the underestimated and the decrease of overestimated incorrect responses were the general rule. In the 1st series, subseries 1a and in the pilot experiment as well the minimum limit and the modulus were the same. In these experiments the same tendency was observed: only the quantity of overestimated responses decreased significantly whereas the underestimated responses increased to a small degree. As a result of this, in the 1st series and in the pilot experiment, the quantity of overestimated responses continued to exceed the quantity of underestimated responses. Only in subseries 1a did both of these quantities become equal. Such a similarity between the 1st series and the pilot experiment was observed in spite of the significant difference between these experiments in the level of conjugate pairs recognition.

2. During the 2nd time signal presentation, an increase in the level of the conjugate pair recognition in all series of the primary experiment, and in the pilot experiment was observed to different degrees.

3. In both the 2nd B series and in the pilot experiment the same number of weakest, i.e. unrecognized conjugate pairs were observed (on five unrecognized pairs in each of them).

Differences

Examining the differences between the data of the pilot and the primary experiments, should be focused on the difference between the data of the pilot experiment and the 2nd B series of the primary experiment.

As mentioned above, during the 1st time signal presentation in both the 2nd B series and in the pilot experiment, it was observed that the same number of weakest conjugate pairs, the indices of the level of conjugate pairs recognition, and quantity of correct responses in the 2nd B series were lower than in the remaining series of the primary experiment and approached such indices in the pilot experiment. Despite this similarity, the pilot experiment was considerably different from the 2nd B series and from the remaining series of the primary experiment in the peculiarities of all processes of the correlation enumerated above. These differences are as follows:

1. The difference in the level of conjugate pairs recognition, and in the quantity of correct responses;

(a) It should be noted that in the pilot experiment the range of presented stimuli was less than such a range in all series of the primary experiment, 13 and 14 respectively. A smaller quantity of stimuli would seem to have facilitated their recognition. However as shown above, the quantity of correct responses in the 2nd B series significantly exceeded the quantity of correct responses in the pilot experiment during both the 1st and 2nd time signal presentation. Moreover, during the 2nd time signal presentation the gap between these indices became greater. Correct responses in the 2nd B series became predominant. The indices of correct responses observed in the pilot experiment, were also observed during the second time signal presentation, so they did not reach the indices which were observed during the first time signal presentation in the 2nd B series of the primary experiment.

(b) During the 1st time signal presentation in the pilot experiment, there were no cases of the absolute recognition[4] or strong conjugate pairs, whereas in the 2nd B series such cases had already been observed. During the 2nd time signal presentation in the pilot experiment, only two strong pairs appeared on the 1st and 2nd points of the scale and four relatively strong conjugate pairs appeared mainly on the last three points of the scale. Strong and relatively strong pairs constituted only 46% of responses to all presented stimuli in this experiment. Two pairs out of thirteen remained weakest or unrecognized pairs[5]. During the 2nd time signal presentation in the 2nd B series, only strong conjugate pairs were 50% and together with relatively strong pairs became 85% of all responses to stimuli presented in this experiment. No weakest or unrecognized pairs were observed in the 2nd B series as well as in the other series of the primary experiment during the 2nd time signal presentation.

2. The difference in the response distribution along the scale:

---

[4] **Absolute correct recognition -** were among the responses given to the presentation of a specific signal. Correct responses amounted to upwards of fifty per cent and most of the remaining responses were the approximately correct ones.
**Relatively correct recognition -** were among the responses given to the presentation of a specific signal. Correct and approximately correct responses together amounted to upwards of fifty per cent.
**Absence of recognition -** were among the responses given to the presentation of a certain signal. Correct and approximately correct responses together were less than fifty per cent.

[5] **Strong pairs -** (SP) more than 50% of correct responses
**Relatively strong pairs -** (RSP) more than 40 % of correct responses close to the level of 50 %, and put together with approximately correct responses, more than 50 and higher percents
**Weak pairs** - (WP) lower than 40% of correct responses and put together with approximately correct responses 50% and higher.
**Weakest pairs -** (WtP) lower than 30 % of correct responses and put together with approximately correct responses, less than 50 percents.

During the 1st time signal presentation in all series of the primary experiment, either one or both of the limit conjugate pairs were strong. In the pilot experiment both limit conjugate pairs were not strong. As shown above in the 2nd B series during the 1st time signal presentation, the minimum limit pair was weak whereas the maximum limit pair was very strong. The quantity of correct responses to the longest time signal was more than double that of the shortest time signal. Most of strong and relatively strong pairs were observed on half of the scale which belonged to the maximum limit pair; whereas the absolute majority of weakest pairs were observed on half of the scale which belonged to the minimum limit pair. Unlike the 2nd B series in the pilot experiment, the maximum limit pair was weak and the minimum limit pair was relatively strong.

During the 2nd time signal presentation in the 2nd B series as well as in the 1st and 2nd A series, a sharp decrease in the quantity of correct responses to the maximum limit and the sharp increase in the quantity of such responses to the minimum limit occurred. In the pilot experiment this phenomenon was not observed. The quantity of correct responses in this experiment increased in both the minimum and maximum limit signals. The minimum limit pair, 1 second – 1 cm and the adjacent to it the pair, 2 seconds – 2 cm became strong. However, the maximum limit pair 13 seconds – 13 centimeters and adjacent pairs became only relatively strong.

3. The difference in the responses distribution dynamics:

As mentioned, observers were presented two types of scales of stimuli. The 1st of them contained a conjugate pair which was both a modulus and a minimum limit of this scale. This type of the scale was used in the 1st series, the subseries 1a of the primary experiment, and in the pilot experiment as well. The 2nd type of scale contained a minimum limit which was not a modulus at the same time. This type of the scale was used in the 2nd A and the 2nd B series of the primary experiment. A significant difference in the dynamics and the ratio between underestimated and overestimated responses between experiments which used the 1st type of the scale and experiments which was used the 2nd type of the scale was observed.

During the 2nd time signal presentation in the 1st series, only overestimated responses decreased significantly whereas underestimated responses increased to a small degree, and the quantity of overestimated responses exceeded the quantity of underestimated responses. In the subseries 1a which was conducted under a lower uncertainty condition, only the quantity of overestimated responses was decreased but the quantity of underestimated responses was not changed. The same tendency was observed in the pilot experiment. (see Table 10).

Unlike the aforementioned tendency which was observed in the 1st series, subseries 1a and the pilot experiment, in the 2nd A and B series a quantity of overestimated responses sharply decreased, and at the same time a quantity of underestimated responses increased sharply. In the 2nd A series a quantity of overestimated responses decreased by more than twice, and a quantity of underestimated responses increased significantly as well, and began to prevail over overestimated ones. In the 2nd B series, the quantity of overestimated responses decreased by twice as much, and the quantity of underestimated ones increased by twice as much as well. Thus both of these quantities became equal.

4. During the first or the second time signal presentation in all series of the primary experiment and especially in the 2nd B series, which was conducted under the high level of uncertainty condition, Outburst of the Recognition Activity in the Middle part of the Scale Phenomenon – ORAMPSP was observed. In the pilot experiment this phenomenon did not appear during any such presentations. Thus, on the one hand, the pilot experiment resembled the 2nd B series by the quantity of unrecognized conjugate pairs but on the other hand, it was similar to the 1st series of the primary experiment by the ratio between overestimated and underestimated responses and by their dynamics during both the 1st and the 2nd time signal presentation. However, the level of recognition of the conjugate pairs and the development of this recognition during the first, and especially during the second time signal presentation in the pilot experiment were considerably lower than in the previously mentioned series of the primary experiment and were comparable to these series in the early stage of development.

**The analysis of auxiliary parts of the experimental data, facts, and phenomena and their relations to presented intervals correlation processes in the main part of the primary experiment**

The experimental data analysis showed differences between the pilot experiment and all series of the primary experiment, and between these series in the correlation processes, and levels and dynamics of the conjugate pairs recognition.

In regard to these two questions:

1. Why, in the pilot experiment during the first and the second time signal presentations was the recognition of conjugate pairs considerably worse than in all series of the primary experiment, and particular in the 2nd B series, which was conducted under the high uncertainty condition?

2. What processes of the correlation between perceived durations and lengths lead to the successful recognition of conjugate pairs in all series, and especially in the 2nd B series of the primary experiment?

The first question relates to the problem of the conjugate pair recognition in the pilot experiment. The second question is connected to problems of the correlation between lengths and durations in human perception.

**The accuracy of the estimation of presented space and time intervals and the success of the correlation of these intervals by observers according to the experimental task**

Based on the comparative analysis of the pilot experiment and the primary experiment and especially in the 2nd B series data, one can suggest two factors which prevent or delay successful recognition of conjugate pairs. The first factor is the one-to-one principle of the structure of presented stimuli scales. The second factor, which was connected with the first factor, is the high possibility of applying the social metrics for correlation purposes. In other words an observer could calculate lengths and durations by centimeters or inches and seconds for their commensuration. Observers could do this only with respect to two conjugate pairs (1 second – 1 cm, 2 seconds – 2 cm), which consisted of the shortest lengths and durations in the sets of presented stimuli. During the second time signal presentation, these conjugate pairs became strong. The remaining longer intervals seemingly could not be measured in seconds and centimeters with the same accuracy as the first two in the process of their correlation. For the correct correlation of longer intervals, other measuring systems, which were impeded by the action of the social metrics, were apparently required. This hypothesis can be supported by the results of correct responses analysis, which was conducted on the basis of the data in the primary experiment. As mentioned above (see page 6), in the primary experiment, after observers had performed the task of correlation of presented durations and lengths in the main part of the experiment, they were asked to evaluate the same intervals separately; verbally the durations in seconds, and the lengths in centimeters. Observers had to reproduce nonverbally the duration of these time signals by pressing a timer button, and by drawing the length of space intervals on paper as well. The results of the data analysis and phenomena of the auxiliary experiment caused a number of questions which refer to the processes of the correlation of perceived spatial and temporal intervals, and the recognition of conjugate pairs. One of these questions was as follows: Was the capability of the observers to make a precise estimate of space and time intervals in centimeters and seconds, connected with the success of the correlation of these intervals according to the experimental task? To answer to this question, the observers' estimations of space and time intervals which served in the main part of the experiment as elements of certain conjugate pairs were divided into eight types:

1. Accurate verbal and non-verbal estimates of conjugate temporal (in seconds) and spatial (in cm) intervals. - **VNSC** (verbal, non-verbal seconds, centimeters)
2. Accurate verbal estimates of conjugate temporal (in seconds) and spatial (in cm) intervals while non-verbal estimates of both of these intervals were incorrect – **VSC** (verbal, seconds, centimeters)
3. Accurate non-verbal estimates of conjugate temporal and spatial intervals, while verbal estimates of both of these intervals were incorrect. - **NSC** (non-verbal, seconds, centimeters)

4. Accurate verbal (in seconds) and non-verbal estimates of temporal intervals only, in both verbal and non-verbal estimates of their conjugate space intervals were incorrect - **VNS** (verbal, non-verbal, seconds)
5. Accurate verbal estimates of temporal intervals only, in non-verbal estimates of this interval as well as both verbal and non-verbal estimates of their conjugate space intervals, were incorrect - **VS** (verbal, seconds)
6. Accurate non-verbal estimates of temporal intervals only, in verbal estimates of this interval as well as both verbal and non-verbal estimates of their conjugate space intervals were incorrect - **NS** (non-verbal, seconds)
7. Accurate estimates of space intervals only, in both verbal and non-verbal estimates of their conjugate time intervals were incorrect - **C** (centimeters)[6]

---

[6] The difference between indices of the verbal and nonverbal correct estimations of the space intervals can be found only in the 2nd A series, where the number of verbal estimations of space intervals exceeded the quantity of nonverbal ones. However, in the 1st and in the 2nd B series this difference was not found. For convenience in the analysis of the correct responses, which were obtained in the main part of the primary experiment, the indices of the verbal and nonverbal estimations of the space intervals in this estimation classification were not isolated separately. However, the special features of the verbal and nonverbal estimations of the space intervals, and particularly in the 2ndA series will be taken into account with the consideration of others data of the experiment.

The indices of the level of making more active scale points which represented the conjugate pairs as indices of the conjugate pairs strength itself (see p. 18) are the quantity of correct responses to signals which refer to elements of these pairs. The greater the number of correct responses to one signal or another, the higher the level of making a more active scale with corresponding points and the strength of those conjugate pairs which were physically presented to observers. The difference between the terms “activation of points of the scale” and “strength of the conjugate pairs” consists of the following: The term “activation points of the scale” refers to all points on the scale of possible stimuli i.e., for such points at which the relations between the assumed, but not presented stimuli are represented. The level of the activation of such points can be seen revealed only indirectly. Whereas the term “strength of the conjugate pairs” refers only to the relations between the actual presented physical stimuli.
Note (end)

8. Absence of any accurate verbal and non-verbal estimates of both conjugate temporal and spatial intervals - **AEA** (accurate estimate absence)

Correct responses made in the main part of each series of the primary experiment were distributed among above indicated types, depending on how an observer estimated both elements of a conjugate pair, to the presentation to which he or she gave correct responses in the main part of the experiment.

Table 13

The distribution of the quantity of correct responses with different types of evaluations of the conjugate space and time intervals (% of the total number of correct responses)

| ## | Types of the evaluation | **SERIES OF THE PRIMARY EXPERIMENT** | | |
|---|---|---|---|---|
| | | 1st series | 2nd A series | 2nd B series |
| 1. | □NVC | 0.18 | 1.6 | 0 |
| 2. | VC | 0.97 | 2.3 | 0 |
| 3. | □ NSC | 4.7 | 10.4 | 4.4 |
| 4. | NV | 0.45 | 2.3 | 0 |
| 5. | V | 1.5 | 3.9 | 3.7 |
| 6. | N | 8 | 17.1 | 15.3 |
| 7. | C | 29 | 15.4 | 11.5 |
| 8. | AEA | 55.1 | 46.7 | 64.7 |

Legends

NVC - Nonverbal and verbal estimate in seconds and centimeters
VC - Verbal estimate in seconds and centimeters
NSC - Nonverbal estimate of time and space intervals
NV - Nonverbal and verbal estimate in seconds
V - Verbal estimate in seconds
N - Nonverbal estimate of time intervals
C - Nonverbal and verbal estimate in centimeters
AEA - Accurate estimate absence

As can be seen from Table 13 and Figure 3, in all series of the primary experiment the majority of correct responses were given when observers estimated both elements of the conjugate pairs incorrectly. The quantity of such responses depended on the level of uncertainty conditions under which an experiment was conducted. The higher this level, the more correct responses were given under AEA conditions, or the absence of accurate verbal and non-verbal estimates of both conjugate temporal and spatial intervals. The most correct responses were therefore given in the 2nd B series. Cases of conjugate pair's recognition, when observers estimated correctly both elements of a recognized conjugate pair, were rarely observed and in the 2nd A series which was conducted under a low uncertainty condition. And these cases were connected only with nonverbal estimations of conjugate pairs elements.

**Figure 3**

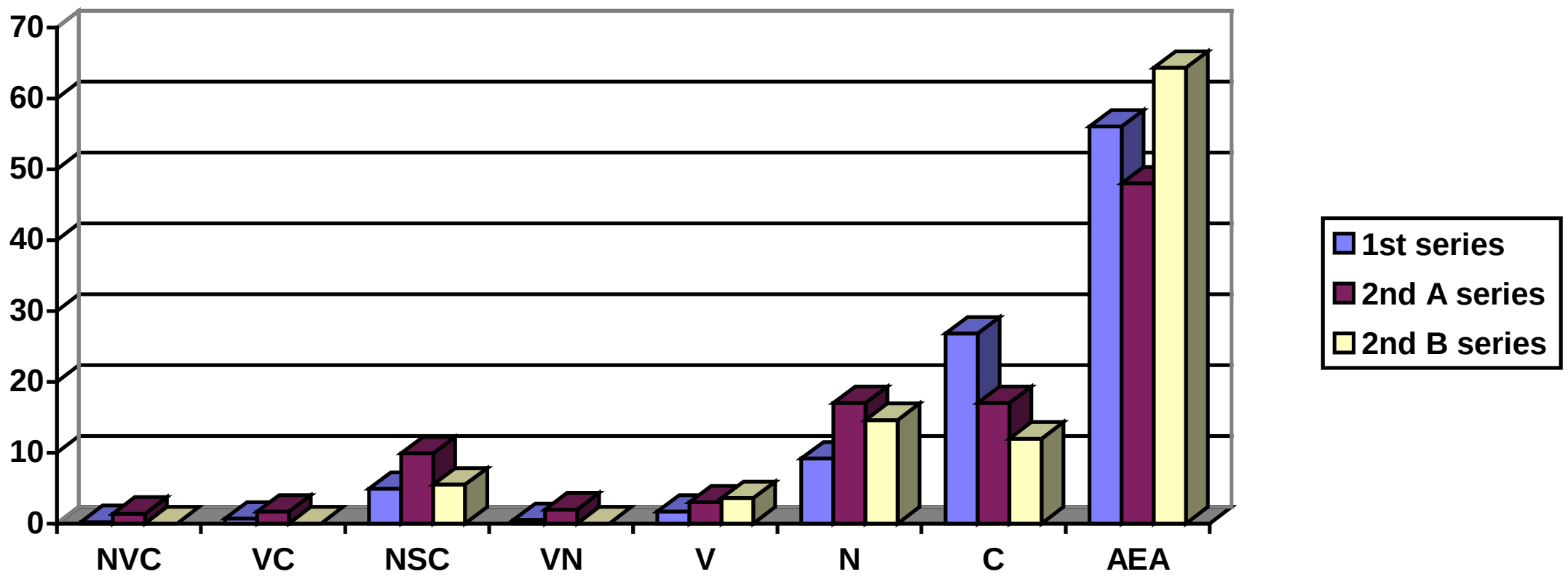

More frequently, were the cases of the conjugate pair recognition when an observer could make a correct estimation only of one element of these pairs either a time interval, which was generally estimated nonverbally or a space interval which was estimated verbally or non-verbally.

Thus the cited data allowed arriving at the following conclusions:

1).The correlation between time intervals and space intervals and the evaluation of these intervals in particular on the basis of the social metrics were two independent cognitive processes. In each of these processes, different measuring systems were involved. In other words, the success of the conjugate pair recognition did not depend on the accuracy when observers estimated space and time intervals of which these conjugate pairs consisted.

2).The interaction and mutual influence of the afore mentioned measuring systems in the processes of the correlation of space and time intervals in the human perception was not a subject of study in psychology. Based on the data of the present experiment, it is possible to make an assumption that the measuring system used in the processes of the correlation of spatial and temporal intervals, unlike the social metrics, did not use centimeters and seconds, but other units of measurement, biorhythms for example.

The results of this system activity, especially in conditions of a high level of uncertainty, and deficiency of the time decision, that is to say making correct correlations between space and time intervals would not be successful if: 1) observers began to apply social metrics, i.e. the correlation of space and time intervals that measured them in centimeters and seconds, 2) the simplified situation of perception of these interval combinations which were created and built according to the one – to – one principle. For example, 7 seconds – 7 centimeters. The quantity of correct responses which related to this conjugate pair was two or more times less, especially during the second time signals presentation, than the quantity of such responses concerning the conjugate pair of 4.5 seconds – 7.5 centimeters, which was not built according to this principle though the numerical value of centimeters in both pairs was approximately identical (see Table 14)

Table 14

The quantity of correct responses in the pilot experiment and three series of the primary experiment related to two conjugate pairs 7 sec – 7 cm and 4.5 sec – 7.5 cm. (% of all responses)

| **Conjugate pairs** | **7 seconds - 7 centimeters** | 4.5 seconds – 7.5 centimeters The Primary Experiment | | |
|---|---|---|---|---|
| Types of the experiment | **The Pilot Experiment** | 1st series | 2nd A series | 2nd B series |
| 1st presentation | 20 | 39.8 | 54.5 | 41 |
| 2nd presentation | 21 | 47.6 | 59.1 | 57.7 |

One can suggest that the measuring system underlying the correlation of space and time intervals refers to the estimation of the motion of perceived objects. Accordingly, ratios between space and time intervals, which lie at the bases of speed are also different, and change. These ratios are not constructed according to the one – to – one principle, but have more complex varying structures. A perceptive measuring system is adopted genetically to adequately estimate precisely such structures in particular under conditions of time deficiency.

The transfer of the one – to – one principle in attempts to apply the social metrics to the perceptive estimation of the correlation of lengths and durations can lead to the errors, which we called **the errors of simplification**. These errors appeared when observers in their responses to a time signal of specific duration, selected a space interval in which the quantity of centimeters corresponded to the quantity of seconds of a signal which had been calculated by them during its perception. However, this selection can not correspond to the correlation of intervals assigned in the experiment. For example, in response to the time signal of 4.5 seconds an observer selected not the length of 7.5 centimeters according to the correlation 4.5 sec – 7.5 cm (the modulus pair 0.3 s – 0.5 cm) which was assigned in this experiment, but the length of 4.5 centimeters. In other words he or she constructed the relationship between these intervals according to the one – to – one principle. Only during the subsequent presentations of time signals did these observers replace this principle with another corresponding to the experimental task.

As can be seen from table 13, in contrast to the 1st and especially the 2nd B series, in the 2nd A series, the cases with AEA did not compose an absolute majority. A large part of the correct responses in this series were connected with the different forms of nonverbal or verbal correct evaluations of one and even both elements of the conjugate pairs.

This fact makes it possible to assume that the accuracy of the nonverbal and verbal estimations of the perceived intervals of space or time rises separately, if both of these intervals were correlated proportionally with each other under conditions of a low level of uncertainty.

**Tendencies in the erroneous estimations of sizes of the space and time intervals after the accomplishment by the observers of the task for these intervals correlation**

As previously mentioned in the primary experiment, numerical values in seconds of time intervals (for instance, 1.8 seconds) were considerably less than the numerical values in centimeters (for instance, 3 centimeters) of spatial intervals which were conjugated with these time intervals.

Table 15

Estimations of time and space intervals performed separately after the implemented task to correlate them (% of the quantity of presentations)

Table 15 (a)

Estimations of time intervals (% of responses to all signals)

| Estimation types | Non-verbal estimations | | | Verbal estimation | | |
|---|---|---|---|---|---|---|
| Estimation sorts | Under estimation | Over estimation | Exact responses | Exact Responses | Over estimation | Under estimation |
| 1st series | 44.4 | 54.9 | 0.6 | 2.2 | 81.4 | 16.3 |
| 2nd A series | 49.6 | 36.4 | 13 | 3.8 | 83.1 | 12.8 |
| 2nd B series | 31.8 | 55 | 12.9 | 1.64 | 81.5 | 17.6 |

Table 15 (b)

Estimations of space intervals (% of responses to all signals)

| Estimation types | Non-verbal estimations | | | Verbal estimation | | |
|---|---|---|---|---|---|---|
| Estimation sorts | Under estimation | Over Estimation | Exact responses | Exact Responses | Over estimation | Under estimation |
| 1st series | 72.8 | 11.6 | 11.6 | 22.4 | 16.2 | 61.4 |
| 2nd A series | 79.3 | 11 | 9.25 | 16.5 | 15.9 | 67 |
| 2nd B series | 84.4 | 10 | 5.1 | 9.2 | 21.1 | 69.3 |

Adjunct information to Tables 15 a and 15 b

The conjugate pair 1.8 sec – 3 cm as an example of

time intervals (a)

| Series | # of the scale's point | Duration in seconds | Non-verbal estimations | | | Verbal estimation | | |
|---|---|---|---|---|---|---|---|---|
| | | | Underestimation | Over estimation | Exact Responses | Exact responses | Over estimation | Under Estimation |
| 1st series | 6 | **1.8** | 42.7 | 56.4 | 0.9 | 0.9 | 85.5 | 13.6 |
| 2nd A series | 4 | **1.8** | 33.3 | 48 | 19 | 4.7 | 95 | 0 |
| 2nd B series | 4 | **1.8** | 35 | 54 | 11 | 0 | 88 | 12 |

and as an example of space intervals (b)

| Series | # of the scale's point | Length in cm | Non-verbal estimations | | | Verbal estimation | | |
|---|---|---|---|---|---|---|---|---|
| | | | Under estimation | Over estimation | Exact responses | Exact responses | Over estimation | Under Estimation |
| 1st series | 6 | **3** | 71.4 | 16.2 | 12.4 | 18.2 | 14.5 | 67.3 |
| 2nd A series | 4 | **3** | 66 | 14 | 19 | 24 | 19 | 57 |
| 2nd B series | 4 | **3** | 69 | 15 | 15 | 19 | 19 | 62 |

As can be seen from Tables 15 a and b when evaluated, those intervals in seconds and those in centimeters, after implementing the task to correlate them, observers overestimated (especially verbally) time intervals, and underestimated space intervals. It was as if they were trying to equalize numerical values of both elements of a certain conjugate pair. This tendency can be seen in the data response to all presented signals as a whole, and the chosen example of conjugated intervals 1.8 sec – 3 cm in all series of the primary experiment with the exception of nonverbal estimates of time intervals in the second A series. There the underestimated of these intervals prevailed over the overestimated. One should be reminded of the fact that in this series, observers were preliminarily informed of minimum and maximum limit pairs. Thus it seems that one of the important stages of the correlation processes was done beforehand and could have had an influence on the non-verbal estimates of time intervals.

Such phenomenal widening and contracting of the sizes of conjugated spatial and temporal signals in non-verbal and verbal estimations of observers was observed to the greatest degree in the $2^{nd}$ B series. This series was carried out under the condition of a high level of uncertainty.

**Phenomenological leveling off or equalization of the numerical values of seconds and centimeters of both elements of the conjugate pairs in the estimations of observers.**

The above mentioned phenomenal widening and contracting of conjugated spatial and temporal signals in observers estimations in most cases lead to the phenomenological leveling off of numerical values of both conjugate pairs elements. The analysis of these estimations showed different kinds of such an equalization and corresponding phenomena related with it, altogether they numbered 13. These phenomena hint at a connection which an observer established between elements of a certain conjugate pair even though he did not express this connection directly by a correct response in the main part of the experiment. One can consider such phenomena in two ways. On the one hand as indirect evidence of the establishment of a connection between both elements of a certain conjugate pair, and on the other hand as a method of correlation between space and time intervals.

All the phenomena and inferences resulting from this analysis will not be dwelled upon here. Some of the observed phenomena taken as an example response to the presented stimuli putting together the “weak” conjugate pair 1.8 sec – 3 cm. will be described here however.

## Direct and indirect evidence of the establishment of a connection between both elements of a certain conjugate pair

### Direct evidences

Table 16

Direct evidences of establishment of a connection between both elements of a certain conjugate pair

| Parts of the experiment | Presented signals and their numerical values | Evidences | |
|---|---|---|---|
| | | Correct responses | Approximately correct responses |
| Main part | Duration 1.8 second | 3 cm | 2.5 – 3.5 cm |
| 1st auxiliary part | Length 3 centimeters | 1.8, 1.9 seconds | 1.7 – 2 seconds |

It should be noted that in a number of cases observers gave correct responses not in the main experiment, but only in the first auxiliary part of the experiment. As mentioned above (see p.8 and Table 3) observers were presented space intervals served before as stimuli in the main part to evaluate them by pressing the button of a timer, as if they had to “give back” these space interval times.

## Indirect evidences

Table 17

The phenomenal equalizing of the numerical values of conjugate pairs elements by the widening and contracting of conjugated spatial and temporal signals (the conjugate pair 1.8 sec – 3 cm)

| Parts of the experiment | Presented signals and their numerical values | The kind of an estimate | | Numerical value of a conjugated with this signal interval |
|---|---|---|---|---|
| | | Overestimate numerical value | Underestimate numerical value | |
| The 2nd auxiliary part | Duration | Evaluated as | | Centimeters |
| | 1.8 seconds | **3**, 2.9, 3.1 sec | | **3** centimeters |
| The 3rd auxiliary part | Length | | Evaluated as | Seconds |
| | 3 centimeters | | 1.7, **1.8**, 1.9 centimeters | **1.8** seconds |

As shown on table 17, when estimating the time signal duration of 1.8 seconds, an observer magnified this duration into 3 seconds thus equalizing its numerical value with the numerical value of centimeters of this conjugate pair second element. On the other hand, in estimating the space interval length of 3 centimeters an observer "minimized" it into 1.8 centimeters thus equalizing its numerical value with the numerical value in seconds of this conjugate pair first element. Of interest is the fact that an observer made such a reduction when he drew this length on paper while seeing it on the screen of the apparatus.

## The Hypothetical Mechanism of Processes of the Correlation between Perceived Lengths and Durations which was based on the present experimental data

In connection with the aforementioned, the following questions arise.
1. As a result of what appears to be the leveling off (equalization) of the numerical values of the conjugate pairs elements is it a.) A result of the correlation of these elements or b.) Is it the prerequisite of correct correlation?
2) What function does this leveling off provide in the processes of the correlation of space and time in human perception?
3) As shown in the previous paragraph (see p. 87, table 17), the phenomenon of the leveling off of numerical values of both elements of the conjugate pairs served as indirect evidence of the fact that the observer recognized the connection between the conjugate intervals, although he or she could not always express this connection in the main part of the experiment by a correct response. Why does this difference appear between how an observer established the connection of the presented temporal signal with a certain space interval, and how does he or she respond to this signal? 4) What kinds of measurement systems participate in the work of the mechanism of the correlation between perceived lengths and durations?
Answers to these questions require special interdisciplinary studies in the areas of psychology, linguistics, the physiology of higher nervous activity, biology, physics, chemistry, and mathematics, since these questions can concern not only the mechanisms of the correlation of space and time in the human perception, but also other cognitive processes. These processes include thinking, memory, and speech.

Based on the results of the pilot and primary experiments, and in particular on the data of the 2nd B series of the primary experiment, which was carried out under conditions of a high level of uncertainty, the following hypothesis regarding the mechanism of the correlation of perceived space and time intervals can be offered.

**Measurement systems that participated in the correlation processes**

Based on the data analysis of the experiment one can suggest an existence of two interacting measurement systems that participated in the correlation processes: the **biological metrics** and the **social metrics**

**Biological metrics is** the innate metric system which is used by both people and animals for orientation and organization of their motion in space and time. It can be suggest that the basis of biological metrics is the rhythmical processes of the organism, especially endogenous rhythms, which are relatively independent of the surrounding influences and support a homeostasis and stationary proportional relationship between rhythmical activities of the organism's different parts.

**Social metrics is** a generally accepted system of measures and standards which people learned from a society where they live and use in their daily cognitive activity. Social metrics is acquired and applied jointly with different logical operations and methods (generalization, limitation, classification, categorization, analysis, synthesis) by persons on the basis of and by means of a native language or other systems of social symbols such as the deaf – and – dumb alphabet they have mastered.

**Processes and the hypothetical mechanism of the of the correlation between perceived lengths and durations**

Results of the primary experiment series showed that the correlation of space intervals with time intervals will be successful if observers possess information about the limits of the sets of presented intervals and the modules which can serve as the measurement unit of these intervals, and also about other components and parameters of such sets and conditions of their presentation. The more complete this information is, the lower the level of signals presentation uncertainty will be, and the more success observers will have in recognizing conjugate pairs of perceived space and time intervals and the more they will give correct responses already during the first presentation of these intervals.

In subseries A of the primary experiment, which was carried out under a very low level of uncertainty, the quantity of correct responses of observers in the first signals presentation considerably exceeded this quantity in all the rest of the series and in particular in the 2nd B series of the primary experiment which was carried out under conditions of a high level of uncertainty. It was only during the second signals presentation that observers who participated in the 2nd B series of the experiment reached the same quantity of correct responses that were already reached during the first presentation of these signals observers participated in the subseries 1 a (See Table 11). This fact indicates that in the 2nd B series of the primary experiment, during the first time signals presentation, a preparatory process had taken place which laid the basis for the successful recognition of conjugate pairs in the second presentation of these signals. Based on this fact, in the processes of the correlation of temporal and spatial intervals under a high level of uncertainty conditions, **one can isolate two stages: the adaptation and the recognition or activation stages**. During the adaptation stage an observer searches for a corresponding modulus, scale, its limits, correlation methods, and constructs a mental scale of possible stimuli. During the recognition or activation stage the above are applied for correlation purposes. We suggested that a certain conjugate pair would be recognized if a corresponding point on the scale would be activated by the influence of a corresponding time signal. Processes of the adaptive stage occur in essence during the first presentation of signals and are accomplished in the following sequence.

1. **Standard Internal Rhythm**.

Internal rhythm appears after the presentation of the first signal. This rhythm brings to a rhythmical state the perceptive images of temporal signals and spatial intervals, placed on the screen of the apparatus. Continuous stimuli - temporal and spatial signals are divided into discrete elements according to frequency characteristics of the emergent internal rhythm. The presented temporal and spatial intervals are ranked and are correlated by a value and a quantity of elements, which arose as a result of this division. In the process of this ranking, the search and the determination of the upper limits of the presented sets of stimuli occur, i.e. the longest intervals of space and time in these sets. As shown in Table 5 (1) (appendix 1), in the 2nd B series of the experiments, the greatest quantity of correct responses was given to the presentation of longest temporal intervals and to those intervals which were closest to them by their length.

The idea of the standard internal rhythm is consonant with contemporary ideas of a central internal clock based on either an oscillatory processes

(M. R. Jones, 1976; M. R. Jones & Boltz, 1989; Large & Jones, 1999; Barnes & Jones, 2000; Large, 2008; McAuley & Jones, 2003; Church and Broadbent, 1990; Treisman, Faulkner, Naish, and Brogan, 1990; Treisman, M., Faulkner, A., Naish, P. L. N., & Brogan, D., 1990) or on a pacemaker–counter device (Creelman, C. D. 1962; Rammsayer & Ulrich, 2001; Treisman, 1963; Treisman, M., Faulkner, A., Naish, P. L. N., & Brogan, D., 1990; Allan and Kristofferson, 1974; Gibbon, 1977, 1991, 1992; Gibbon, J., 1977; Gibbon, J., 1992; Wearden, 2003;Meck, 2003; H. Eisler, 1975, 1976, 2003;  Eisler, H., 1976; H. Eisler et al., 2008; Killeen &Weiss, 1987; Church, 2003; Block & Zakay, 2008; Grondin, 1993, 2003; Grondin, S., 2003; Gibbon, Church, & Meck, 1984). The difference between the first and second ideas is that the function of a standard internal rhythm is to divide both for correlation purposes, temporal and spatial signals into discrete elements whereas a central internal clock is intended for a time calculation

2**. Creation of the modular mental formation**

During the adaptive stage of the correlation of spatial and temporal intervals, observers in their responses to temporal signals of different duration selected spatial intervals whose length was longer than that which according to the condition for the experiment must be conjugated with these signals.

Acting thus, an observer searched among the signals presented to him or her the time interval which more closely approximated according to its size, a maximum limit of the set of spatial intervals. The purpose of the motion of this selection to the maximum limit is the scope of the entire presentation of this set as a whole. However, as can be seen in Confusion Matrices: Tables 1(3), 2(4), 3(4), 4(4), and especially 5(4) from Appendix 1, the absolute majority of observers in their responses to short and even to middle duration temporal signals, observers selected those spatial intervals which were considerably less than the maximum limit interval and intervals nearest to it along their length. At the same time, in their responses to the presentation of the longest signal from the presented set of time intervals, and also the closest signals to it in their duration, observers selected the longest space interval and intervals closet to it in their length. As it was shown in Table 5 (1) (Appendix 1) in the 2 B series of the experiments the nearer a presented time signal was to the maximum limit the greater number of correct responses was received to its presentation. The greatest quantity of correct responses was given to the presentation of a very prolonged time signal of 6.3 seconds. Such an increase in the quantity of correct responses to the presented minimum limit time interval, and intervals which were adjacent to it by their duration was not observed during the adaptive stage.

Taking into account the aforementioned information, it is possible to make the assumption that in the processes of the correlation of the intervals of space and time, **the innate mechanism of proportionality** comes into play already during their first presentation. This mechanism does not allow one to correlate, for example, the shortest temporal signal with the longest spatial interval or intervals which are close to it by their length even if this temporal signal was produced at the very beginning of the experiment before the presentation of other longer signals. First of all, the mechanism of proportionality makes a search, then correlates and establishes the connection between the longest time interval and the longest interval of space from the presented sets of intervals. Secondly, such connections are established between two other intervals which are close to the first by their duration and length. The establishment of the connection between maximum limits of sets of space and time intervals, and intervals which are close to these limits by their values leads to the creation in the observers **mental formations which does not reflect the stimuli themselves but the relationship** between them. In other words these mental formations are not the simple reflection of the combination of the space and the time intervals assigned in the experiment as conjugate pairs, but are the integral undifferentiated structure with their specific features. In contrast to the assigned conjugate pairs where each of the elements could have a different numerical value of their sizes, the numerical value of sizes of the elements of the mental formation is one and the same. The estimations of the sizes of conjugate pair **elements** were produced by using units of *different measuring systems.* The estimations of the sizes of **mental formations** are produced by means of *the units of one and the same measuring system*. This assumption can be confirmed by the above-mentioned results of the analysis of observers' estimations of intervals of space and time, after the accomplishment of their task for the correlation of these intervals. It also answered the questions stated on a special questionnaire that was presented to observers at the end of experiment. As shown in Table 15 a and b, and Table17, in observer's estimations of space and time intervals when presented separately, there was observed overestimation of time intervals and the underestimation of space intervals. As a result, the numerical values of conjugate pairs elements were equalized. Time and space seemingly moved towards each other and the mental process of observers attempted to merge one with another.

During  the experiment observers produced different mental and locomotor actions both implicitly and explicitly while he/she chose a corresponding length to response to a presented certain time signal. Such information was obtained by two ways: 1. by observation of observers' behavior while they performed experimental tasks; 2. by a mention above special questionnaire The observation showed that while performing correlation between perceived durations and lengths, many observers accompanied their search for the corresponding interval of space in response to the time signal by rhythmical rapping by foot on the floor, tapping by palm on the table or by rocking (oscillation) their bodies from the one side to another.

Answering questions about the questionnaire, the majority of observers indicated that during the correlation of time and space intervals they produced a rhythmical calculation which referred to the determination neither of the quantity of seconds nor of the quantity of centimeters of the presented intervals. As an example let us give the answers of one of them.

Questions

1. Did you count when the signal was produced? (The 1st variant is meant when the signal's duration was compared to the one of the pipes.)

a) Was the count free? What form did it take? What type of a count was it? Describe it in detail.

b) Was there an attempt to count seconds?

c) Was there an attempt to define lengths in centimeters done?

d) If there were such attempts, how did you manage to do it? Possible answer: made use of imaginary watch and ruler or some other objects? Or did not use them. Then what means did you use?

Answers

James H.

1. Yes. Free count, no relation to seconds or centimeters or watch or ruler

a) Free-internal counting and comparison of signals to establish a norm

b) No

c) No

d) Does not apply

The mentioned mental formation in which space and time were integrated together and the estimation of sizes which were determined by means of units of one and the same measuring system, which we called the ***mental extent***.

A quantity of correct responses to one or another time signal can serve as one of the indices in the priority of forming the mental extents from the assigned set of conjugate pairs. The more correct responses to a certain time signal that were obtained, the earlier a mental extent was formed on the basis of that conjugate pair which comprised this signal.

During the first presentation of signals, more than 50% of correct responses were obtained to the presentation in only three of fourteen signals. A quantity of correct responses to the remaining 11 signals was below 50%. As has been said above, the greatest quantity of correct responses were given to the presentation of the signal of 6.3 seconds (79.5%), which was the maximum limit of the presented set of signals. After that a large number of responses followed the near-limit signal of 6 seconds or 62.8% and the signal of 5.4 seconds or 51.3%. (Table 5 (1) in appendix1 shows these data). On the basis of these data, it is possible to determine that first of all, the mental extent appears to be on the basis of the established connection between the maximum limits of presented sets of space and time intervals (the conjugate pair of 6.3 seconds - 10.5 centimeters). And secondly, this extent appears to be on the basis of the established connection between the intervals which in their length- duration are nearest to the maximum limits (conjugate pair of 6 seconds - 10 centimeters). The mental extent that has arisen on the basis of the established connection between the maximum limits of the presented sets of space and time intervals such as the conjugate pair of 6.3 seconds - 10.5 centimeters let us name the maximum limit of mental extents. Now let us name the minimum limit of mental extents such as the conjugate pair of 0.3 seconds – 0.5 centimeters, that was formed on the basis of the established connection between the minimum limits of the presented sets of space and time intervals. (Dimensions of this conjugate pair are indicated below).

Formed as a result of the first correlation, mental extents were compared with each other. It has been determined that there was a difference in values between the maximum limit of mental extents and the mental extent which was formed secondarily after the formation of the maximum limit and its nearest to this limit value.

This difference was separate and took shape as an independent mental count that did not have an analog in the presented physical stimuli; however it became a measure for correlation nevertheless. This measure has been named a **modular mental formation** which possesses the following special features:

a) It served as a constant of proportionality in the processes of the correlation of space intervals with time intervals.

b) It has something in general to do with respect to the relation between both space and time intervals. Reports made by observers attested to the fact that they determined the solution to the problem that produced the rhythmical calculation, which were determined by neither the quantity of centimeters nor the quantity of seconds, while they correlated presented space and time intervals.

c) It appears to be the incremental divider of perceived time signals into equal intervals.

d) It is the smallest interval that serves as a minimum limit on the mental scale of possible mental extents. It is the starting point in the process of establishing an incremental scale of intervals that are separate from other relevant time signals. On this scale a certain number of step intervals are located in accordance with a quantity of intervals that are isolated from the evaluation of a certain presented time signal.

e) It is the principle of organizing a mental predictive space / time structure of a set of possible conjugate pairs of space and time intervals, according to which any conjugate pair is a combination of a certain quantity of one and the same single step measure.

**3. Creation of a temporary mental scale of possible stimuli**

Influenced by the modular mental formation, a change in the period of initial internal rhythm occurred after the presentation of the first signal. This period became equal to the value of the period of the formed module rhythm. Thus a transformed internal rhythm was superimposed on the received temporal signals and on the spatial intervals which were located on the panel on which the observer was tested. This allowed the possibility for an observer to calculate lengths and durations. Modular mental formation divides the maximum limit of a set of mental extents into many equal intervals. Because of this division a continuous limit was converted into the discrete mental formation which was used by observers for the commensuration of those compared in the process of correlation of space and time intervals that were different in their values. The indicated mental formation used by observers for the correlation of the intervals of space and time, have been named as a temporary mental scale of possible stimuli. The modular mental formation which does not have an analog in the presented physical stimuli became the starting point, and the beginning of this scale - its minimum limit. The last division or point of this scale designates the maximum limit of mental extents which appears on the basis of interaction of the longest interval of space and longest time interval from sets of presented physical stimuli. Each subsequent point, after the starting point towards to the maximum limit is longer in comparison with previous mental extents. The quantity of all points of a temporary mental scale of possible stimuli exceeds a quantity of physical stimuli in the presented sets of signals.

The indirect confirmation of this proposed hypothesis can be found in the answers which were given by observers to questions in the questionnaire about their mental actions which they performed during the correlation of space and time intervals presented to them. Observers, who described these mental actions in detail, indicated that during the correlation of the presented intervals they used as ruler for measuring all remaining shorter intervals as compared to the longest of them (See appendix 4)

A temporary mental scale of possible stimuli formed in the process of the correlation of intervals possesses the following special features:

a) It contained information about the quantity of all possible conjugate pairs which can be presented to an observer on the basis of the relationship assigned in the experiment.

b) It was common for both spatial and temporal signals as a modular mental formation.

c) It could be the function of both with the support of the actual perceived physical stimuli or without the support of them.

The hypothesis about the forming and functioning of a temporary mental scale of possible stimuli can also be confirmed by facts and phenomena which will be discussed in the next paragraph.

## The activation and testing of points on the scale of possible stimuli
## The Phases of Conjugate Pairs Recognition

The results of the experiments showed that a quantity of correct responses to the signals of different durations from the presented set of stimuli was distributed unevenly. According to the above mentioned classification (see page 19) conjugate pairs were divided by a quantity of correct responses and a ratio between them and approximately correct responses to presented signals into strong, relatively strong, weak, and weakest pairs. The first three of these were placed in the category of different degrees of recognized conjugate pairs. The weakest conjugate pairs were designated as the absence of recognition. It was as if the signals which were elements of these pairs, were not presented completely to observers. As stated above (see p. 45) in the $2^{nd}$ B series of the primary experiment, which was carried out under conditions of a high level of uncertainty during the first signals presentation, five of fourteen presented conjugate pairs remained unrecognizable. The majority of strong and relative strong pairs (4 of 6) arose on half of the scale of the presented stimuli which adjoined the maximum limit of this scale. The absolute majority of the very weak, i.e., unrecognized pairs (4 of 5) were located on half of the scale which adjoined the minimum limit (see table 8 e). Nevertheless on this part of the scale, together with the unrecognized pairs, also recognized pairs were observed: two relatively strong (1.2 seconds - 2 cm; 2.4 seconds - 4 cm) and one weak pair (0.9 second - 1.5 cm – the minimum limit of the scale of the presented signals). As has been stated earlier, on this half of the scale the presented signals corresponded with the conjugate pairs that were arranged evenly step by step. The size of each step corresponded to the size of the module. Among the signals located on the regular part of the scale which adjoined the minimum limit of the presented set of stimuli, the greatest quantity of correct responses was obtained to the signal 2.4 seconds. This signal was the element of the conjugate pair of 2.4 seconds - 4 cm that according to the classification of levels of the recognition of conjugate pairs was designated as a relatively strong pair. Conjugate pairs of 2.1sec - 3.5 cm and 2.7 sec - 4.5 cm, which were located on the scale of the presented stimuli before and after the conjugate pair 2.4 sec - 4 cm, during the first signals presentation remained unrecognized. This phenomenon was named by us as an Outburst of Recognition Activity in the Middle Part of the Scale - (ORAMPS phenomenon (p. 45). As it was shown in Table 8 (e), the ORAMPS phenomenon appeared in the sixth place of the scale of the presented stimuli or in the eighth place of the scale of possible stimuli in the middle of that part of the scale, where five unrecognized pairs were arranged. It was as if observers could establish the connection between elements of the conjugate pair 2.4 sec - 4 cm and could not establish connections between elements of five unrecognized conjugate pairs (1.5 sec - 2.5 cm, 1.8 sec - 3 cm, 2.1 sec - 3.5 cm, 2.7 sec - 4.5 cm, 3.3 sec - 5.5 cm).

All these conjugate pairs which were not recognized during the first presentation of signals were recognized during their second presentation. There were different degrees of recognition. Facts and phenomena described above cause two questions: 1. Why were conjugate pairs, which during the first presentation of signals not recognized by observers, but were in different degrees recognized during the second presentation of these signals? When was the connection between the elements of these conjugate pairs established? Was it during the first or only during the second presentation of temporal signals? 2. Why during the first presentation of signals does the intensive recognition of the conjugate pair of 2.4 seconds - 4 cm, located on the scale of the presented stimuli appear in the middle of the zone of unrecognized conjugate pairs? According to the hypothesis stated above (p.93 Creation of the modular mental formation), the innate mechanism of proportionality formed the mental modular formation and the mental measuring scale of the correlation of possible stimuli during the first presentation of signals in the minds of observers. Each point of this scale reflected the relation between the elements of the specific conjugate pair. The modular mental formation became a starting point of the scale and its minimum limit. The last incremental point of this scale was the maximum limit of mental extents. On the basis of this hypothesis, the connections between elements of all possible conjugate pairs were already established during the first presentation of signals in this experiment. Nevertheless, observers gave the greatest number of correct responses during the second signals presentation. In other words, the mass recognition of the conjugate pairs did not begin immediately, but with some retardation after the formation of connections between their elements. The following hypothesis, in our view, can serve as the most probable explanation to this fact. The formation of the connection between elements of the specific conjugate pair itself is insufficient for its recognition. So that this recognition could take place, it was necessary that the point of the scale which presented this connection should be activated. The activation of a scale point brings this point into the state of selective readiness to react to only the specific signal which coincided in its size with one of the elements of the recognized conjugate pair. Thus, it was possible to isolate three phases of the recognition of conjugate pairs –

1.The formation of a connection between two elements of the conjugate pairs in the process of the forming of a mental measuring scale of the correlation of possible stimuli. –

2. The activation of points of the scale which presents these connections i.e., they bring this point into the state of the selective readiness to react to only the specific signal which coincides in its size with one of the elements of the recognized conjugate pair –

3. The correct or approximately correct response i.e., the cognitive action which identified the established connection between the elements of the specific conjugate pair in the form of motor reaction, speech or other forms of the expression of this identification.

**Processes of the activation of points of the scale of possible stimuli after the completion of this scale creation**

In the first phase, the formation of the connection between elements of conjugate pairs is the basis not only of processes of these pairs recognition, but also as a one of the means of establishing and making more precise the calibration of the mental measuring scale for the correlation of possible stimuli. This connection can exist both in the passive (hidden) and in the active mode. In the first case the response to the appropriate signal will be incorrect, although the connection between the two elements of the conjugate pair which refers to this signal will be correct i.e. will correspond to the relationship of the intervals assigned in the experiment. This difference between the correct establishment of the connection between elements of the assigned conjugate pair and the incorrect response to corresponding to this pair signal was shown in the analysis of the estimations by observers of the sizes of such elements in the auxiliary parts of the experiment. The quantitative values of time and space intervals that were elements of one and the same assigned conjugate pair were equalized in these estimations. On the one hand, this phenomenon indicates specific peculiarities in the correlation of the time and space mechanism in the human psyche, which should be studied. On the other hand, it is the indirect confirmation of the fact that the correct establishment of the connection between the elements of the assigned conjugate pair in the experiment took place in spite of the incorrect response of the observer to the appropriate signal. Passage into the active mode of connections between elements of conjugate pairs (second phase of recognition) occurred unevenly and to the different degrees both during the first and second presentation of signals. [7]

The process of activating points on the scale can be divided into two stages. The basic elements of the mechanism of correlation of intervals during the first stage are activated such as maximum and the minimum limits of the scale of possible stimuli, and a modular mental formation. As was shown above, modular mental formation is the unit of measurement of the presented signals that establishes the relations between them, and at the same time it is the minimum limit of the mental scale of possible mental extents and the primary reference point. From this point on the scale, a certain quantity of step intervals, in accordance with a quantity of intervals isolated with the evaluation of the certain presented time signal is located. The first stage in the majority of observers was observed during the first presentation of signals.

---

7

A comparative analysis of the results of the conjugate pairs recognition in three series of the primary experiment during the first presentation of signals, showed the essential difference in the structure of the distribution of correct responses on scales of presented stimuli of these series. The quantity of stimuli presented in each series of the primary experiment was identical: 14 spatial and 14 temporal intervals. These intervals were selected for the presentation from sets of possible stimuli which consisted of 21 spatial and 21 temporal intervals. Thus in all series of the primary experiment, observers were presented with “shortened” sets of intervals each of which lacked seven physical stimuli for their completeness. Such “shortened” sets of intervals as mentioned above, underlie the partially complete scales of the presented stimuli. One should remember that these scales consisted of two parts: uniform and non-uniform scales.

The uniform part – in the first part of the partially complete scale, conjugate pairs were placed regularly step by step on each point of this scale. The non-uniform part – in the second part of the partially complete scale, conjugate pairs were located irregularly where adjacent conjugate pairs could be located several steps away from each other. In the primary experiment, two types of partially completed scales were used. In the first type of the scale the uniform part began from the first point of the scale of possible stimuli. This point was, at the same time, a minimum limit and a modular conjugate pair 0.3 sec - 0.5 cm. Such a type of partially complete scale was used in the 1$^{st}$ series of the primary experiment. The second type of the partially complete scale contained the uniform part that began from the third point of the scale of the possible stimuli (conjugate pair 0.9 sec - 1.5 cm) and served as the minimum limit of the scale of presented stimuli. This type of scale was used in series 2 A and B of the primary experiment.

Thus, the essential difference between the first and second type of the partially complete scale of the presented intervals consisted of the following. Although a quantity of presented stimuli in both sets of presented intervals was equal, a basis of the first type of the partially complete scale was the set of stimuli which contained intervals that served as a basis for the formation of three basic elements of the correlation process: the maximum and minimum limits of the scale of possible stimuli, and a modular mental formation. The set of stimuli that contained the interval which formed only one element of the correlation process, the maximum limit of the scale of possible stimuli was the basis for the second type of scale. Intervals which served as the basis for the formation of the minimum limit and a modular mental formation, in this set were absent. Observers that participated in the 1st and 2nd B series of the experiments where different types of this scale were used were given preliminary information about neither limits of these scales nor about a modulus as a unit of measurement of presented signals. Only observers that participated in 2nd A series and in the subseries 1a possessed such information. The comparison of data of the 1st and 2nd B series of the primary experiment showed the essential difference between them in the distribution of correct responses along the scale of presented stimuli during the first signals presentation. In the 1st series where the first type of the partially completed scale of the presented intervals was used, there contained signals that served as the basis for the formation of three basic elements of the correlation process. They were the maximum and minimum limits of the scale of possible stimuli and modular mental formation. The greatest quantity of correct responses was given to the presentation of precisely these signals, and also to signals which adjoin them.

The greatest quantity of correct responses was given to the presentation of the signal 0.3 seconds of one of the elements of the conjugate pair 0.3 sec - 0.5 cm. That was the same time as the minimum limit of the scale of possible stimuli and as a modular mental formation. The quantity of correct responses to presented signals became less the further away these signals were located from both limits of the scale of presented stimuli.
However, this decrease was not uniform.
The quantity of correct responses could be of greater statistical significance to the signal located nearer to the middle of the scale than to a signal located nearer to the limit of this scale. Such a signal is, for example, the time interval of the 2.4 seconds (conjugate pair of 2.4 seconds - 4 cm). This conjugate pair is located in 8th place both on the scale of presented stimuli which consists of 14 points, and on the scale of possible stimuli which consists of 21 points. As was shown on page 37, a quantity of correct responses to the signal of 2.4 seconds was exceeded by a statistical significance the quantity of such responses both to the previous signal of 2.1 seconds, that was located on the side of the minimum limit, and to the subsequent signal of 3.3 seconds, located on the side of the maximum limit. This signal was located in the second non-uniform part of the scale of presented intervals where all seven possible signals of 2.7, 3.0, 3.9, 4.2, 4.8, 5.1, and 5.7 seconds could be placed. In the 1st series on the scale of the presented intervals between signals of 2.4 seconds and 3.3 seconds, the signals of 2.7 and 3.0 seconds could be placed. The conjugate pair of 3.3 seconds – 5.5 cm during the first presentation of signals remained unrecognized (a weakest pair), i.e., a quantity of correct and approximately correct responses comprised less than 50% of the total.

Observers responded to this signal of 3.3 seconds as if the interval of 5.5 cm was absent on the apparatus screen. It was as if in reality there were no intervals of 4.5, 5, 6.5, 7, 8, 8.5, and 9.5 cm. Observers' overestimated responses to the signal of 3.3 seconds were considerably larger, exceeding 69% (see p.17 and Table 3 (4) from appendix 1) more than to the other signals. As a whole, in the 1st series the overestimated tendency, i.e., the motion of the response distribution to the side of the maximum limit exceeded the underestimated tendency more than three times as large. This overcompensation (predominance) although in considerably smaller sizes, remained also during the 2nd time signals presentation despite the fact that a maximum quantity of correct responses from observers were given to the presentation of the signal of 0.3 seconds, i.e., the minimum limit of the scale. As it was shown on page 77, during the second signals presentation in the 1st series, a quantity of correct responses to the presentation of the maximum limit sharply decreased while these responses to the presentation of the minimum limit sharply increased. Thus, in the 1st series of experiments during two time signal presentations, three phenomena were observed which have relevance to the organization of the correlation of conjugate pairs elements and the conjugate pairs recognition.

1. During the first time signals presentation, the quantity of correct responses to the minimum signal of 0.3 seconds was greater than to the signals of other lengths and even to the longest signal of 6.3 seconds which was the maximum limit of the presented signals. During the second presentation of signals, a quantity of correct responses to the presentation of the shortest signal sharply increased to 88.2%, and sharply decreased to the presentation of the longest signal, 50.8%. At the same time the quantity of correct responses to all remaining signals also sharply increased including the signal of 6 seconds (65.9%) adjacent to the maximum limit.

2. During the first presentation of signals, a quantity of correct responses to the signal of 2.4 seconds was considerably more than to the adjacent signals located also in the middle part of the scale of the presented stimuli. Such a difference in the quantity of correct responses to these signals indicated that the appearance of the internal limit divided the scale into two parts. One of them adjoined the minimum limit that was simultaneously the element of a modular pair and consisted of eight points. Another part adjoined the maximum limit. A quantity of points on this part of the scale can be considered in two ways. On the scale of presented stimuli in the 1st series, this part was non-uniform, and consisted of six points. On the scale of possible stimuli there were thirteen points. A question about the quantity on points of the scale which observers used for the correlation of presented intervals will be examined below.

3. There was a predominance of the overestimated tendency or the motion of the distribution of incorrect responses to the side of the maximum limit over and above the underestimated tendency or the motion of the distribution of such responses to the side of the minimum limit. This predominance was observed more than three times during the first signals presentation, and remained although in smaller sizes during the second presentation. This occurred in spite of the maximum concentration of correct responses to the presentation of the signal of 0.3 seconds, or the minimum limit of the scale which was the element of the modular pair of 0.3 seconds - 0.5 cm.

Thus, in the 1st series of experiments during the first presentation of signals, three basic elements of the mechanism of the correlation of these signals were formed and activated: the minimum limit, the modular pair, and the maximum limit of the scale of presented intervals. Furthermore, inside this scale the limit was formed which divided the scale into two unequal parts, one of which adjoins the minimum limit and the other the maximum limit. The comments made by observers mentioned earlier in this paper described their mental actions during the experiment and the experimental data allowed one to make the following assumption: The scale which observers used for the correlation of intervals presented to them in a quantity of points considerably **exceeded** the scale of the intervals presented to these observers and coincided with the scale of possible stimuli.

The predominance of the greater tendency which also remained during the second presentation of signals can be explained by the tendency of the observer's perceptive system to activate those points which were used by them, and which did not coincide with the points of the scale of presented intervals. The points of the scale used by observers that did not coincide with points of the scale of the presented intervals could be located only in that part of the scale which adjoined its maximum limit. In that case the part of the scale which adjoined the maximum limit contained not six, but thirteen points, six of those were supported by the physically presented stimuli and seven not supported by such stimuli. (See above data). The part of the scale which adjoined the minimum limit contained only eight points supported by the physically presented stimuli. Thus, having appeared during the 1$^{st}$ time signals presentation, internal limit divided the formed mental scale into two parts according to the principle of the golden section a 13/8 ratio. This assumption became more convincing if one compared the similarity and differences in the data of the 1$^{st}$ and 2$^{nd}$ B series of the primary experiments.

As previously stated, the structure of the scale of presented signals in the 2nd A and B series was different than in the 1st series. In the 2nd series the minimum limit of the scale of the presented stimuli was conjugate pair 0.9 sec - 1.5 cm which began the uniform part of this scale. In the 1st series, this conjugate pair occupied third place on the scale of presented stimuli, and likewise on the scale of possible stimuli. In the set of time and space intervals in the 2nd A and B series, two minimum conjugate signals 0.3 sec - 0.5 cm and 0.6 sec - 1 cm were removed and substituted by longer conjugate signals 2.7 sec - 4.5 cm and 3.9 sec - 6.5 cm. Thus, in the non-uniform part of the scale used in the 2nd series it could be placed not 7 possible conjugate pairs, as occurred in the 1st series, but only five (3 sec - 5 cm, 4.2 sec - 7 cm, 4.8 sec - 8 cm, 5.1 sec - 8.5 cm and 5.7 sec - 9.5 cm). Two possible minimum pairs of 0.3 sec - 0.5 cm and 0.6 sec - 1cm would be possible, placed only before the minimum limit of the scale of the presented stimuli 0.9 sec - 1.5 cm. The uniform part of the scale of the presented stimuli in the 2nd series was located inside the non-uniform part. We may count possible conjugate pairs as elements of this part which were not reinforced by the physically presented stimuli. The basis of the scale in the 2nd series of the experiments was the set of stimuli that contained the interval which formed only one element of the process of the correlation: the maximum limit of the scale of possible stimuli. The intervals served as the basis for the formation of the minimum limit and the modular mental formation were absent in this set.

## Table 18

The comparison of results of recognition in the 1st and 2nd B and A series of the primary experiment (% of responses to all signals)

### Table 18 a

The 1st time signal presentation

The assumed complete scale of conjugate pairs in the main experiment

| | | Numbers of places on the scale of possible stimuli | | | | | | | | | | | | |
|---|---|---|---|---|---|---|---|---|---|---|---|---|---|---|
| Stimulus categories | | 1 | 2 | 3 | 4 | 5 | 6 | 7 | 8 | 9 | 10 | 11 | 12 | 13 |
| Duration | Seconds | 0.3 | 0.6 | 0.9 | 1.2 | 1.5 | 1.8 | 2.1 | 2.4 | 2.7 | 3 | 3.3 | 3.6 | 3.9 |
| Length | Centimeters | 0.5 | 1 | 1.5 | 2 | 2.5 | 3 | 3.5 | 4 | 4.5 | 5 | 5.5 | 6 | 6.5 |
| The 1st series | | **71.5** | **56** | 36.6 | 39.8 | 27.7 | 28 | 28 | 37 | | | 24 | 28.5 | |
| The 2nd B series | | | | 37.2 | **46.1** | 23.1 | 23.1 | 16.7 | **47.4** | 25.6 | | 30.7 | 25.6 | 32.1 |
| The 2nd A series | | | | **65.2** | **60.6** | **45.5** | 31.8 | 31.8 | 34.8 | 25.8 | | 30.3 | 34.8 | 37.9 |

Continuation

| | | Numbers of places on the scale of pos. stimuli | | | | | | | | |
|---|---|---|---|---|---|---|---|---|---|---|
| Stimulus categories | | 14 | 15 | 16 | 17 | 18 | 19 | 20 | 21 | |
| Duration | Seconds | 4.2 | 4.5 | 4.8 | 5.1 | 5.4 | 5.7 | 6 | 6.3 | |
| Length | Centimeters | 7 | 7.5 | 8 | 8.5 | 9 | 9.5 | 10 | 10.5 | |
| The 1st series | | | 39.8 | | | **44.3** | | **63.8** | **68.3** | |
| The 2nd B series | | | **41** | | | **51.3** | | **62.8** | **79.5** | |
| The 2nd A series | | | **54.5** | | | **53** | | **68.2** | **75.8** | |

Table 18 b

The 2nd time signal presentation

The assumed complete scale of conjugate pairs in the main experiment

| | | Numbers of places on the scale of possible stimuli | | | | | | | | | | | | |
|---|---|---|---|---|---|---|---|---|---|---|---|---|---|---|
| Stimulus categories | | 1 | 2 | 3 | 4 | 5 | 6 | 7 | 8 | 9 | 10 | 11 | 12 | 13 |
| Duration | Seconds | 0.3 | 0.6 | 0.9 | 1.2 | 1.5 | 1.8 | 2.1 | 2.4 | 2.7 | 3 | 3.3 | 3.6 | 3.9 |
| Length | Centimeters | 0.5 | 1 | 1.5 | 2 | 2.5 | 3 | 3.5 | 4 | 4.5 | 5 | 5.5 | 6 | 6.5 |
| The 1st series | | **88.2** | **68.7** | **59.3** | **50.8** | **43.1** | 32.9 | **45.1** | **40.7** | | | **45.5** | **50.4** | |
| The 2nd B series | | | | **65.4** | **64.1** | **59** | 33.3 | **39.7** | **57.7** | **48.7** | | **46.1** | **48.7** | **47.1** |
| The 2nd A series | | | | **84.8** | **66.7** | **68.2** | **39.4** | **50** | **57.6** | **39.4** | | **56.1** | **53** | **56.1** |

| Continuation | | Numbers of places on the scale pos. stimuli | | | | | | | | |
|---|---|---|---|---|---|---|---|---|---|---|
| Stimulus categories | | 14 | 15 | 16 | 17 | 18 | 19 | 20 | 21 | |
| Duration | Seconds | 4.2 | 4.5 | 4.8 | 5.1 | 5.4 | 5.7 | 6 | 6.3 | |
| Length | Centimeters | 7 | 7.5 | 8 | 8.5 | 9 | 9.5 | 10 | 10.5 | |
| The 1st series | | | **47.6** | | | **50.4** | | **65.9** | **50.8** | |
| The 2nd B series | | | **57.7** | | | **46.2** | | **53.8** | **65** | |
| The 2nd A series | | | **59.1** | | | **59.1** | | **62.1** | **54.8** | |

The comparison of the 1st and the 2nd B series data showed the essential difference and the individual cases of similarity which confirmed this difference in the structure of the distribution of correct responses along the scales of the presented stimuli. As can be seen from Table 18 a, during the first time signal presentation in the 2nd B series, in contrast to the 1st series, the maximum concentration of correct responses to the presentation of conjugate pair 6.3 sec - 10.5 cm took place which is the maximum limit of the scale of presented stimuli. The quantity of correct responses to the presentation of the maximum limit was more than twice the quantity of such responses to the presentation of the minimum limit. The quantity of correct responses during the 1st time signals presentation to signals located on the first half of scale that adjoined the maximum limit, considerably exceeded the quantity of correct responses to signals located on the second half of the scale which adjoined the minimum limit of this scale. In the first series, the quantity of correct responses to signals located on both parts of the scale were practically equal (see p.17, Appendix I, Tables 3 (3), 5 (3)).

The quantity of unrecognized conjugate pairs during the first presentation of signals in the 2nd B series considerably exceeded the quantity of unrecognized conjugate pairs in the 1st series of the experiment (5 and 2 of 14 presented in these series conjugate pairs respectively). Moreover in the 2nd B series, 3 out of 5 unrecognized pairs were arranged on half of the scale which adjoined its minimum limit. In the 1st series, each half of the scale contained only one such a pair. The overestimated tendency (the motion of the responses distribution to the side of the maximum limit) in the 2nd B series was like the 1st series but exceeded the underestimated tendency (the motion of the responses distribution to the side of the minimum limit) by almost four times. However, unlike the 1st series in the 2nd B series this excess during the second time signals presentation disappeared and both tendencies were equalized. This was due to a sharp decrease in the level of the overestimated tendency, and a sharp increase in the level of the underestimated tendency.

During the 1st time signal presentation in the 2nd B series, a quantity of correct responses to signals of 1.2 seconds and especially of 2.4 seconds (elements of conjugate pairs 1.2 sec - 2 cm and 2.4 sec -4 cm respectively) were located on the part of the scale which adjoined the minimum limit and was considerably more than to other signals located on the same part of the scale. The conjugate pairs 1.2 sec - 2 cm and 2.4 sec - 4 cm were arranged respectively on the 2nd and 6th places of the scale of the presented stimuli and on the 4th and 8th places of the scale of possible stimuli. The conjugate pair 2.4 sec - 4 cm in the 2nd B series was just the same as in the 1st series of the experiments. This conjugate pair was the internal limit of the scale which divided it into two parts, one of which adjoined to the minimum limit and another to the maximum limit. However, this internal limit was more clearly expressed in the 2nd B series. The quantity of correct responses to the signal of 2.4 seconds in the 2nd B series considerably exceeded the quantity of correct responses to this signal in the 1st series. There is an important difference between the ratio of the quantity of correct responses to signals which represent the internal limit and minimum limits of the scales of the presented stimuli in the 1st and 2nd B series. In the 1st series a quantity of correct responses to the signal of 2.4 seconds which represented the internal limit of the scale was considerably less than to the signal of 0.3 seconds which represented the minimum limit of this scale. In the 2nd B series the opposite was found. The quantity of correct responses to the signal of 2.4 seconds was considerably more than to the signal 0.9 seconds which in this case represented the minimum limit of the scale of presented stimuli. The differences in the ratio between quantities of correct responses to signals which represented the internal (2.4 sec) and the maximum (6.3 sec) limits of the scales of presented stimuli between the 1st and 2nd B series did not prove out in actual fact.

In both series, a quantity of correct responses to the signal of 6.3 seconds considerably exceeded a quantity of such responses to the signal of 2.4 sec. (see Table 18 a). This amount nearly exceeded the same percentage (in the 1st series 31.3% and in the 2nd B series 32.1%). The similarity in the relation of quantities of correct responses to the signals which represented the internal and maximum limits of the scales in the 1st and 2nd B series showed that the specific functions of these limits for organizing the mental scale of possible stimuli and the recognition of conjugate pairs in both of these series were identical. The detail of these functions will be examined below. However, the aforementioned differences between relations of the quantities of correct responses to the signals which represented the internal and the minimum limits of the scales of the presented stimuli in the 1st and the 2nd B series indicated the essential difference of such functions at the minimum limits of the scales of these series of experiments. As it was noted above, the minimum limit of the scale of presented stimuli in the 1st series was at the same time the minimum limit of the scale of possible stimuli, and the basis of modular mental formation. It should be noted that this modular formation served as the unit of measurement of the received signal, and as the method of calibrating and forming a mental scale of possible stimuli. Therefore this modular pair was activated under the effect of the real physical stimulus (a time signal) faster, and to a larger degree than other conjugate pairs located on the scale of presented stimuli. The minimum limit of the scale of presented stimuli in the 2nd B series was not the minimum limit of the scale of possible stimuli and did not serve as a basis for the modular mental formation. Therefore, although this limit was strengthened by a real physical stimulus, it was considerably weaker than conjugate pair 2.4 sec - 4 cm that served as an internal limit of the scale. During the first presentation of signals this conjugate pair was located in the middle of two nearest unrecognized, i.e., not activated conjugate pairs 2.1 sec - 3.5 cm and 2.7 sec - 4.5 cm. As to the conjugate pair 0.9 sec - 1.5 cm, the minimum limit of the scale of presented stimuli, the quantity of correct responses to the presentation of the signal 0.9 seconds in the 2nd B series was the same as the signal in the 1st series where the conjugate pair 0.9 sec - 1.5 cm was not the minimum limit of the scale but occupied it only in the third place (see Table 18 a). This fact made it possible to arrive at the assumption that an *observer in the 2nd B series considered the pair 0.9 sec - 1.5 cm not as the beginning of the scale of the presented stimuli, but as the continuation of the scale of possible stimuli which was formed at the beginning of the experiment, and was used by observers for the correlation of received signals.* The beginning of the scale of possible stimuli had to be conjugate pairs 0.3 sec - 0.5 cm and 0.6 sec - 1 cm. Those elements were presented in the 1st series of the experiment. Since in the set of stimuli of the 2nd B series, such intervals were absent their existence and the degree of their activation on the scale of possible stimuli could be revealed indirectly in the dynamics of the conjugate pairs recognition during the 1st and 2nd time signals presentation. As mentioned above in the 2nd B series during the 1st time signals presentation the top level of recognition was observed with respect to the conjugate pair of 6.3 seconds - 10.5 cm which was the maximum limit of the scale of the presented stimuli. The quantity of correct responses to the presentation of this pair was considerably more (79.5%) than to the presentation of the adjacent conjugate pair 6 sec - 10 cm. which was the second on the level of the conjugate pairs recognition (62.8 %). See Table 18 a for all data. A question arose as to why there was such a large difference (16.7%, $t = 3.7$, $p < .001$) in a quantity of correct responses between conjugate pairs which were so very close to their length duration? The difference in the length / duration between them composed 0.3 sec - 0.5 cm.

The following could be the most probable answer. In the 2nd B series the maximum limit fulfilled two functions.

1.The maximum limit participated in the shaping of a mental modular pair. The extent of the maximum limit was compared with the extent of other presented conjugate intervals. As a result of this comparison, the difference between the extent of the maximum limit 6.3 sec - 10.5 cm and the extent of the nearest value to it was the conjugate pair of intervals of 6 sec - 10 cm. that separated them. This difference (0.3 sec - 0.5 cm) became the independent unit of measurement of received signals and the method of calibrating the scale which was used by an observer for the correlation of received signals.

2. The extent of the maximum limit was the basis of the mental scale of possible stimuli. The calibration of this extent was by means of the mental modular pair that divided it into 21 conjugate intervals. As an example:

6.3 sec: 0.3 sec = 21

10.5 cm: 0.5 cm = 21

The mental scale of possible stimuli was obtained in such a way that it contained three basic components of the correlation mechanism: a true minimum limit and corresponding to it, the modular pair 0.3 sec - 05 cm which in 2nd B series unlike the 1st series of experiments <u>was not strengthened by the directly presented physical stimuli and was found in the latent state</u>, and the maximum limit which was strengthened by presented physical stimuli.

Thus the greatest quantity of correct responses to the signal representing the maximum limit, i.e. the maximum activation of this limit pair (in comparison with the rest conjugate pairs) can be explained by the fact that during the 1st time signals presentation the greatest mental energy departed because of the creation of the mental modular formation and the scale of possible stimuli. After its creation, this scale was divided by an internal limit into two parts according to the principle of the golden section. One part adjoined the concealed “true” minimum limit, and another the maximum limit. The purpose of this scale division into two parts, was the creation of conditions which facilitated the activation of the remaining points on the scale. After the formation of the mental scale, the mental modular formation apparently passed into the zone of this scale minimum limit. This was one of the reasons why the overestimation tendency i.e. the motion of the observers’ responses distribution to the side of the maximum limit sharply decreased during the 2nd time signals presentation in the 2nd B series (by more than twice), and the underestimation tendency, i.e., the motion of these responses distribution to the side of the minimum limit sharply increased (also by two times) and both tendencies were equalized. It is interesting that during the 2nd time signals presentation a quantity of correct responses as a whole in the percent ratio in the 1st series and the 2nd B series practically coincided (52.8% and 52.3% respectively).

**Methods of the activation of points of the scale of possible stimuli**

Despite the mentioned similarity in the quantity correct responses between the 1st and 2nd B series during the 2nd time signal presentation, the latent period of the observers’ correct responses from the 1st series of experiments was considerably shorter than in observers from the 2nd B and A series (3.27, 4.42, 4.4 seconds respectively, $t = 4.5$, $p < .01$). This difference in the latent periods of correct responses of observers from the 1st and the 2nd B and A series can be explained as follows. For the correct correlation of received signals and the recognition of conjugate pairs, it was necessary that the formed mental modular formation which measured these signals and correlated them with certain points on the mental scale of possible stimuli must be maximally activated.

One can show three methods of such activation.

1. The direct activation
2. The indirect activation
3. The artificial activation

In contrast to the artificial activation, the first two methods of activation can be named natural or spontaneous. The explanation of the differences between them will be given below.

1**. The direct activation.**

In the 1st set of presented stimuli, there were signals which were one of elements of certain conjugate pairs located on the mental scale of possible stimuli and exerted direct influence upon them. In the same set of presented stimuli of the 1st series of the experiments there were signals that composed the basis of three basic elements of correlation - minimum and maximum limits of the scale of possible stimuli and the modular mental formation.

**2. The indirect activation**

In the set of presented stimuli there were signals which were not one of the elements of a certain conjugate pair located on the mental scale of possible stimuli and directly exerted influence upon them. In particular there were no signals which composed the basis of two out of three basic elements of correlation – the minimum limit and the modular mental formation. Such a set of stimuli occurred in the 2nd A and B series. The mental modular formation was activated and operated indirectly through the scale of possible stimuli via the presentation of other signals which referred to other points of this scale. As the index of this action, a sharp increase of the underestimated tendency (the motion of the responses distribution to the side of the minimum limit) during the 2nd time signal presentation in the 2nd B and especially in the 2nd A series was shown. The direct influence of the signals which referred to the elements of a modular pair accelerated its activation as this was in the 1st series, whereas the indirect influence of signals which did not have relation to these elements, slowed the rate of this activation. Therefore the latent period of correct responses in the 1st series was considerably shorter than in the 2nd B and A series. In the processes of the recognition of conjugate pairs, direct and indirect activations appeared spontaneously, selecting points on the scale for their activation in a determined order.

**3. The artificial activation**

The preliminary information about the composition of the set of presented stimuli that were presented to an observer before the beginning of the experiment is one of the forms of artificial activation.

In the 2nd A series observers were given preliminary information about the minimum and maximum limits of the scale of the presented stimuli (in this case, the conjugate pairs 0.9 seconds - 1.5 cm and 6.3 seconds - 10.5 cm.). As a result, in the 2nd A series in contrast to the 2nd B series, a quantity of correct responses to the signal of 0.9 seconds grew sharply, artificially disguising the spontaneous activation of the conjugate pair 2.4 sec - 4 cm which was observed in experimental conditions of the 2nd B series. Recall that the activation of the conjugate pair 2.4 sec – 4 cm during the 1st presentation of signals in the 2nd series created an internal limit on the scale of possible stimuli dividing it into two parts according to the principle of the golden section. Regardless, the signs of the formation of this limit in the less distinct form could be observed during the first presentation of signals; also in the 2nd A series of the experiments. The quantity of correct responses to the signal 2.4 seconds exceeded the quantity of correct responses to the adjacent signal of 2.7 seconds by a statistically significant value and to the signal of 2.1 seconds by a statistically insignificant value. The quantity of correct responses to the signal of 2.4 seconds grew sharply during the 2nd time signals presentation, and this quantity became greater by a statistically significant value than a quantity of such responses to both these adjacent signals (See Table 18 b). Moreover, quantities of correct responses to the signal of 2.4 seconds during 2nd time signals presentation in the 2nd A and B series in a percentage ratio coincided completely (See Table 18 b). This attests to the fact that the processes of forming the mental scale of possible stimuli and its internal limit in the 2nd A and B series are identical. The difference consists in a degree of the observability of these processes in the 2nd A and B series during the1st and the 2nd presentation of temporal signals. The formation and observation of the internal limit of the scale of possible stimuli in the 2nd A series during the first presentation of signals was considerably weaker than in the 2nd B series. This occurred because of the artificial activation of the conjugate pair 0.9 seconds - 1.5 cm - that was the minimum limit of the scale of presented stimuli - in the 2nd A series at the beginning of the experiment which led to a weakening of the spontaneous activation of the internal limit of the formed scale of possible stimuli. During the 2nd presentation of signals the influence of this artificial activation disappeared. As a result, the spontaneous activation of the internal limit of the conjugate pair 2.4 seconds - 4 cm was sharply strengthened. The observability of the formation of this limit in the 2nd A series became the same as in the 2nd B series. The similarity of the processes of forming the mental scale of possible stimuli used by observers for the recognition of conjugate pairs in the 2nd A and B series was also observed in the dynamics of the distribution of correct and incorrect responses, and in their latency. Despite this fact, observers were given preliminary information about both limits of the scale of presented stimuli, the quantity of correct responses to the presentation of the maximum limit in the 2nd A series during the 1st time signal presentation considerably exceeded the quantity of such responses to the presentation of the minimum limit. The quantity of correct responses to the presentation of the maximum limit in the 2nd A and the 2nd B series during the 1st time signal presentation nearly coincided (see Table 18 a). This meant that the perceptive system of observers in the 2nd A series used for the recognition of the maximum limit, was not about the preliminary information concerning the limits of the set of intervals, but the methods and operations whose sequence was spontaneously assigned by the mechanism of correlation in both the 2nd B and 2nd A series. The change in the dynamics tendencies of the distribution of observers' responses was one of such methods. In the 2nd A and B series the overestimated tendency, that is the motion of the distribution of the responses to the side of the maximum limit, was observed during the 1st signals presentation. During the 2nd time signal presentation the overestimated tendency was sharply changed by the underestimated tendency, or the motion of the distribution of responses to the side of the minimum limit. During the first presentation of temporal signals, a quantity of overestimated responses in these series considerably exceeded the quantity of underestimated responses. A quantity of correct responses to the presentation of the maximum limits in these series was also considerably more than to the presentation of the minimum limits. However, in the 2nd B series such overestimated was considerably greater than in the 2nd A series. During the 2nd time signals presentation in these series there was a sharp decrease in the quantities of overestimated responses plus correct responses to the presentation of the maximum limit. A sharp increase in the quantities of underestimated responses plus the correct responses to the presentation of minimum limits occurred. This change in the tendencies of the distribution of responses during the 2nd time signal presentation in both series was accompanied by a sharp increase in the level of the recognition of conjugate pairs (see Tables 4 (2) and 5 (2) Appendix 1). However regarding ratio of overestimated and underestimated, incorrect responses and the quantities of correct responses to the presentation of the maximum and minimum limits of the scale of stimuli during the 2nd time signal presentation, a difference could be noted between the 2nd A and the 2nd B series. In the 2nd B series the quantity of underestimated and overestimated responses and also quantities of correct responses to the presentation of the maximum and minimum limits of the scale of stimuli during the 2nd time signal presentation became equalized. See Tables 5 (1) and 5 (2) in Appendix 1. In the 2nd A series during the 2nd time signal presentation, a quantity of underestimated responses became considerably more than overestimated responses ($t = 3$, $p < .1$). The quantity of correct responses to the presentation of the minimum limit became considerably more than to the presentation of the maximum limit. The difference between them was 30.3%, $t = 4.1$, $p < .001$, (see Tables 4 (1) and 4 (2) in Appendix 1). Furthermore, a quantity of correct responses as a whole, and especially a quantity of strong conjugate pairs during the 2nd time signal presentation in the 2nd A series was considerably more not only than in the 2nd B series, but also in the 1st series where the module was represented in the presented signals. See Tables 3 (2), 4 (2), 5 (2) in Appendix 1; Table 2, 3, 4 in Appendix 3. At the same time as shown above, the latency of correct responses of observers in the 1st series of experiments was shorter than in the observers of 2nd A and B series (3.27, 4.44, 4.2 seconds respectively, $t = 4.5$, $p < 0.01$; (see table 6, Appendix 1). The difference in the latent periods of the correct responses of observers in the 1st series and the 2nd A series can be explained by the same hypothesis that was expressed above with respect to this difference in the correct responses of observers of the 1st series and the 2nd B series (see page 126). However, the difference in the levels of recognition of conjugate pairs between responses of observers of the 2nd series on the one hand, and of the 1st series on the other hand requires further explanations.

It is quite understandable that the level of recognition of conjugate pairs in the 2nd A series was higher than in other series of the experiment because it was connected to the artificial activation. In this case it was connected to the preliminary information about limits of the scale of the presented stimuli given to observers at the beginning of the experiment. However, the question about how this activation affected the processes of recognition of conjugate pairs remains open.

The answer to this question can be partially explained by the comparison of the special features of the dynamics of the observers' responses distribution in the three series of the experiment. As was shown above, during the first time signal presentation in observer's responses in all series of the experiment was that an overestimated tendency predominated over the underestimated tendency. During the second time signal presentation the overestimated tendency was reduced, and the underestimated tendency increased being accompanied by the growth of the level of the conjugate pair's recognition. However, the ratios between overestimated and underestimated responses during the first and especially the second presentation of signals in each of the series of experiments were different. During the first signals presentation the predominance of the quantity of overestimated responses over the quantity of underestimated responses was a 3.9 ratio difference in the 2nd B series, and a 3.3 ratio difference in the 1st series, and only a 1.9 ratio difference in the 2nd A series. During the second time signal presentation this predominance remained only in the 1st series of the 1.9 ratio. In the 2nd B series a quantity of overestimated and underestimated responses became equal, and in the 2nd A series there appeared the predominance of the underestimated responses above the overestimated to a ratio of 1.6 (See Tables 3 (2), 4 (2), 5 (2), the appendix I). Thus, the underestimated tendency was manifested in the responses of observers in the 2nd A series, and to a considerably larger degree than in responses of observers from other series of the primary experiment. This indicated that the artificial activation intensified the process of moving the observer's responses distribution to the side of the minimum limit of the mental scale of possible stimuli. It was said, that this motion was accompanied by an increase in the level of the recognition of conjugate pairs. The motion of the responses distribution to the sides of maximum and minimum limits is connected with the processes of direct and indirect activations. Under the conditions of artificial activation these processes were strengthened.

The process of direct activation maximally activated those points of the mental scale of the possible stimuli that coincided with the points of the scale of the presented stimuli. In other words, physically presented signals coincided with the conjugate pairs elements located on the mental scale. The process of indirect activation strengthened the activation of those points of the mental scale of possible stimuli that did not coincide with points on the scale of presented stimuli. The points of the mental scale which represented the minimum limit and the modular mental formation were activated more frequently than in the 2nd B series and started to be used for the commensuration of the presentation of temporal and spatial intervals. Nevertheless, the indirect activation of the modular mental formation did not reach such strength as was observed during the direct activation. In spite of an increase in the frequency of recalling this formation to the measuring actions, the speed of correct responses remained the same as in the 2nd B series. The aforementioned makes it possible to assume the following: The spontaneous direct and indirect activation was one of the basic components of the innate mechanism of the correlation of space and time intervals, and the recognition of their conjugate pairs.

**Two systems of a mechanism for the correlation of space and time intervals**

On the basis of the comparative data analysis of the experiment, this mechanism made the commensuration (having the same measure) of received intervals, of the organization of the modular mental formation, and a temporary mental scale of possible stimuli, plus the assignment of a certain sequence of activation of points on the scale. The measurement and also other processes of the correlation of intervals occur on the basis of the innate measuring system named by us as “biological metrics”. The artificial activation was one of the external factors which can reflect or exert influence differently upon the processes of correlation, but they cannot change their natural structure and order.

Such factors are either assigned intentionally before the beginning of some cognitive action, or they are formed in the process of vital activity of an observer. The artificial activation purposefully used in the 2nd A series of the primary experiment, as shown above, accelerated the processes of correlation, but it did not change their structure and order. Another factor was the possibility of a person to evaluate the intervals of space and time in centimeters and seconds, that was formed in the process of education and training, and was used in the pilot experiment. The results of the pilot experiment showed the low level of conjugate pairs recognition during not only the 1st but also during the 2nd time signal presentations. Also in the pilot experiment conditions, the possibility to evaluate the intervals of space and time in centimeters and seconds was the factor that slowed the processes of their correlation. On the other hand, verbal and nonverbal estimations by observers of the presented intervals in centimeters and seconds reflected some processes and methods of correlation of these intervals which could not be observed directly in responses of observers in the main part of the experiment. The phenomenon of leveling off of numerical values of elements of certain conjugate pairs in observers' estimations could serve as an example. The numerical value of centimeters of a space interval became equal to the numerical value in seconds of a time interval or vice versa, which were elements of one and the same conjugate pair. This phenomenon, in particular, indicated that an observer correctly established the connection between the elements of the assigned conjugate pair although his straight response to this signal in the main part of the experiment could be incorrect. Other external factors which reflected and influenced the processes of correlation in the majority of cases were based on the measuring system, which a person had acquired in the society where he or she lived. Unlike the innate biological metrics, such a measuring system was named by us as "social metrics".

On the basis of results of the comparative analysis of phenomena and the data of carried out experiments, one can assume the existence of two systems of mechanism for the correlation of space and time intervals: a **measuring** system and an **expressive** system.

**The Measuring system -** (1) this system finds and establishes the modular mental formation, the principle of the organization of proportional relations between these intervals, (2) this system forms the mental scale of possible stimuli which an observer used for the correlation of intervals. These actions of the measuring system (which underlies the innate mechanism of proportionality) occurred in the process of the perception of spatial and temporal signals that are received from the specific situation of an observer's vital activity. In this case a situation is described and analyzed in this experiment.

**The Expressive system -** records, shapes, arranges, and transfers the correlation measuring system activity results for their use in the different behavioral, cognitive, communicative, and social actions of a person. The expressive system has a complex structure that comprises mental imagery, linguistic and locomotive spheres of the human mind from where it selects material for the designation of correlation results. This selection is determined by the specific character of an observer individual experience that consists of both innate and acquired different kinds of predispositions, attitudes, norms, value orientations, knowledge and other methods of their acquisition. Among such methods, one should note that what we call social metrics, are i.e., systems of measurement manufactured by a society, and mastered by an individual in the process of his/her socialization. The social metrics refers directly to the record of results of the correlation measuring system activity. The adequacy of cognitive and other behavioral actions to these results depend on how they will be recorded, designed and transmitted by the expressive system for their use. Behavioral actions can correspond to the results of correlation completely, partially or not correspond at all. In other words, man can express the obtained knowledge correctly, incorrectly or not express it at all. In the cases of our experiment, observers having correctly established the given in the experiment relations between presented space and time intervals frequently gave incorrect responses. They indicated other responses not appropriate to those correct relationships of these intervals established by them. The analysis of verbal and nonverbal observers' estimations of intervals of space and time separately after these intervals were presented in the experiments for their correlation testifies to the fact of such divergences. It was noticed that in these estimations of space and time it was as if they came to meet each other. The duration of time intervals increased, and the length of the space intervals conjugated with them decreased in such a way that the number of seconds of the time interval became equal to the quantity of centimeters of conjugated space intervals. This fact showed that although an observer in the experiment for the correlation of these intervals gave incorrect responses, in reality the connection between elements of these conjugate pairs was established.

In our view, the activity of the measuring system and results of this activity (the forming of modular mental formation, the mental scale of possible stimuli, and processes of activating points of this scale) are amodal. The activity of the expressive system and the results of this activity are multimodal in nature. An amodal measuring system separates the general content from the information obtained from different sensory organs. For example, the observers separated and measured in the perceived temporal signals of different modalities such as light, sound, light-sound signals had one general property, their duration. Correlating and being commensurate in length and duration of perceived intervals, observers separated their proportional relations and considered these relations as united time-spatial extents. The multimodal expressive system reflected and converted the intermediate and final results of the amodal measuring system activity into the mental imagery of a different modality, into the verbal or locomotive reactions. The modality of this reflection depended on a given person's individual specific features of their cognitive system: the specific character of his/her memory, thinking, speech, and imagination. Reports by observers concerning the experiment carried out with them can serve as an illustration about the aforementioned work of the expressive system (See appendix 4).

**The present experiment and future research of the correlation processes between space and time in human perception**

The detailed analysis of auxiliary parts of the primary experiment in this work was not represented. Some facts and phenomena, which were observed in these parts were described here in connection with their relation to processes of the recognition of conjugate pairs.

The detailed analysis of the phenomena that were observed in the parts of the completed experiment can in particular show specifically: 1. How and in what forms does the transformation of perceived multimodal signals into amodal mental constructions take place. 2. What methods does the innate measuring system use for this conversion?

As was said above, the formation of such mental constructions allowed an observer to recognize correctly conjugate pairs in a given experimental situation, or in his/her practical life to establish correctly the temporal-spatial relations of elements of the surroundings of their vital activity at any given moment.

The phenomenal equalization of numerical values of the elements of conjugate pairs described above, can serve in further experiments as the basic object of the indicated analysis. For the study of the processes of correlation of space and time intervals in the represented experiment, pairs of different length and duration intervals,  the basis of which only two modular relationships were used 0.3 seconds - 0.5 centimeters, and also 1 second - 1 centimeter. In future experiments, the pairs of intervals used will be the basis that underlie other modular relationships such as 0.5 seconds 0,7 centimeters and so on. This will make it possible to deepen and to use as practical knowledge of the mechanism of the correlation of space and time in human perception.

However, the facts and phenomena discovered in the course of the experiments represented here and the hypotheses built on their basis, can reflect general regularities for the correlation of intervals as the basis which underlie different modular relationships. Therefore these facts, phenomena, and hypotheses can serve as a guideline for studying the different key problems of this correlation. Their number includes, in particular, the problems of interaction and mutual influence of the measuring and expressive systems of the correlation mechanism. These problems on the one hand are connected to the problems of interaction of such cognitive processes as perception, memory, thinking, speech, and imagination. On the other hand, they also concern the interaction of innate biological and acquired social measuring systems that include nonverbal and verbal intellects, unconscious and conscious, processes of the circulation of information between the right and left hemisphere, functioning by appropriate cerebral mechanisms. The research of the problems indicated has a high scientific and practical value, and requires a wide interdisciplinary study.

**Individual differences in the correlation of perceived space and time intervals**

Different people participating in the experiment performed the same task differently under the same experimental conditions.

These differences refer to:

(1) The quantity of correct and approximately correct responses by observers

(2) A observer's time reaction

(3) The modality of signals to which an observer gave the most correct responses

(4) The tendency in the responses distribution in the first and second presentations of signals

(5) The conditions under which an observer gave the most correct responses

(6) A method by which an observer performed the experimental task

Different observers had different levels of success in recognizing conjugate pairs. Depending upon this success one can divide them into four shooting categories: 1) snipers, 2) good riflemen, 3) bad riflemen and 4) terrible riflemen.

**Snipers -** gave more than 50 percent correct responses to 84 presented temporal signals. This put together with approximately correct responses produced a 90 to 96 percent level of correctness. (27% of the total number of 150 observers participated in this experiment).

**Good riflemen -** gave more than 40 percent correct responses. This put together with approximately correct responses produced more than a 70 percent level of correctness. (35% of the total number of 150 observers participated in this experiment)

**Bad riflemen -** gave less than 40 percent of correct responses. This put together with approximately correct responses produced a 50 to 55 percents level of correctness. (27% of the total number of 150 observers participated in this experiment)

**Terrible riflemen** - less than 30 percent correct responses. This put together with approximately correct responses produced less then a 50 percent level of correctness. (11% of the total number of 150 observers participated in this experiment). However, as the analysis shows the responses of most **bad riflemen** and **terrible riflemen** contained indirect evidences that they nevertheless established connections between elements of almost all presented conjugate pairs.

From answers to questions presented in the special questionnaire (the fourth auxiliary part of the primary experiment, and conversations with observers) one can learn some common personal peculiarities of cognitive activity and behavior of people pertaining to each of the listed categories. For example, snipers unlike bad riflemen had a high level of mental performance and curiosity, a tendency to abandon standard opinions, and looked for new ways to solve a problem. They also had the ability to refute their earlier mistaken stated hypothesis, and quickly replace them with new more probable ones. They quickly adapted to changing situations, and were able to orient themselves in an unfamiliar area. In reference to the decision time (reaction time), observers were divided in the three groups.

The first group arrived at correct decisions more quickly whereas when they took more time, more incorrect decisions were made. With the second group, the opposite was true. They needed more time to make correct decisions. The third group had no such distinctions in decision times. Depending on the peculiarities of their memory, observers made the most correct correlations while perceiving either the light or the sound signals, or the light and sound signals together. In responding to time signals, some observers chose only the shortest length rather than a length necessary to perform the experimental task. The tendency of the response distribution was towards the minimal limit pair.

Whereas most of the others observers were quite to the opposite. The tendency of responses distribution was towards the maximal limit pair. Some observers could successfully perform the experimental task under conditions of a high level of uncertainty (2 B series), whereas others observers could do the task only with the support of the preliminary knowledge about the minimum and maximum limit conjugated pairs (2 A series).

It is possible to show several methods which observers used for correlation purposes. They reported this in their answers to questions in the special questionnaire. One of the most used methods was a free internal count without relation to seconds, centimeters, watch, or ruler. Another example was imagining a moving mercury thermometer.

Sex, age, education, and profession had no noticeable impact on the precision of the implementation of a given task. Some people participated in this experiment twice with an interruption of 5 to 9 years, but gave about the same results, and showed the same individual peculiarities in their ability to correlate. For instance, one of them was tested for the first time in 1992 in Tashkent, when he was a six years old boy. And after nine years, in 2001 in Chicago, he was tested again using the same device and method. His results and level of successfulness were the same as those given in 1992.

He was a real **sniper**.

## Possible areas of application

Phenomena observed during our experiment are important for the understanding of cognitive processes mechanisms, the interaction between social and biological spheres, and consciousness and unconsciousness in the human psyche. This also includes the diagnostics of a person's particular cognitive activity and the modeling of the cognitive processes in different cybernetic devices as well.

## The summary

To study the processes and mechanisms of the correlation between space and time, particularly between lengths and durations in human perception, a special method (device and procedure) to conduct this experiment was designed and called LDR (Length Duration Relation)

In the present study a pilot and three series of the primary experiment were conducted under conditions of different levels of uncertainty.

The primary experiments consisted of the main part and six auxiliary parts.

In the main part of all the types of experiments, signals of a certain duration and modality were presented twice in random order to the observers.

Observers had to respond to time signals of different durations by choosing a corresponding space interval. In other words, a subject had to recognize a connection given by the experimenter between a definite length of space and a definite duration of time based on a certain coefficient of proportionality. This relation in this experiment was called a conjugate pair.

Any feedback was eliminated. Observers did not know if their responses were correct or not. They could not be aware of the fact that a certain signal was presented to them the first time or the second time. The data which were obtained during the 1st and the 2nd time signal presentations were examined separately. In the auxiliary parts of the experiment subjects had to estimate length and duration of the same stimuli separately without correlating them.

The comparative data analysis of the experiment showed significant differences between the 1st and 2nd presentation of signals in the quantity of correct responses, the responses distribution along the scale of stimuli, the phenomena which occurred during the experiment.

The clearness of these differences depended on the level of uncertainty condition under which a certain type of the experiment was conducted. The higher this level of uncertainty, the more clearly this difference was manifested. As the data analysis of auxiliary parts of the experiment showed subjects underestimated numerical values of space intervals and overestimated numerical values of time intervals in such a way that numerical values of both elements of a certain conjugate pair would become equal. We supposed that this phenomenon reflected one of the functions of the perceptive mechanism which was used by a subject in the process of correlation and commensuration of presented intervals of space and time.

Based on results of the experiments comparative data analysis, one can suppose that this perceptive mechanism, named by us as an **innate mechanism of proportionality**, performed the correlation of these intervals into two stages: adaptation and activation.

In the **adaptation stage** (which took place during the 1st time signals presentation) observers searched for a corresponding modulus, limits of the scale of presented stimuli, and created a temporary mental scale of possible stimuli. In the experiment conducted under the high level of uncertainty this scale was divided by an internal limit into two parts according to the principle of the golden section and detected locations of other conjugate pairs even though he or she did not recognize all of them. The purpose of this scale division into two parts was the creation of conditions which facilitated the activation of the remaining points on the scale. In the **activation or recognition stage** (which took place during the 2nd time signals presentation) the above were applied for correlation purposes.

We suggested that a certain conjugate pair would be recognized if a corresponding point on the scale would be activated by the influence of a corresponding time signal. As a result during the 2nd time signal presentation in comparison with the 1st time signal presentation the quantity of correct responses increased more than double. Based on the data analysis of the experiment one can suggest an existence of two interacting measurement systems that participated in the correlation processes: the **innate biological metrics** that underlie the organization of both human and animal physiological, behavioral and mental processes and the **acquired social metrics** which people learned from their society where they live and use it in their daily cognitive activity. In addition to that one can assume as well the existence of two systems of mechanism for the correlation of space and time intervals: the **amodal measuring** system and the **multimodal expressive** system. The amodal measuring system measures, correlates, and proportionates the length and duration of presented stimuli of different modalities. The multimodal expressive system records, shapes, arranges, and transfers the amodal measuring system activity results for their use in the different behavioral, cognitive, communicative and social actions of a person.

Individual differences were found in the cognitive actions of subjects while they performed experimental tasks. Therefore, one of the areas of practical applications of these investigative results can be used in the diagnosis of some peculiarities of a person cognitive activity.

I am very grateful to the professor of psychology at Washington University in St. Louis, I.J. Hirsh (now deceased) – a well known psychologist by his works on perception of duration and speech and the professor of chemistry G.S. Yablonsky – well known in the field of chemistry of temporal processes for their support and help.

# Appendixes

# Appendix 1

## The pilot experiment

1. Data

## Table 1

Table 1 (1)

Dynamics of the conjugate pairs recognition

| Presentations | 1/1 | 2/2 | 3/3 | 4/4 | 5/5 | 6/6 | 7/7 | 8/8 | 9/9 | 10/10 | 11/11 | 12/12 | 13/13 | |
|---|---|---|---|---|---|---|---|---|---|---|---|---|---|---|
| **1st pr** | **45.7** | **44.8** | **25.7** | **20** | **17.2** | **16.2** | **8.6** | **20** | **20** | **22.9** | **25.7** | **37.1** | **38** | |
| 2nd pr | **69.5** | **58** | **49.5** | **24.8** | **29.5** | **17.2** | **18** | **17.2** | **21** | **28.6** | **41** | **43.8** | **44.6** | |
| | Approximately + correct responses | | | | | | | | | | | | | |
| **1st pr** | ***77.1*** | ***81.9*** | ***51.4*** | ***47.6*** | ***39.1*** | ***55.3*** | ***25.7*** | ***41.9*** | ***49.5*** | ***52.5*** | ***66.6*** | ***79*** | ***69.4*** | |
| **2nd pr** | ***96.1*** | ***86.6*** | ***79*** | ***66.6*** | ***62.9*** | ***51.4*** | ***45.6*** | ***53.4*** | ***48.6*** | ***61.8*** | ***66.8*** | ***74.3*** | ***65.6*** | |

Table 1(2)

**Dynamics in shifting tendencies to both limits of the scale and correct and approximately correct responses (average data)**

| PRESENTATIONS | Shifting tendencies to | | Correct responses | Approximately + correct Responses |
|---|---|---|---|---|
| | Max. limit | Min. limit | | |
| The 1st presentation | 55 | 24.8 | **26.3** | ***56.7*** |
| The 2nd presentation | 40.2 | 29.4 | **35.7** | ***66*** |

## 2. Confusion matrices

Table 1 (3)

**1st presentation**

| - | 0 | 6.6 | 13.4 | 19 | 21 | 33.4 | 15.2 | 24.7 | 19.9 | 27.6 | 21.9 | 32.3 | 62 |
|---|---|---|---|---|---|---|---|---|---|---|---|---|---|
| Sec/ cm | 1 | 2 | 3 | 4 | 5 | 6 | 7 | 8 | 9 | 10 | 11 | 12 | 13 |
| 1 | **45.7** | 6.6 | 2.9 | | | | | | | | | | |
| 2 | 31.4 | **44.8** | 10.5 | 6.6 | 2.9 | | | | | | | | |
| 3 | 11.4 | 30.5 | **25.7** | 12.4 | 5.7 | 2.9 | 1.9 | 0.95 | 0.95 | | | | |
| 4 | 4.8 | 8.6 | 15.2 | **20** | 12.4 | 8.6 | 2.9 | 3.8 | 2.9 | 1.9 | 0.95 | 0.95 | |
| 5 | 3.8 | 5.7 | 17.1 | 15.2 | **17.2** | 21.9 | 6.6 | 11.4 | 0.95 | 3.8 | 2.9 | 1.9 | 0.95 |
| 6 | 0.95 | 1.9 | 9.5 | 8.6 | 9.5 | **16.2** | 3.8 | | 1.9 | 1.9 | 0.95 | 2.9 | 4.8 |
| 7 | 0.95 | 0.95 | 7.6 | 11.4 | 15.2 | 17.2 | **8.6** | 8.6 | 6.6 | 6.6 | 3.8 | 2.9 | 3.8 |
| 8 | | | 3.8 | 4.8 | 10.5 | 8.6 | 13.3 | **20** | 6.6 | 4.8 | 6.6 | | 0.95 |
| 9 | | 0.95 | 3.8 | 14.3 | 16.2 | 16.2 | 17.2 | 13.3 | **20** | 8.6 | 2.9 | 6.6 | 4.8 |
| 10 | 0.95 | | 0.95 | 2.9 | 3.8 | 4.8 | 6.6 | 15.2 | 22.9 | **22.9** | 3.8 | 5.7 | 2.9 |
| 11 | | | 2.9 | 2.9 | 3.8 | 3.8 | 15.2 | 14.3 | 16.2 | 21 | **25.7** | 11.4 | 12.4 |
| 12 | | | | 0.95 | 2.9 | | 10.5 | 9.5 | 17.2 | 21 | 37.1 | **37.1** | 31.4 |
| 13 | | | | | | | 13.3 | 2.9 | 3.8 | 7.6 | 15.2 | 30.5 | **38** |
| + | 54.25 | 48.6 | 60.9 | 61.05 | 61.9 | 50.6 | 76.1 | 55.2 | 60.1 | 49.6 | 52.3 | 30.5 | **-** |
| appr | **77.1** | **81.9** | **51.4** | **47.6** | **39.1** | **55.3** | **25.7** | **41.9** | **49.5** | **52.5** | **66.6** | **79** | **69.4** |
| Sec/ cm | 1 | 2 | 3 | 4 | 5 | 6 | 7 | 8 | 9 | 10 | 11 | 12 | 13 |

right responses 341.9 = 26.3
approximately right 737 = 56.7 no approximate 5
no domination 5

“Right” responses 341.9 = 26.3%
“Approximately right” responses 737 = 56.7%

# Table 1 (4)

## 2nd presentation

| **-** | - | 7.6 | 14.25 | 23.7 | 25.8 | 31.4 | 26.7 | 33.3 | 33.3 | 34.3 | 32.6 | 36.2 | 53.3 |
|---|---|---|---|---|---|---|---|---|---|---|---|---|---|
| Sec/ Cm | 1 | 2 | 3 | 4 | 5 | 6 | 7 | 8 | 9 | 10 | 11 | 12 | 13 |
| 1 | **69.5** | 7.6 | 0.95 | | | | | | | | | | |
| 2 | 26.6 | **58** | 13.3 | 5.7 | 2.9 | | | | | | | | |
| 3 | 3.8 | 21 | **49.5** | 18 | 5.7 | 2.9 | 0.95 | | | | | | |
| 4 | | 10.5 | 16.2 | **24.8** | 17.2 | 10.5 | 3.8 | 2.9 | 1.9 | 1.9 | | | |
| 5 | | 1.9 | 6.6 | 23.8 | **29.5** | 18 | 8.6 | 8.6 | 2.9 | 1.9 | 0.95 | 1.9 | |
| 6 | | | 4.8 | 7.6 | 16.2 | **17.2** | 13.3 | 6.6 | 9.5 | 4.8 | 2.9 | | 0.95 |
| 7 | | | 3.8 | 8.6 | 7.6 | 16.2 | **18** | 15.2 | 11.4 | 5.7 | 4.8 | 1.9 | 1.9 |
| 8 | | | 0.95 | 3.8 | 6.6 | 9.5 | 14.3 | **17.2** | 7.6 | 4.8 | 2.9 | 3.8 | 3.8 |
| 9 | | | 2.9 | 5.7 | 5.7 | 13.3 | 14.3 | 21 | **21** | 15.2 | 12.4 | 9.5 | 12.4 |
| 10 | | | | 0.95 | 4.8 | 4.8 | 18 | 11.4 | 20 | **28.6** | 8.6 | 8.6 | 6.6 |
| 11 | | | | | 0.95 | 5.7 | 4.8 | 10.5 | 15.2 | 18 | **41** | 10.5 | 8.6 |
| 12 | | 0.95 | 0.95 | 0.95 | 1.9 | 0.95 | 2.9 | 4.8 | 9.5 | 13.3 | 17.2 | **43.8** | 19 |
| 13 | | | | | 0.95 | 0.95 | 0.95 | 1.9 | 0.95 | 5.7 | 9.5 | 20 | **46.6** |
| + | 30.4 | 34.35 | 36.2 | 51.4 | 44.7 | 51.4 | 55.3 | 49.6 | 45.7 | 37 | 26.7 | 20 | - |
| Appr | **96.1** | **86.6** | **79** | **66.6** | **62.9** | **51.4** | **45.6** | **53.4** | **48.6** | **61.8** | **66.8** | **74.3** | **65.6** |
| Sec/ Cm | 1 | 2 | 3 | 4 | 5 | 6 | 7 | 8 | 9 | 10 | 11 | 12 | 13 |

Table 1 (5)

(Average data)

1 2 3 4 5 6 7 8 9 10 11 12 13

| sec<br>cm | 1 | 2 | 3 | 4 | 5 | 6 | 7 | 8 | 9 | 10 | 11 | 12 | 13 |
|---|---|---|---|---|---|---|---|---|---|---|---|---|---|
| 1 | **57.1** | 7.1 | 1.9 | | | | | | | | | | |
| 2 | 30 | **51.4** | 11.4 | 6.2 | 2.9 | | | | | | | | |
| 3 | 7.1 | 25.7 | **37.1** | 15.2 | 5.7 | 3.8 | 1.4 | 1 | 0.5 | | | | |
| 4 | 2.4 | 9.5 | 15.7 | **22.4** | 14.8 | 10 | 2.9 | 3.3 | 2.4 | 1.4 | 0.5 | | |
| 5 | 1.9 | 3.3 | 13.3 | 18.5 | **23.8** | *19.5* | 8 | 9 | 1.9 | 2.9 | 1.9 | 1.9 | 0.5 |
| 6 | 0.5 | 1 | 6.2 | 8.1 | 11.9 | **16.6** | 8.6 | 3.8 | 5.7 | 3.3 | 1.4 | 1.9 | 2.9 |
| 7 | 0.5 | 1 | 5.7 | 10 | 11.4 | 16.2 | **13.3** | 11.9 | 9.5 | 6.6 | 4.3 | 2.4 | 2.9 |
| 8 | - | - | 2.4 | 4.8 | 9 | 9.5 | 13.3 | **18.6** | 7.1 | 4.8 | 4.8 | 1 | 2.4 |
| 9 | - | 0.5 | 3.8 | 10 | 10 | 13.8 | *15.7* | 16.2 | **20.5** | 11.9 | 7.1 | 8.6 | 8 |
| 10 | 0.5 | - | 0.5 | 2.4 | 3.8 | 4.8 | 12.4 | 13.3 | 21.4 | **24.3** | 6.6 | 8 | 5.7 |
| 11 | | - | 1.4 | 1.4 | 3.3 | 4.8 | 10 | 12.9 | 16.2 | 20.5 | **33.3** | 9.5 | 10.5 |
| 12 | | 0.5 | 0.5 | 1 | 2.8 | 0.5 | 6.6 | 6.6 | 11.9 | 17.1 | 27.6 | **40.5** | 25.2 |
| 13 | | | | | 0.5 | 0.5 | 7.6 | 3.3 | 2.9 | 7.1 | 12.4 | 26.2 | **41.9** |

approx. 87.1 84.2 64.2 56.1 50.5 52.3 35.2 46.7 49 56.7 67.5 76.2 67.1

## The subseries 1 a

### 1. Data

Table 2

Table 2 (1)

Dynamics of the conjugate pairs recognition

| Presentations | 0.3<br>0.5 | 0.6<br>1 | 0.9<br>1.5 | 1.2<br>2 | 1.5<br>2.5 | 1.8<br>3 | 2.1<br>3.5 | 2.4<br>4 | 3.3<br>5.5 | 3.6<br>6 | 4.5<br>7.5 | 5.4<br>9 | 6<br>10 | 6.3<br>10.5 |
|---|---|---|---|---|---|---|---|---|---|---|---|---|---|---|
| **1st pr** | **77.8** | **72.2** | **72.2** | **55.6** | **22.2** | **44.4** | **61.1** | **66.6** | **33.3** | **33.3** | **44.4** | **50** | **50** | **44.4** |
| 2nd pr | **88.9** | **94.4** | **77.7** | **61.1** | **44.4** | **55.6** | **38.8** | **55.6** | **55.6** | **33.3** | **44.4** | **44.4** | **72.2** | **66.6** |
| | Approximately + correct responses | | | | | | | | | | | | | |
| **1st pr** | ***100*** | ***88.8*** | ***77.7*** | ***83.4*** | ***55.9*** | ***77.9*** | ***88.9*** | ***88.9*** | ***61*** | ***61.1*** | ***72.2*** | ***88.9*** | ***94.4*** | ***62*** |
| **2nd pr** | ***100*** | ***100*** | ***88.9*** | ***88.8*** | ***72.2*** | ***83.3*** | ***72.1*** | ***88.8*** | ***61.2*** | ***72.2*** | ***61*** | ***94.3*** | ***83.3*** | ***77.7*** |

Table 2 (2)

**Dynamics in shifting tendencies to both limits of the scale and correct and approximately correct responses (average data)**

| PRESENTATIONS | Shifting tendencies to | | Correct responses | Approximately + correct Responses |
|---|---|---|---|---|
| | Max. limit | Min. Limit | | |
| The 1st presentation | 30.3 | 23.2 | **52** | ***78.7*** |
| The 2nd presentation | 22.6 | 22.7 | **59.5** | ***81.7*** |

Table 2 (3)

The quantity of correct responses to the time signals which were placed on the left and right halves of the stimuli scale (% of quantity of correct responses)

| The half of the scale belong to the | Presentations | |
|---|---|---|
| | The 1st presentation | The 2nd presentation |
| Minimum limit | **55.7** | **55.3** |
| Maximum limit | **44.3** | **44.7** |

## 2. Confusion matrices

Table 2 (4)

1st presentation

Responses to the given signals (subseries 1a)

| - | 0 | 0 | 5.5 | 11.1 | 22.2 | 27.8 | 11.2 | 22.2 | 16.6 | 33.4 | 27.9 | 16.6 | 27.7 | 55.6 |
|---|---|---|---|---|---|---|---|---|---|---|---|---|---|---|
| | 0.3<br>0.5 | 0.6<br>1 | 0.9<br>1.5 | 1.2<br>2 | 1.5<br>2.5 | 1.8<br>3 | 2.1<br>3.5 | 2.4<br>4 | 3.3<br>5.5 | 3.6<br>6 | 4.5<br>7.5 | 5.4<br>9 | 6<br>10 | 6.3<br>10.5 |
| 0.3<br>0.5 | **77.8** | | | | | | | | | | | | | |
| 0.6<br>1 | | **72.2** | 5.5 | | | | | | | | | | | |
| 0.9<br>1.5 | 22.2 | 16.6 | **72.2** | 11.1 | | 5.6 | | | | | | | | |
| 1.2<br>2 | | | | **55.6** | 22.2 | | | | | | | | | |
| 1.5<br>2.5 | | 11.1 | 5.5 | 16.7 | **22.2** | 22.2 | 5.6 | | | | | | | |
| 1.8<br>3 | | | 5.5 | | 11.1 | **44.4** | 5.6 | 11.1 | | | | | | |
| 2.1<br>3.5 | | | 5.5 | 11.1 | 16.7 | 11.1 | **61.1** | 11.1 | | 5.6 | | | | |
| 2.4<br>4 | | | | | 16.7 | 5.6. | 22.2 | **66.6** | 16.6 | 22.2 | 5.6 | | | |
| 3.3<br>5.5 | | | | 5.5 | | 11.1 | 5.6 | 5.6 | **33.3** | 5.6 | 5.6 | | | |
| 3.6<br>6 | | | | | | | | 5.6 | 11.1 | **33.3** | 16.7 | 5.5 | | 5.5 |
| 4.5<br>7.5 | | | 5.5 | | 11.1 | | | | 33.3 | 22.2 | **44.4** | 11.1 | 5.5 | 16.7 |
| 5.4<br>9 | | | | | | | | | | 5.6 | 11.1 | **50** | 22.2 | 16.7 |
| 6<br>10 | | | | | | | | | | 5.6 | 5.5 | 27.8 | **50** | 16.7 |
| 6.3<br>10.5 | | | | | | | | | 5.6 | | 11.1 | 5.5 | 22.2 | **44.4** |
| + | 22.2 | 27.7 | 22 | 33.3 | 55.6 | 27.8 | 27.8 | 11.2 | 50 | 33.4 | 27.7 | 33.3 | 22.2 | 0 |
| Appr | 100 | 88.8 | 77.7 | 83.4 | 55.5 | 77.7 | 88.9 | 83.3 | 61 | 61.1 | 72.2 | 88.9 | 94.4 | 61.1 |

## Table 2 (5), **2nd presentation**

Responses to the given signals (subseries 1a)

| | | | | | | | | | | | | | | |
|---|---|---|---|---|---|---|---|---|---|---|---|---|---|---|
| **-** | 0 | 0 | 5.6 | 22.2 | 33.4 | 22.2 | 27.8 | 22.2 | 16.8 | 22.3 | 16.7 | 22.2 | 27.8 | 33.3 |
| | 0.3<br>0.5 | 0.6<br>1 | 0.9<br>1.5 | 1.2<br>2 | 1.5<br>2.5 | 1.8<br>3 | 2.1<br>3.5 | 2.4<br>4 | 3.3<br>5.5 | 3.6<br>6 | 4.5<br>7.5 | 5.4<br>9 | 6<br>10 | 6.3<br>10.5 |
| 0.3<br>0.5 | **88.9** | | | | | | | | | | | | | |
| 0.6<br>1 | | **94.4** | 5.6 | 5.6 | | | | | | | | | | |
| 0.9<br>1.5 | 11.1 | 5.6 | **77.7** | 16.6 | 5.6 | | | | | | | | | |
| 1.2<br>2 | | | 5.6 | **61.1** | 27.8 | 11.1 | 5.6 | | | | | | | |
| 1.5<br>2.5 | | | 5.6 | 11.1 | **44.4** | 11.1 | 11.1 | 5.6 | 5.6 | | | | | |
| 1.8<br>3 | | | 5.6 | 5.6 | | **55.6** | 11.1 | | 5.6 | 5.6 | | | | |
| 2.1<br>3.5 | | | | | 16.6 | 16.6 | **38.8** | 16.6 | | | | | | |
| 2.4<br>4 | | | | | 5.6 | | 22.2 | **55.6** | 5.6 | 11.1 | 5.6 | | | 5.6 |
| 3.3<br>5.5 | | | | | | 5.6 | 5.6 | 16.6 | **55.6** | 5.6 | 11.1 | 5.6 | | |
| 3.6<br>6 | | | | | | | 5.6 | 5.6 | | **33.3** | | | 5.6 | |
| 4.5<br>7.5 | | | | | | | | | 27.7 | 33.3 | **44.4** | 16.6 | 11.1 | 16.6 |
| 5.4<br>9 | | | | | | | | | | 11.1 | 16.6 | **44.4** | 11.1 | |
| 6<br>10 | | | | | | | | | | | 22.2 | 33.3 | **72.2** | 11.1 |
| 6.3<br>10.5 | | | | | | | | | | | | | | **66.6** |
| + | 11.1 | 5.6 | 16.8 | 16.7 | 22.2 | 22.2 | 33.4 | 22.2 | 27.7 | 44.4 | 38.8 | 33.3 | 0 | 0 |
| Appr | **100** | **100** | **88.7** | **88.8** | **72.2** | **83.3** | **72.1** | **88.8** | **61.2** | **72** | **61** | **94.3** | **83.3** | **77.7** |

# The 1st series

## 1. Data

Table 3

Table 3 (1)

Dynamics of the conjugate pairs recognition

| Presentations | 0.3<br>0.5 | 0.6<br>1 | 0.9<br>1.5 | 1.2<br>2 | 1.5<br>2.5 | 1.8<br>3 | 2.1<br>3.5 | 2.4<br>4 | 3.3<br>5.5 | 3.6<br>6 | 4.5<br>7.5 | 5.4<br>9 | 6<br>10 | 6.3<br>10.5 |
|---|---|---|---|---|---|---|---|---|---|---|---|---|---|---|
| **1st pr** | **71.5** | **56** | **36.6** | **39.8** | **27.7** | **28** | **28** | **37** | **24** | **28.5** | **39.8** | **44.3** | **63.8** | **68.3** |
| 2nd pr | **88.2** | **68.7** | **59.3** | **50.8** | **43.1** | **32.9** | **45.1** | **40.7** | **45.5** | **50.4** | **47.6** | **50.4** | **65.9** | **50.8** |
| | Approximately + correct responses | | | | | | | | | | | | | |
| **1st pr** | ***89.4*** | ***88.1*** | ***54.8*** | ***63.4*** | ***48*** | ***51.5*** | ***56.4*** | ***61.4*** | ***39*** | ***54.9*** | ***70.7*** | ***84.1*** | ***92.7*** | ***84.1*** |
| **2nd pr** | ***98.4*** | ***94.3*** | ***80*** | ***80.5*** | ***72*** | ***64.2*** | ***79.7*** | ***65.1*** | ***69.1*** | ***87.4*** | ***79.7*** | ***79.3*** | ***94.5*** | ***73.9*** |

Table 3 (2)

**Dynamics in shifting tendencies to both limits of the scale and correct and approximately correct responses (average data)**

| PRESENTATIONS | Shifting tendencies to | | Correct responses | Approximately + correct Responses |
|---|---|---|---|---|
| | Max. limit | Min. Limit | | |
| The 1st presentation | 48.4 | 14.7 | **42.4** | ***67*** |
| The 2nd presentation | 34 | 18.2 | **52.8** | ***79.8*** |

Table 3 (3)

The quantity of correct responses to the time signals which were placed on the left and right halves of the stimuli scale (% of quantity of correct responses)

| The half of the scale belong to the | Presentations | |
|---|---|---|
| | The 1st presentation | The 2nd presentation |
| Minimum limit | **48.5** | **52.4** |
| Maximum limit | **51.5** | **47.6** |

## 2. Confusion matrices

Table 3 (4)

1st presentation

Responses to the given signals. Modulus pair, MP ( 0.3 s – 0.5 cm) presents in the set of signals In this and next tables, the value in any cell represents the percentage of the corresponding response. Bold and underlined values correspond the “absolute right” responses. It is clear the these responses to all time signals are related to the maximum of distribution (% of the number signals presented)

| - | 0 | 0 | 2.8 | 6.1 | 11.8 | 13 | 14.6 | 19.9 | 6.9 | 16.6 | 15.4 | 20.4 | 17 | 31.6 |
|---|---|---|---|---|---|---|---|---|---|---|---|---|---|---|
| | 0.3<br>0.5 | 0.6<br>1 | 0.9<br>1.5 | 1.2<br>2 | 1.5<br>2.5 | 1.8<br>3 | 2.1<br>3.5 | 2.4<br>4 | 3.3<br>5.5 | 3.6<br>6 | 4.5<br>7.5 | 5.4<br>9 | 6<br>10 | 6.3<br>10.5 |
| 0.3<br>0.5 | **71.5** | | | 0.4 | | | | | | | | | | |
| 0.6<br>1 | | **56** | 2.8 | | | | | | | | | | | |
| 0.9<br>1.5 | 17.9 | 32.1 | **36.6** | 5.7 | 3.7 | 0.8 | 0.4 | | 0.4 | | | | | |
| 1.2<br>2 | 5.7 | 8.9 | 15.4 | **39.8** | 8.1 | 3.7 | 2 | 2 | | | | | | |
| 1.5<br>2.5 | 2 | 1.6 | 10.2 | 17.9 | **27.7** | 8.5 | 4.5 | 4.5 | 0.4 | 0.4 | | | | |
| 1.8<br>3 | 0.4 | 1.2 | 5.7 | 6.9 | 12.2 | **28** | 7.7 | 6.9 | 0.4 | 1.2 | | | | |
| 2.1<br>3.5 | 2 | | 6.9 | 10.2 | 10.2 | 15 | **28** | 6.5 | | 1.6 | 0.8 | | | |
| 2.4<br>4 | | | 9.3 | 7.3 | 11.4 | 20.3 | 20.7 | **37** | 5.7 | 7.7 | 1.2 | 1.7 | 0.4 | |
| 3.3<br>5.5 | 0.4 | | 5.3 | 6.5 | 8.1 | 13 | 17.9 | 17.9 | **24** | 5.7 | 5.7 | 0.8 | 1.2 | 0.4 |
| 3.6<br>6 | | | 2.8 | 1.2 | 4 | 5.7 | 6.1 | 7.7 | 9.3 | **28.5** | 7.7 | 4.5 | 2.8 | 0.8 |
| 4.5<br>7.5 | | | 3.2 | 2.8 | 4.9 | 3.3 | 8.5 | 15.9 | 24 | 20.7 | **39.8** | 13.4 | 2.8 | 5.3 |
| 5.4<br>9 | | | 0.8 | 0.8 | 3.6 | 1.6 | 3.3 | 1.2 | 15.4 | 19.9 | 23.2 | **44.3** | 9.8 | 9.3 |
| 6 | | | 0.4 | 0.4 | 3.6 | | 0.8 | 0.4 | 15 | 11 | 16.7 | 26.4 | **63.8** | 15.8 |
| 10 | | | | | | | | | | | | | | |
| 6.3<br>10.5 | | | 0.4 | | 2.4 | | | | 5.3 | 3.2 | 4.9 | 8.9 | 19.1 | **68.3** |
| + | 28.4 | 43.8 | 60.4 | 54 | 60.4 | 58.9 | 57.3 | 43.1 | 69 | 54.8 | 44.8 | 35.3 | 19.1 | 0 |
| Appr | **89.4** | **88.1** | **54.8** | **63.4** | **48** | **51.5** | **56.4** | **61.4** | **39** | **54.9** | **70.7** | **84.1** | **92.7** | **84.1** |
| | 0.3<br>0.5 | 0.6<br>1 | 0.9<br>1.5 | 1.2<br>2 | 1.5<br>2.5 | 1.8<br>3 | 2.1<br>3.5 | 2.4<br>4 | 3.3<br>5.5 | 3.6<br>6 | 4.5<br>7.5 | 5.4<br>9 | 6<br>10 | 6.3<br>10.5 |

Table 3 (5), 2nd presentation

Responses to the given signals

Modulus pair, MP ( 0.3 s – 0.5 cm) presents in the set of signals in this and next tables, the value in any cell represents the percentage of the corresponding response. The bold and underlined values correspond the "absolute right" responses. It is clear the these responses to all time signals are related to the maximum of distribution (% of the number signals presented)

| **-** | 0 | 0 | 0.4 | 10.2 | 14.6 | 18.2 | 23.6 | 11.7 | 13 | 17.9 | 16.2 | 18.7 | 24.3 | 49.2 |
|---|---|---|---|---|---|---|---|---|---|---|---|---|---|---|
| | 0.3<br>0.5 | 0.6<br>1 | 0.9<br>1.5 | 1.2<br>2 | 1.5<br>2.5 | 1.8<br>3 | 2.1<br>3.5 | 2.4<br>4 | 3.3<br>5.5 | 3.6<br>6 | 4.5<br>7.5 | 5.4<br>9 | 6<br>10 | 6.3<br>10.5 |
| 0.3<br>0.5 | **88.2** | | | | | | | | | | | | | |
| 0.6<br>1 | | **68.7** | 0.4 | 0.4 | | | | | | | | | | |
| 0.9<br>1.5 | 10.2 | 25.6 | **59.3** | 9.8 | 2.8 | 0.4 | | | | | | | | |
| 1.2<br>2 | 1.6 | 4.9 | 20.3 | **50.8** | 11.8 | 4 | 3.6 | 1.2 | | 0.4 | | | | |
| 1.5<br>2.5 | | 0.8 | 9.8 | 19.9 | **43.1** | 13.8 | 4.5 | 2.4 | 0.8 | | 0.4 | | | |
| 1.8<br>3 | | | 4.1 | 10.2 | 17.1 | **32.9** | 15.5 | 2.8 | 1.2 | 0.8 | 0.4 | | | |
| 2.1<br>3.5 | | | 3.6 | 5.3 | 13.8 | 17.5 | **45.1** | 5.3 | 3.3 | 1.2 | 1.2 | | | |
| 2.4<br>4 | | | 0.8 | 2.4 | 7.7 | 13.4 | 19.1 | **40.7** | 7.7 | 4.9 | 1.6 | 0.8 | | |
| 3.3<br>5.5 | | | 1.2 | 0.8 | 1.6 | 13 | 8.1 | 19.1 | **45.5** | 10.6 | 5.7 | 3.2 | 0.8 | 0.8 |
| 3.6<br>6 | | | 0.4 | 0.4 | 1.2 | 3.3 | 2.4 | 17.1 | 15.9 | **50.4** | 6.9 | 4.9 | 1.6 | 4.9 |
| 4.5<br>7.5 | | | | | 0.8 | 1.6 | 1.6 | 9.3 | 15.9 | 26.4 | **47.6** | 9.8 | 4 | 4.5 |
| 5.4<br>9 | | | | | | | | 1.2 | 6.9 | 4.5 | 25.2 | **50.4** | 17.9 | 15.9 |
| 6<br>10 | | | | | | | | 0.8 | 2.8 | 0.8 | 9.8 | 19.1 | **65.9** | 23.1 |
| 6.3<br>10.5 | | | | | | | | | | | 1.2 | 11.8 | 9.8 | **50.8** |
| + | 11.8 | 31.3 | 40.2 | 39 | 42.2 | 48.8 | 31.2 | 47.5 | 41.5 | 31.7 | 36.2 | 30.9 | 9.8 | 0 |
| Appr | **98.4** | **94.3** | **80** | **80.5** | **72** | **64.2** | **79.7** | **65.1** | **69.1** | **87.4** | **79.7** | **79.3** | **94.5** | **73.9** |
| | 0.3<br>0.5 | 0.6<br>1 | 0.9<br>1.5 | 1.2<br>2 | 1.5<br>2.5 | 1.8<br>3 | 2.1<br>3.5 | 2.4<br>4 | 3.3<br>5.5 | 3.6<br>6 | 4.5<br>7.5 | 5.4<br>9 | 6<br>10 | 6.3<br>10.5 |

Table 3 (6)

(average data)

| **Res-pon-ses (in cm)** | Given signals (in seconds) | | | | | | | | | | | | | |
|---|---|---|---|---|---|---|---|---|---|---|---|---|---|---|
| | 0.3 | 0.6 | 0.9 | 1.2 | 1.5 | 1.8 | 2.1 | 2.4 | 3.3 | 3.6 | 4.5 | 5.4 | 6 | 6.3 |
| 0.5 | ***82*** | | | | | | | | | | | | | |
| 1 | | ***58.6*** | 1.4 | 0.45 | | | | | | | | | | |
| 1.5 | 13.5 | 30.5 | ***48.3*** | 7.4 | 3.3 | 1 | 0.3 | | 0.15 | | | | | |
| 2 | 2.5 | 6.8 | 18 | ***45.6*** | 10 | 5.9 | 2.4 | 1.4 | - | 0.3 | | | | |
| 2.5 | 0.8 | 2 | 9.4 | 18.8 | ***35.3*** | 13.8 | 4.7 | 3 | 0.6 | 0.15 | 0.3 | | | |
| 3 | 0.15 | 0.6 | 5.4 | 7.9 | 12.9 | ***29.2*** | 10.5 | 4.7 | 1 | 1 | 0.15 | | | |
| 3.5 | 0.8 | 0.9 | 5.8 | 8.2 | 12.3 | 16 | ***35.3*** | 8 | 1.5 | 1.4 | 0.8 | | 0.15 | |
| | 0.15 | 0.6 | 5.3 | 4.7 | 10 | 13.9 | 19.5 | ***41.2*** | 8.2 | 6.8 | 1.7 | 1 | 0.15 | 0.3 |
| 4 | | | | | | | | | | | | | | |
| 5.5 | 0.15 | | 2.5 | 3.8 | 6 | 12.1 | 14.2 | 17.6 | ***33.2*** | 8.8 | 5 | 1.8 | 0.8 | 0.8 |
| 6 | | | 1.4 | 1.2 | 2.3 | 3.6 | 3.3 | 12.1 | 15 | ***38.8*** | 7.9 | 4.7 | 2.4 | 2.4 |
| 7.5 | | | 1.7 | 1.5 | 3.5 | 3 | 5.4 | 10.9 | 19.7 | 23.3 | ***45.9*** | 10.9 | 4.8 | 5.6 |
| 9 | | | 0.45 | 0.3 | 2.1 | 1.2 | 2.9 | 1.2 | 9.8 | 11.6 | 22.6 | ***48.3*** | 12.1 | 11.2 |
| 10 | | | 0.15 | 0.15 | 1.4 | - | 1 | 0.3 | 7.9 | 5.6 | 13 | 22.7 | ***64.7*** | 19.5 |
| 10.5 | | | 0.15 | | 0.9 | 0.15 | 0.3 | | 2.9 | 2.1 | 2.7 | 10.5 | 14.9 | ***60.2*** |
| appr | **95.5** | **89.1** | **67.7** | **71.8** | **58.2** | **59** | **65.3** | **66.8** | **56.4** | **70.9** | **76.4** | **81.9** | **91.7** | **79.7** |

# The 2nd A series

## 1. Data

### Table 4

### Table 4 (1)

Dynamics of the conjugate pairs recognition

| Presentations | 0.9<br>1.5 | 1.2<br>2 | 1.5<br>2.5 | 1.8<br>3 | 2.1<br>3.5 | 2.4<br>4 | 2.7<br>4.5 | 3.3<br>5.5 | 3.6<br>6 | 3.9<br>6.5 | 4.5<br>7.5 | 5.4<br>9 | 6<br>10 | 6.3<br>10.5 |
|---|---|---|---|---|---|---|---|---|---|---|---|---|---|---|
| **1st pr** | **65.2** | **60.6** | **45.5** | **31.8** | **31.8** | **34.8** | **25.8** | **30.3** | **34.8** | **37.9** | **54.5** | **53** | **68.2** | **75.8** |
| 2nd pr | **84.8** | **66.7** | **68.2** | **39.4** | **50** | **57.6** | **39.4** | **56.1** | **53** | **56.1** | **59.1** | **59.1** | **62.1** | **54.5** |
| | Approximately + correct responses | | | | | | | | | | | | | |
| **1st pr** | ***80.4*** | ***86.3*** | ***68.2*** | ***53*** | ***51.5*** | ***60.5*** | ***51.6*** | ***48.5*** | ***59.1*** | ***72.7*** | ***80.3*** | ***83.3*** | ***92.5*** | ***91*** |
| **2nd pr** | ***95.4*** | ***90.9*** | ***94*** | ***78.8*** | ***71.2*** | ***80.3*** | ***74.3*** | ***71.3*** | ***72.7*** | ***80.3*** | ***84.8*** | ***87.9*** | ***93.9*** | ***80.3*** |

### Table 4 (2)

**Dynamics in shifting tendencies to both limits of the scale and correct and approximately correct responses (average data)**

| PRESENTATIONS | Shifting tendencies to | Correct responses | Approximately + correct |
|---|---|---|---|

| | Max. Limit | Min. limit | | Responses |
|---|---|---|---|---|
| The 1st presentation | 37.8 | 19.8 | **46.4** | ***69.9*** |
| The 2nd presentation | 16 | 26.3 | **57.5** | ***82.5*** |

Table 4 (3)
The quantity of correct responses to
the time signals which were placed on
the left and right halves of the stimuli scale
(% of quantity of correct responses)

| The half of the scale belong to the | Presentations | |
|---|---|---|
| | The 1st presentation | The 2nd presentation |
| Minimum limit | **45.4** | **50.4** |
| Maximum limit | **54.5** | **49.6** |

## 2. Confusion matrices

Table 4 (4)
The 1st presentation

Responses to the given signals. MP is absent but subjects were informed previously with the “limit” pairs (0.9 s - 1.5 cm and 6.3 s - 10.5 cm ) without information of the dimensions of these space and time intervals in centimeters and seconds.

| - | 0 | 12.1 | 16.6 | 22.8 | 18.2 | 40.9 | 16.6 | 13.7 | 24.3 | 16.6 | 22.8 | 13.7 | 15.1 | 24.2 |
|---|---|---|---|---|---|---|---|---|---|---|---|---|---|---|
| | 0.9<br>1.5 | 1.2<br>2 | 1.5<br>2.5 | 1.8<br>3 | 2.1<br>3.5 | 2.4<br>4 | 2.7<br>4.5 | 3.3<br>5.5 | 3.6<br>6 | 3.9<br>6.5 | 4.5<br>7.5 | 5.4<br>9 | 6<br>10 | 6.3<br>10.5 |
| 0.9 | **65.2** | 12.1 | 4.5 | 1.5 | | | | | | | | | | |
| 1.5 | | | | | | | | | | | | | | |
| 1.2<br>2 | 15.2 | **60.6** | 12.1 | 6.1 | | 1.5 | 1.5 | | | | | | | |
| 1.5<br>2.5 | 6.1 | 13.6 | **45.5** | 15.2 | 10.6 | 7.6 | 1.5 | | | | | | | |
| 1.8<br>3 | 4.5 | 3 | 10.6 | **31.8** | 7.6 | 7.6 | 3 | 1.5 | | | | | | |
| 2.1<br>3.5 | 3 | 7.6 | 9.1 | 6.1 | **31.8** | 24.2 | 3 | | | | | | | |
| 2.4<br>4 | 3 | 3 | 6.1 | 10.6 | 12.1 | **34.8** | 7.6 | 6.1 | 6.1 | | 1.5 | | 1.5 | |
| 2.7<br>4.5 | 3 | | 4.5 | 13.6 | 7.6 | 1.5 | **25.8** | 6.1 | 1.5 | | | | | |
| 3.3<br>5.5 | | | 1.5 | 4.5 | 9.1 | 13.6 | 18.2 | **30.3** | 16.7 | 3 | 6.1 | 1.5 | 1.5 | |
| 3.6<br>6 | | | 4.5 | 4.5 | 7.6 | 9.1 | 10.6 | 12.1 | **34.8** | 13.6 | 7.6 | 6.1 | | 4.5 |
| 3. 9<br>6.5 | | | | | | | | 10.6 | 7.6 | **37.9** | 7.6 | | | |
| 4.5<br>7.5 | | | | 3 | 6.1 | | 21.2 | 13.6 | 10.6 | 21.2 | **54.5** | 6.1 | 4.5 | 3 |
| 5.4<br>9 | | | 1.5 | 3 | 3 | | 6.1 | 12.1 | 13.6 | 18.2 | 18.2 | **53** | 7.6 | 1.5 |
| 6<br>10 | | | | | 4.5 | | | 6.1 | 6.1 | 4.5 | 4.5 | 24.2 | **68.2** | 15.2 |
| 6.3<br>10.5 | | | | | | | 1.5 | 1.5 | 3 | 1.5 | | 9.1 | 16.7 | **75.8** |
| + | 34.8 | 27.2 | 37.8 | 45.3 | 50 | 24.2 | 57.6 | 56 | 41.1 | 45.4 | 22.7 | 33.3 | 16.7 | 0 |
| appr | **80.4** | **86.3** | **68.2** | **53.1** | **51.5** | **60.5** | **51.6** | **48.5** | **59.1** | **72.7** | **80.3** | **83.3** | **92.5** | **91** |
| | 0.9<br>1.5 | 1.2<br>2 | 1.5<br>2.5 | 1.8<br>3 | 2.1<br>3.5 | 2.4<br>4 | 2.7<br>4.5 | 3.3<br>5.5 | 3.6<br>6 | 3.9<br>6.5 | 4.5<br>7.5 | 5.4<br>9 | 6<br>10 | 6.3<br>10.5 |

492.1= + 35.2 no approximate - 1
257.6 = - 18.4
no domination – no
“Right” responses 650 = 46.4%

“Approximately right” responses 979 = 69.9%

Table 4 (5)
2nd presentation

Responses to the given signals. MP is absent but subjects were informed previously with the “limit” pairs (0.9 s - 1.5 cm and 6.3 s - 10.5 cm ) without information of the dimensions of these space and time intervals in centimeters and seconds.

| - | 0 | 12.1 | 16.7 | 36.4 | 36.3 | 27.2 | 39.4 | 24.2 | 34.8 | 27.2 | 24.2 | 21.1 | 22.8 | 45.4 |
|---|---|---|---|---|---|---|---|---|---|---|---|---|---|---|
| | 0.9<br>1.5 | 1.2<br>2 | 1.5<br>2.5 | 1.8<br>3 | 2.1<br>3.5 | 2.4<br>4 | 2.7<br>4.5 | 3.3<br>5.5 | 3.6<br>6 | 3.9<br>6.5 | 4.5<br>7.5 | 5.4<br>9 | 6<br>10 | 6.3<br>10.5 |
| 0.9<br>1.5 | **84.8** | 12.1 | 1.5 | | | | | | | | | | | |
| 1.2<br>2 | 10.6 | **66.7** | 15.2 | 7.6 | 4.5 | | | | | | | | | |
| 1.5<br>2.5 | 3 | 12.1 | **68.2** | 28.8 | 13.6 | 1.5 | 3 | | | | | | | |
| 1.8<br>3 | 1.5 | 7.6 | 10.6 | **39.4** | 18.2 | 3 | 6.1 | | 1.5 | | | | | |
| 2.1<br>3.5 | | 1.5 | 3 | 10.6 | **50** | 22.7 | 12.1 | 3 | | 1.5 | | | | |
| 2.4<br>4 | | | | 9.1 | 3 | **57.6** | 18.2 | 12.1 | 3 | 1.5 | 1.5 | | | |
| 2.7<br>4.5 | | | | | 4.5 | | **39.4** | 9.1 | 10.6 | | | | | 1.5 |
| 3.3<br>5.5 | | | 1.5 | 4.5 | 4.5 | 3 | 16.7 | **56.1** | 19.7 | 12.1 | 3 | 1.5 | | 1.5 |
| 3.6<br>6 | | | | | 1.5 | 9.1 | 3 | 6.1 | **53** | 12.1 | 6.1 | 4.5 | | 3 |
| 3. 9<br>6.5 | | | | | | 3 | | 6.1 | | **56.1** | 13.6 | 1.5 | | |
| 4.5<br>7.5 | | | | | | | 1.5 | 7.5 | 10.6 | 12.1 | **59.1** | 13.6 | 6.1 | 3 |
| 5.4<br>9 | | | | | | | | | 1.5 | 3 | 12.1 | **59.1** | 16.7 | 10.6 |
| 6<br>10 | | | | | | | | | | 1.5 | 4.5 | 15.2 | **62.1** | 25.8 |
| 6.3<br>10.5 | | | | | | | | | | | | 4.5 | 15.1 | **54.5** |
| + | 15.1 | 21.2 | 15.1 | 24.2 | 13.5 | 15.1 | 21.2 | 19.7 | 12.1 | 16.6 | 16.6 | 19.7 | 15.1 | 0 |
| Appr | **95.4** | **90.9** | **94** | **78.8** | **71.2** | **80.3** | **74.3** | **71.3** | **72.7** | **80.3** | **84.8** | **87.9** | **93.9** | **80.3** |
| | 0.9<br>1.5 | 1.2<br>2 | 1.5<br>2.5 | 1.8<br>3 | 2.1<br>3.5 | 2.4<br>4 | 2.7<br>4.5 | 3.3<br>5.5 | 3.6<br>6 | 3.9<br>6.5 | 4.5<br>7.5 | 5.4<br>9 | 6<br>10 | 6.3<br>10.5 |

Table 4 (6)
(average data)

| **Res-pon-ses** | Given signals (in seconds) | | | | | | | | | | | | | |
|---|---|---|---|---|---|---|---|---|---|---|---|---|---|---|
| **(in cm)** | 0.9 | 1.2 | 1.5 | 1.8 | 2.1 | 2.4 | 2.7 | 3.3 | 3.6 | 3.9 | 4.5 | 5.4 | 6 | 6.3 |
| 1.5 | ***74.2*** | 10.6 | 3 | 0.8 | | | | | | | | | | |
| 2 | 13.6 | ***63.6*** | 13.6 | 6.8 | 2.3 | 0.8 | 0.8 | | | | | | | |
| 2.5 | 4.5 | 14.4 | ***56.8*** | 20.5 | 12.1 | 3 | 2.3 | | | | | | | |
| 3 | 3 | 5.3 | 10.6 | ***35.6*** | 12.9 | 8.3 | 4.5 | 0.8 | 0.8 | | | | | |
| 3.5 | 1.5 | 4.5 | 6 | 9.8 | ***40.9*** | 22 | 8.3 | 1.5 | - | 1.5 | | | | |
| 4 | 1.5 | 1.5 | 2.3 | 9.8 | 7.6 | ***46.2*** | 12.9 | 9 | 4.5 | 0.8 | 1.5 | | 0.8 | |
| 4.5 | 1.5 | | 2.3 | 6.8 | 5.3 | 0.8 | ***32.6*** | 8.3 | 6 | - | - | | - | 0.8 |
| 5.5 | | | 2.3 | 4.5 | 7.6 | 8.3 | 16.6 | ***43.2*** | 17.4 | 6.8 | 4.5 | 1.5 | 0.8 | 0.8 |
| 6 | | | 2.3 | 2.3 | 4.5 | 9 | 6.8 | 9.8 | ***43.9*** | 14.4 | 6.8 | 5.3 | - | 2.3 |
| 6.5 | | | | - | - | 1.5 | - | 8.3 | 5.3 | ***47*** | 10.6 | 0.8 | - | - |
| 7.5 | | | - | 1.5 | 3 | | 11.4 | 9 | 8.3 | 15.1 | ***56.8*** | 9 | 5.3 | 3.8 |
| | | | 0.8 | 1.5 | 1.5 | | 3 | 6 | 9.8 | 11.3 | 14.4 | ***56.8*** | 12.1 | 6 |
| 9 | | | | | | | | | | | | | | |
| 10 | | | | | 2.3 | | - | 3 | 2.3 | 2.3 | 5.3 | 19.7 | ***65.2*** | 21.2 |
| 10.5 | | | | | | | 0.8 | 0.8 | 1.5 | 0.8 | | 6.8 | 15.9 | ***65.2*** |
| Appr | **87.8** | **88.6** | **81** | **65.9** | **61.4** | **69** | **62.1** | **61.3** | **66.6** | **76.5** | **81.8** | **85.5** | **93.2** | **86.4** |

# The 2nd B series

## 1. Data

### Table 5

### Table 5 (1)

Dynamics of the conjugate pairs recognition

| Presentations | 0.9<br>1.5 | 1.2<br>2 | 1.5<br>2.5 | 1.8<br>3 | 2.1<br>3.5 | 2.4<br>4 | 2.7<br>4.5 | 3.3<br>5.5 | 3.6<br>6 | 3.9<br>6.5 | 4.5<br>7.5 | 5.4<br>9 | 6<br>10 | 6.3<br>10.5 |
|---|---|---|---|---|---|---|---|---|---|---|---|---|---|---|
| **1st pr** | **37.2** | **46.1** | **23.1** | **23.1** | **16.7** | **47.4** | **25.6** | **30.7** | **25.6** | **32.1** | **41** | **51.3** | **62.8** | **79.5** |
| 2nd pr | **65.4** | **64.1** | **59** | **33.3** | **39.7** | **57.7** | **48.7** | **46.1** | **48.7** | **47.4** | **57.7** | **46.2** | **53.8** | **65** |
| | Approximately + correct responses | | | | | | | | | | | | | |
| **1st pr** | ***59*** | ***60.2*** | ***39.7*** | ***42.3*** | ***32.1*** | ***56.3*** | ***40.9*** | ***42.2*** | ***51.2*** | ***55.1*** | ***69.2*** | ***84.6*** | ***91*** | ***89.7*** |
| **2nd pr** | ***87.2*** | ***87.2*** | ***83.3*** | ***76.9*** | ***75.7*** | ***71.8*** | ***75.6*** | ***69.2*** | ***58.9*** | ***76.8*** | ***83.3*** | ***62.8*** | ***85.8*** | ***78.2*** |

### Table 5 (2)

**Dynamics in shifting tendencies to both limits of the scale and correct and approximately correct responses (average data)**

| PRESENTATIONS | Shifting tendencies to | | Correct responses | Approximately + correct Responses |
|---|---|---|---|---|
| | Max. limit | Min. limit | | |
| The 1st presentation | 52.3 | 13.4 | **38.7** | ***58*** |
| The 2nd presentation | 25.6 | 25.7 | **52.3** | ***76.6*** |

Table 5 (3)

The quantity of correct responses to the time signals which were placed on the left and right halves of the stimuli scale (% of quantity of correct responses)

| The half of the scale belong to the | Presentations | |
|---|---|---|
| | The 1st presentation | The 2nd presentation |
| Minimum limit | **40.4** | **50.2** |
| Maximum limit | **59.6** | **49.8** |

## 2. Confusion matrices

Table 5 (4)

1st presentation

Responses to the given signals. MP (modulus pair) is absent, and subjects were not informed previously with the “limit “ pairs (0.9 s – 1.5 cm and 6.3 s – 10.5 cm)

| - | 0 | 2.6 | 6.4 | 14.2 | 9 | 8.9 | 10.2 | 13 | 14.1 | 16.7 | 23.2 | 19.2 | 15.3 | 20.4 |
|---|---|---|---|---|---|---|---|---|---|---|---|---|---|---|
| | 0.9<br>1.5 | 1.2<br>2 | 1.5<br>2.5 | 1.8<br>3 | 2.1<br>3.5 | 2.4<br>4 | 2.7<br>4.5 | 3.3<br>5.5 | 3.6<br>6 | 3.9<br>6.5 | 4.5<br>7.5 | 5.4<br>9 | 6<br>10 | 6.3<br>10.5 |
| 0.9<br>1.5 | **37.2** | 2.6 | 1.3 | 2.6 | | | | | | | | | | |
| 1.2<br>2 | 21.8 | **46.1** | 5.1 | 2.6 | 2.6 | 1.3 | | | | | | | | |
| 1.5<br>2.5 | 6.4 | 11.5 | **23.1** | 9 | 3.8 | | 1.3 | | 1.3 | 1.3 | | | | |
| 1.8<br>3 | 10.2 | 9 | 11.5 | **23.1** | 2.6 | 3.8 | | 2.8 | | | 1.3 | | | |
| 2.1<br>3.5 | 6.4 | 14.1 | 11.5 | 10.2 | **16.7** | 3.8 | 5.1 | | | 1.3 | 1.3 | | | |
| 2.4 | 6.4 | 5.1 | 11.5 | 12.8 | 12.8 | **47.4** | 3.8 | 6.4 | 6.4 | 1.3 | | | | |
| 4 | | | | | | | | | | | | | | |
| 2.7<br>4.5 | 2.6 | 9 | 6.4 | 11.5 | 9 | 5.1 | **25.6** | 3.8 | | | 1.3 | 1.3 | | |
| 3.3<br>5.5 | 3.8 | 1.3 | 14.1 | 6.4 | 16.7 | 18 | 11.5 | **30.7** | 6.4 | 6.4 | 1.3 | 1.3 | 3.8 | 1.3 |
| 3.6<br>6 | 2.6 | | 6.4 | 6.4 | 7.7 | 7.7 | 2.6 | 7.7 | **25.6** | 6.4 | 9 | 5.1 | 3.8 | 1.3 |
| 3. 9<br>6.5 | | | 2.6 | | | 3.8 | | 3.8 | 19.2 | **32.1** | 9 | | | |
| 4.5<br>7.5 | 1.3 | | 3.8 | 7.7 | 11.5 | 7.7 | 20.5 | 16.6 | 16.6 | 16.6 | **41** | 11.5 | 1.3 | 3.8 |
| 5.4<br>9 | 1.3 | 1.3 | 2.6 | 6.4 | 11.5 | 1.3 | 21.8 | 16.6 | 11.5 | 23.1 | 19.2 | **51.3** | 10.2 | 3.8 |
| 6<br>10 | | | | 1.3 | 3.8 | | 6.4 | 9 | 10.3 | 10.2 | 12.8 | 21.8 | **62.8** | 10.2 |
| 6.3<br>10.5 | | | | | 1.3 | | 1.3 | 2.6 | 2.6 | 1.3 | 3.8 | 7.7 | 18 | **79.5** |
| + | 62.8 | 51.3 | 70.4 | 62.7 | 74.3 | 43.6 | 64.1 | 56.3 | 60.2 | 51.2 | 35.8 | 29.5 | 18 | 0 |
| Appr | **59** | **60.2** | **39.7** | **42.3** | **32.1** | **56.3** | **40.9** | **42.2** | **51.2** | **55.1** | **69.2** | **84.6** | **91** | **89.7** |
| | 0.9<br>1.5 | 1.2<br>2 | 1.5<br>2.5 | 1.8<br>3 | 2.1<br>3.5 | 2.4<br>4 | 2.7<br>4.5 | 3.3<br>5.5 | 3.6<br>6 | 3.9<br>6.5 | 4.5<br>7.5 | 5.4<br>9 | 6<br>10 | 6.3<br>10.5 |

- 173.2 = 13.3 no approximate - 5
+ 680.2 = 52.3
no domination 1
“Right” responses 542.2 = 38.7%
“Approximately right” responses 813.5 = 58.1%

Table 5 (5)

2nd presentation

Responses to the given signals. MP (modulus pair) is absent, and subjects were not informed previously with the “limit “ pairs (0.9 s – 1.5 cm and 6.3 s – 10.5 cm)

| | | | | | | | | | | | | | | |
|---|---|---|---|---|---|---|---|---|---|---|---|---|---|---|
| **-** | 0 | 5.1 | 16.6 | 33.3 | 30.8 | 24.4 | 28.2 | 21.8 | 26.9 | 29.5 | 20.5 | 31.9 | 30.7 | 34.6 |
| | 0.9<br>1.5 | 1.2<br>2 | 1.5<br>2.5 | 1.8<br>3 | 2.1<br>3.5 | 2.4<br>4 | 2.7<br>4.5 | 3.3<br>5.5 | 3.6<br>6 | 3.9<br>6.5 | 4.5<br>7.5 | 5.4<br>9 | 6<br>10 | 6.3<br>10.5 |
| 0.9<br>1.5 | **65.4** | 5.1 | 3.8 | 2.6 | | | | | | | | | | |
| 1.2<br>2 | 21.8 | **64.1** | 12.8 | 5.1 | 3.8 | 2.6 | 1.3 | | | | | | | |
| 1.5<br>2.5 | 5.1 | 18 | **59** | 25.6 | 9 | 5.1 | | | 1.3 | 1.3 | | | | |
| 1.8<br>3 | 3.8 | 1.3 | 11.5 | **33.3** | 18 | 2.6 | 2.6 | 1.3 | | 1.3 | | | | |
| 2.1<br>3.5 | | 1.3 | 6.4 | 18 | **39.7** | 14.1 | 11.5 | 2.6 | 1.3 | 2.6 | 1.3 | | | 1.3 |
| 2.4<br>4 | 2.6 | 5.1 | 2.6 | 7.7 | 18 | **57.7** | 12.8 | 10.2 | 6.4 | 2.6 | 1.3 | | | |
| 2.7<br>4.5 | 1.3 | 3.8 | | | 5.1 | | **48.7** | 7.7 | 7.7 | | | 1.3 | | 1.3 |
| 3.3<br>5.5 | | 1.3 | | 5.1 | 2.6 | 10.2 | 14.1 | **46.1** | 10.2 | 5.1 | 2.6 | 3.8 | | 1.3 |
| 3.6<br>6 | | | 1.3 | 1.3 | 2.6 | 5.1 | 2.6 | 15.4 | **48.7** | 16.6 | 5.1 | 6.4 | 2.6 | 1.3 |
| 3. 9<br>6.5 | | | | 1.3 | 1.3 | 2.6 | | 6.4 | | **47.4** | 10.2 | 3.8 | 6.4 | |
| 4.5<br>7.5 | | | 1.3 | | | | 5.1 | 9 | 20.5 | 12.8 | **57.7** | 16.6 | 5.1 | 5.1 |
| 5.4<br>9 | | | | | | | | 1.3 | 3.8 | 7.7 | 15.4 | **46.2** | 16.6 | 11.5 |
| 6<br>10 | | | 1.3 | | | | | | | 1.3 | 2.6 | 15.4 | **53.8** | 12.8 |
| 6.3<br>10.5 | | | | | | | 1.3 | | | 1.3 | 3.8 | 6.4 | 15.4 | **65.4** |
| + | 34.6 | 30.8 | 24.4 | 33.4 | 29.6 | 17.9 | 23.1 | 32.1 | 24.3 | 23.1 | 21.8 | 21.8 | 15.4 | 0 |
| Appr | **87.2** | **87.2** | **83.3** | **76.9** | **75.7** | **71.8** | **75.6** | **69.2** | **58.9** | **76.8** | **83.3** | **62.8** | **85.8** | **78.2** |
| | 0.9<br>1.5 | 1.2<br>2 | 1.5<br>2.5 | 1.8<br>3 | 2.1<br>3.5 | 2.4<br>4 | 2.7<br>4.5 | 3.3<br>5.5 | 3.6<br>6 | 3.9<br>6.5 | 4.5<br>7.5 | 5.4<br>9 | 6<br>10 | 6.3<br>10.5 |

\- 334.3 = 25.7
\+ 332.3 = 25.6
no approximate no
no domination - no
“Right” responses 733.2 = 52.4 %
“Approximately right” responses 1072.7 = 76.6 %

Table 5 (6)

(average data)

| **Res-pon-ses (in cm)** | Given signals (in seconds) | | | | | | | | | | | | | |
|---|---|---|---|---|---|---|---|---|---|---|---|---|---|---|
| | 0.9 | 1.2 | 1.5 | 1.8 | 2.1 | 2.4 | 2.7 | 3.3 | 3.6 | 3.9 | 4.5 | 5.4 | 6 | 6.3 |
| 1.5 | ***48.7*** | 2.6 | 3.2 | 1.9 | | | | | | | | | | |
| 2 | 23 | ***53.2*** | 8.3 | 3.2 | 1.9 | 0.6 | | | | | | | | |
| 2.5 | 5.1 | 15.4 | ***39.1*** | 13.5 | 5.8 | 2.6 | 0.6 | | 1.3 | 1.3 | | | | |
| 3 | 7.7 | 6.4 | 12.2 | ***26.3*** | 9.6 | 4.5 | 1.3 | 1.9 | - | 0.6 | 0.6 | | | |
| 3.5 | 3.2 | 7.7 | 9.6 | 16.6 | ***26.3*** | 8.3 | 9 | 1.3 | 0.6 | 1.9 | 0.6 | | | 0.6 |
| 4 | 4.5 | 5.1 | 5.1 | 12.8 | 15.3 | ***50*** | 7 | 7 | 7 | 1.9 | 1.3 | | | - |
| 4.5 | 3.2 | 6.4 | 3.8 | 5.8 | 5.8 | 2.6 | ***34.6*** | 7 | 2.6 | - | 0.6 | 1.3 | | 0.6 |
| 5.5 | 1.9 | 2.6 | 7 | 5.8 | 12.8 | 13.5 | 12.8 | ***34.6*** | 8.3 | 4.5 | 1.9 | 2.6 | 1.9 | 1.3 |
| 6 | 1.3 | - | 5.1 | 4.5 | 5.1 | 6.4 | 3.2 | 12.2 | ***34.6*** | 12.8 | 5.8 | 5.1 | 2.6 | 1.3 |
| 6.5 | - | - | 1.3 | 0.6 | 1.3 | 3.8 | - | 5.1 | 10.3 | ***39.1*** | 9 | 1.9 | 3.2 | - |
| 7.5 | 0.6 | 0.6 | 3.8 | 5.8 | 5.8 | 4.5 | 14.7 | 14.7 | 19.2 | 13.5 | ***48.7*** | 10.3 | 3.2 | 3.2 |
| 9 | 0.6 | | 1.3 | 2.6 | 5.8 | 1.9 | 10.9 | 10.3 | 8.3 | 15.4 | 17.9 | ***48*** | 9.6 | 7.7 |
| 10 | | | | 0.6 | 1.9 | | 3.2 | 4.5 | 6.4 | 7 | 9.6 | 20.5 | ***58.3*** | 11.5 |
| 10.5 | | | | | 1.3 | | 1.9 | 1.3 | 1.3 | 1.9 | 3.8 | 10.3 | 21.2 | ***73.7*** |
| appr | **71.7** | **71.2** | **59.6** | **56.4** | **51.2** | **60.9** | **54.4** | **53.8** | **53.2** | **65.4** | **75.6** | **78.8** | **89.1** | **85.2** |

Table 6

The latency of responses to time signal presentations in all series of the primary experiment (in seconds)

| Responses kinds | Correct responses | | | | Incorrect responses | | | | Average | | | |
|---|---|---|---|---|---|---|---|---|---|---|---|---|
| **Series of experiments** | **Subseries 1a** | **1st series** | **2nd A series** | **2nd B series** | **Subseries 1a** | **1st series** | **2nd A series** | **2nd B series** | **Subseries 1a** | **1st series** | **2nd A series** | **2nd B series** |
| **1st presentation** | 3.24 | 3.27 | 4.6 | 4.65 | 4.13 | 3.7 | 4.83 | 4.54 | 3.56 | 3.52 | 4.74 | 4.58 |
| **2nd presentation** | 2.57 | 3.07 | 3.78 | 4.06 | 2.83 | 3.32 | 3.99 | 4.25 | 2.67 | 3.2 | 3.91 | 4.15 |
| **Average** | 2.9 | 3.16 | 4.19 | 4.36 | 3.48 | 3.53 | 4.41 | 4.4 | 3.11 | 3.35 | 4.33 | 4.37 |

## The Time of Decision (Latency)

Table 7

Latency of "right" responses to signals of the different duration (seconds)

| Posi-tions num-bers | Given inter-vals | | Series of experi-Ments | | |
|---|---|---|---|---|---|
| | Se-conds | Cen-time-tres | 1st se-ries | 2nd se-ries A | 2nd se-ries B |
| 1. | 0.3 | 0.5 | 2.61 | | |
| 2. | 0.6 | 1 | 2.45 | | |
| 3. | 0.9 | 1.5 | 3.17 | 3.09 | 3.35 |
| 4. | 1.2 | 2 | 3.43 | 4.63 | 4.22 |
| 5. | 1.5 | 2.5 | 3.6 | 4.4 | 4.84 |
| 6. | 1.8 | 3 | 3.86 | 5.59 | 4.91 |
| 7. | 2.1 | 3.5 | 3.63 | 5.08 | 4.66 |
| 8. | 2.4 | 4 | 3.78 | 4.59 | 5.04 |
| 9. | 2.7 | 4.5 | | 6.12 | 4.91 |
| 10. | 3.3 | 5.5 | 3.45 | 4.39 | 4.48 |
| 11. | 3.6 | 6 | 3.73 | 4.7 | 4.97 |
| 12. | 3.9 | 6.5 | | 3.97 | 4.98 |
| 13. | 4.5 | 7.5 | 3.49 | 4.74 | 4.83 |
| 14. | 5.4 | 9 | 3.4 | 3.9 | 4.27 |
| 15. | 6 | 10 | 2.81 | 3.32 | 3.68 |
| 16. | 6.3 | 10.5 | 2.49 | 3.09 | 2.83 |
| average | | | 3.27 | 4.4 | 4.42 |

# Appendix 2

Comparative tables of conjugate pairs distributions by their strength and the quantity of conjugate pairs observed during the first and the second time signal presentations which did not coincide by their strength with the same conjugate pairs located in the same order but along the scale of ”ideal” “bow” effect distribution.

Table 1

The comparative table of conjugate pairs distribution by their strength

The subseries 1 a

| **Kinds of conjugate pairs distributions by their strength along the scale.** | Numbers of points on the scale of presented stimuli and conjugate pairs placed on them (under number of these points, numbers of seconds and beneath them numbers of centimeters ) | | | | | | | | | | | | | |
|---|---|---|---|---|---|---|---|---|---|---|---|---|---|---|
| | 1<br>**0.3**<br>**0.5** | 2<br>**0.6**<br>**1** | 3<br>**0.9**<br>**1.5** | 4<br>**1.2**<br>**2** | 5<br>**1.5**<br>**2.5** | 6<br>**1.8**<br>**3** | 7<br>**2.1**<br>**3.5** | 8<br>**2.4**<br>**4** | 9<br>**3.3**<br>**5.5** | 10<br>**3.6**<br>**6** | 11<br>**4.5**<br>**7.5** | 12<br>**5.4**<br>**9** | 13<br>**6**<br>**10** | 14<br>**6.3**<br>**10.5** |
| **The “ideal” “bow” effect** | **SP** | **SP** | **RSP** | **RSP** | **WP** | **WP** | **WtP** | **WtP** | **WP** | **WP** | **RSP** | **RSP** | **SP** | **SP** |
| **distribution** | | | | | | | | | | | | | | |
| 1st presentation | SP | SP | SP | SP | WP | RSP | SP | SP | WP | WP | RSP | SP | SP | RSP |
| 2nd presentation | SP | SP | SP | SP | RSP | SP | WP | SP | SP | WP | RSP | RSP | SP | SP |

The quantity of such conjugate pairs during

the 1st presentation - 7

the 2nd presentation - 7

Table 2

The comparative table of conjugate pairs distributions by their strength

1st series

| **Kinds of conjugate pairs distributions by their strength along the scale.** | Numbers of points on the scale of presented stimuli and conjugate pairs placed on them (under number of these points, numbers of seconds and beneath them numbers of centimeters ) | | | | | | | | | | | | | |
|---|---|---|---|---|---|---|---|---|---|---|---|---|---|---|
| | 1<br>**0.3**<br>**0.5** | 2<br>**0.6**<br>**1** | 3<br>**0.9**<br>**1.5** | 4<br>**1.2**<br>**2** | 5<br>**1.5**<br>**2.5** | 6<br>**1.8**<br>**3** | 7<br>**2.1**<br>**3.5** | 8<br>**2.4**<br>**4** | 9<br>**3.3**<br>**5.5** | 10<br>**3.6**<br>**6** | 11<br>**4.5**<br>**7.5** | 12<br>**5.4**<br>**9** | 13<br>**6**<br>**10** | 14<br>**6.3**<br>**10.5** |
| **The "ideal" "bow" effect distribution** | SP | SP | RSP | RSP | WP | WP | WtP | WtP | WP | WP | RSP | RSP | SP | SP |
| 1st presentation | SP | SP | WP | RSP | WtP | WP | WP | WP | WtP | WP | RSP | RSP | SP | SP |
| 2nd presentation | SP | SP | SP | SP | RSP | WP | RSP | RSP | RSP | RSP | RSP | SP | SP | SP |

The quantity of such conjugate pairs during

the 1st presentation - 5

the 2nd presentation - 8

## Table 3

### The comparative table of conjugate pairs distribution by their strength

### 2nd A series

| **Kinds of conjugate pairs distributions by their strength along the scale.** | Numbers of points on the scale of presented stimuli and conjugate pairs placed on them (under number of these points, numbers of seconds and beneath them numbers of centimeters ) | | | | | | | | | | | | | |
|---|---|---|---|---|---|---|---|---|---|---|---|---|---|---|
| | 1<br>0.9<br>1.5 | 2<br>1.2<br>2 | 3<br>1.5<br>2.5 | 4<br>1.8<br>3 | 5<br>2.1<br>3.5 | 6<br>2.4<br>4 | 7<br>2.7<br>4.5 | 8<br>3.3<br>5.5 | 9<br>3.6<br>6 | 10<br>3.9<br>6.5 | 11<br>4.5<br>7.5 | 12<br>5.4<br>9 | 13<br>6<br>10 | 14<br>6.3<br>10.5 |
| **The "ideal" "bow" effect distribution** | SP | SP | RSP | RSP | WP | WP | WtP | WtP | WP | WP | RSP | RSP | SP | SP |
| 1st presentation | SP | SP | RSP | WP | WP | WP | WP | WtP | WP | WP | SP | SP | SP | SP |
| 2nd presentation | SP | SP | SP | WP | SP | SP | WP | SP | SP | SP | SP | SP | SP | SP |

The quantity of such conjugate pairs during

the 1st presentation - 4

the 2nd presentation - 10

## Table 4

### The comparative table of conjugate pairs distribution by their strength

### The 2nd B series

| Kinds of conjugate pairs distributions by their strength along the scale. | Numbers of points on the scale of presented stimuli and conjugate pairs placed on them (under number of these points, numbers of seconds and beneath them numbers of centimeters ) | | | | | | | | | | | | | |
|---|---|---|---|---|---|---|---|---|---|---|---|---|---|---|
| | 1<br>0.9<br>1.5 | 2<br>1.2<br>2 | 3<br>1.5<br>2.5 | 4<br>1.8<br>3 | 5<br>2.1<br>3.5 | 6<br>2.4<br>4 | 7<br>2.7<br>4.5 | 8<br>3.3<br>5.5 | 9<br>3.6<br>6 | 10<br>3.9<br>6.5 | 11<br>4.5<br>7.5 | 12<br>5.4<br>9 | 13<br>6<br>10 | 14<br>6.3<br>10.5 |
| The "ideal" "bow" effect distribution | SP | SP | RSP | RSP | WP | WP | WtP | WtP | WP | WP | RSP | RSP | SP | SP |
| 1st presentation | WP | RSP | WtP | WtP | WtP | RSP | WtP | WtP | WP | WP | RSP | SP | SP | SP |
| 2nd presentation | SP | SP | SP | WP | WP | SP | RSP | RSP | RSP | RSP | SP | RSP | SP | SP |

The quantity of such conjugate pairs during
the 1st presentation - 7, the 2nd presentation - 8

# Appendix 3

## Table and figure 1

## The subseries 1 a

The strength of presented conjugate pairs along the partly complete scale

| | | Numbers of places on the complete scale | | | | | | | | | | | | |
|---|---|---|---|---|---|---|---|---|---|---|---|---|---|---|
| Stimulus categories | | 1 | 2 | 3 | 4 | 5 | 6 | 7 | 8 | 9 | 10 | 11 | 12 | 13 |
| Duration | Seconds | 0.3 | 0.6 | 0.9 | 1.2 | 1.5 | **1.8** | 2.1 | 2.4 | | | 3.3 | 3.6 | |
| Length | Centimeters | 0.5 | 1 | 1.5 | 2 | **2.5** | **3** | **3.5** | 4 | | | 5.5 | 6 | |
| **Numbers of places on the presented scale** | | 1 | 2 | 3 | 4 | 5 | 6 | 7 | 8 | | | 9 | 10 | |
| 1st presentation | | **SP** | **SP** | **SP** | **SP** | **WP** | **RSP** | **SP** | **SP** | | | **WP** | **WP** | |
| 2nd presentation | | **SP** | **SP** | **SP** | **SP** | **RSP** | **SP** | **WP** | **SP** | | | **SP** | **WP** | |

| Continuation | | Numbers of places on the complete scale | | | | | | | | |
|---|---|---|---|---|---|---|---|---|---|---|
| Stimulus categories | | 14 | 15 | 16 | 17 | 18 | 19 | 20 | 21 | |
| Duration | Seconds | | 4.5 | | | **5.4** | | 6 | 6.3 | |
| Length | Centimeters | | **7.5** | | | **9** | | **10** | 10.5 | |
| **Numbers of places on the presented scale** | | | 11 | | | 12 | | 13 | 14 | |
| 1st presentation | | | **RSP** | | | **SP** | | **SP** | **RSP** | |
| 2nd presentation | | | **RSP** | | | **RSP** | | **SP** | **SP** | |

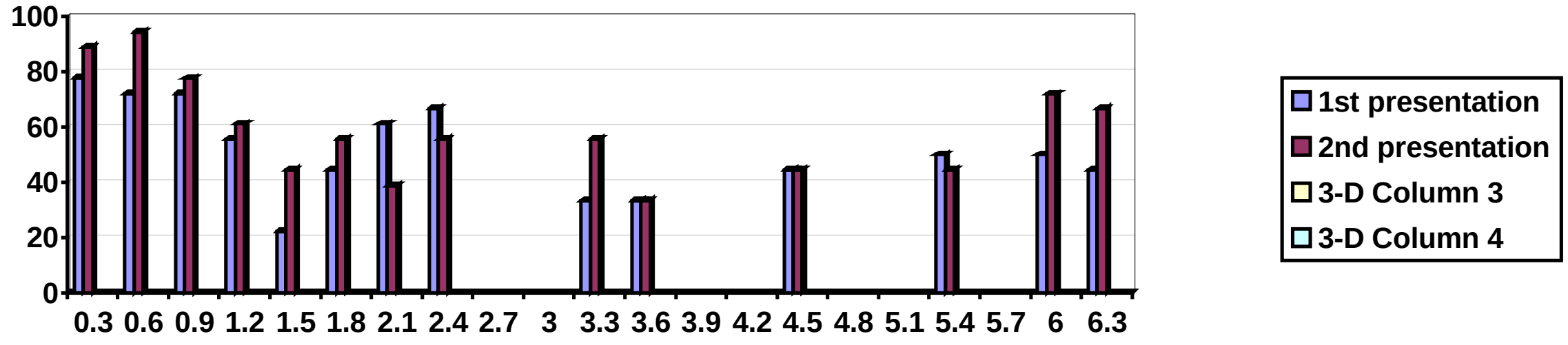

Figure 1. The subseries 1a. The distribution of correct responses along the scale of possible stimuli during the 1st and 2nd time signal presentations.

# Table and figure 2

## The 1st series

### The strength of presented conjugate pairs along the partly complete scale

| | | Numbers of places on the complete scale | | | | | | | | | | | | |
|---|---|---|---|---|---|---|---|---|---|---|---|---|---|---|
| Stimulus categories | | 1 | 2 | 3 | 4 | 5 | 6 | 7 | 8 | 9 | 10 | 11 | 12 | 13 |
| Duration | Seconds | 0.3 | 0.6 | 0.9 | 1.2 | 1.5 | **1.8** | 2.1 | 2.4 | | | 3.3 | 3.6 | |
| Length | Centimeters | 0.5 | 1 | 1.5 | 2 | **2.5** | **3** | **3.5** | 4 | | | 5.5 | 6 | |
| **Numbers of places on the presented scale** | | 1 | 2 | 3 | 4 | 5 | 6 | 7 | 8 | | | 9 | 10 | |
| 1st presentation | | **SP** | **SP** | **WP** | **RSP** | **WtP** | **WP** | **WP** | **WP** | | | **WtP** | **WP** | |
| 2nd presentation | | **SP** | **SP** | **SP** | **SP** | **RSP** | **WP** | **RSP** | **RSP** | | | **RSP** | **RSP** | |

| Continuation | | Numbers of places on the complete scale | | | | | | | | |
|---|---|---|---|---|---|---|---|---|---|---|
| Stimulus categories | | 14 | 15 | 16 | 17 | 18 | 19 | 20 | 21 | |
| Duration | Seconds | | 4.5 | | | **5.4** | | 6 | 6.3 | |
| Length | Centimeters | | **7.5** | | | **9** | | **10** | 10.5 | |
| **Numbers of places on the presented scale** | | | 11 | | | 12 | | 13 | 14 | |
| 1st presentation | | | **RSP** | | | **RSP** | | **SP** | **SP** | |
| 2nd presentation | | | **RSP** | | | **SP** | | **SP** | **SP** | |

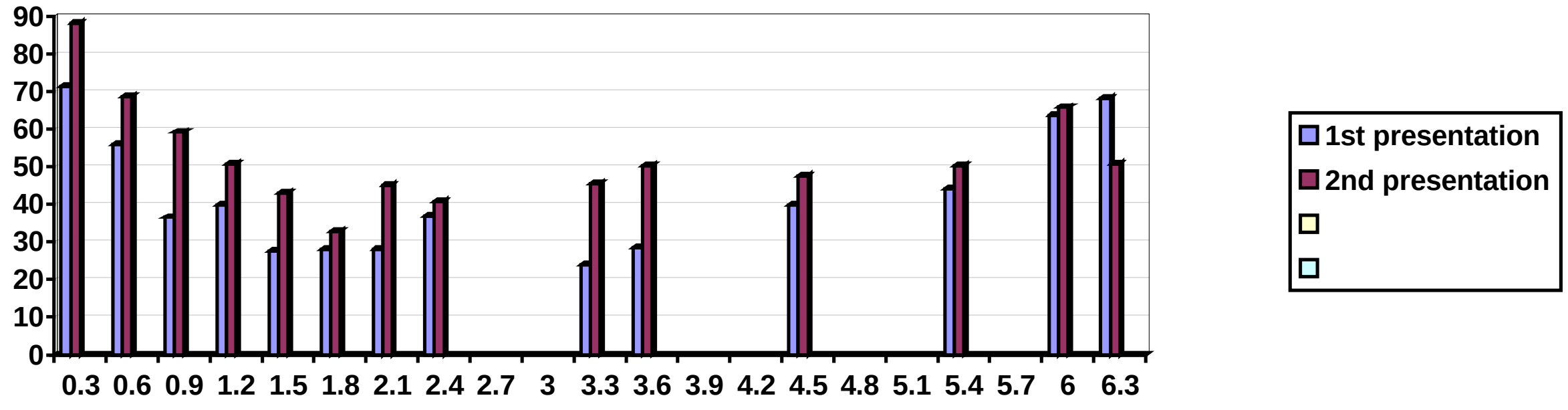

Figure 2. The 1st series. The distribution of correct responses along the scale of possible stimuli during the 1st and 2nd time signal presentations.

## Table and figure 3

## The 2nd A series

## The strength of presented conjugate pairs along the partly complete scale

| | | Numbers of places on the complete scale | | | | | | | | | | | | |
|---|---|---|---|---|---|---|---|---|---|---|---|---|---|---|
| Stimulus categories | | 1 | 2 | 3 | 4 | 5 | 6 | 7 | 8 | 9 | 10 | 11 | 12 | 13 |
| Duration | Seconds | | | 0.9 | 1.2 | 1.5 | **1.8** | 2.1 | 2.4 | 2.7 | | 3.3 | 3.6 | 3.9 |
| Length | Centimeters | | | 1.5 | 2 | 2.5 | **3** | 3.5 | 4 | 4.5 | | 5.5 | 6 | 6.5 |
| **Numbers of places on the presented scale** | | | | 1 | 2 | 3 | 4 | 5 | 6 | 7 | | 8 | 9 | 10 |
| 1st presentation | | | | **SP** | **SP** | **RSP** | **WP** | **WP** | **WP** | **WP** | | **WtP** | **WP** | **WP** |
| 2nd presentation | | | | **SP** | **SP** | **SP** | **WP** | **SP** | **SP** | **WP** | | **SP** | **SP** | **SP** |

## The 2nd A series

| Continuation | | Numbers of places on the complete scale | | | | | | | | |
|---|---|---|---|---|---|---|---|---|---|---|
| Stimulus categories | | 14 | 15 | 16 | 17 | 18 | 19 | 20 | 21 | |
| Duration | Seconds | | 4.5 | | | 5.4 | | 6 | 6.3 | |
| Length | Centimeters | | 7.5 | | | 9 | | 10 | 10.5 | |
| **Numbers of places on the presented scale** | | | 11 | | | 12 | | 13 | 14 | |
| 1st presentation | | | **SP** | | | **SP** | | **SP** | **SP** | |
| 2nd presentation | | | **SP** | | | **SP** | | **SP** | **SP** | |

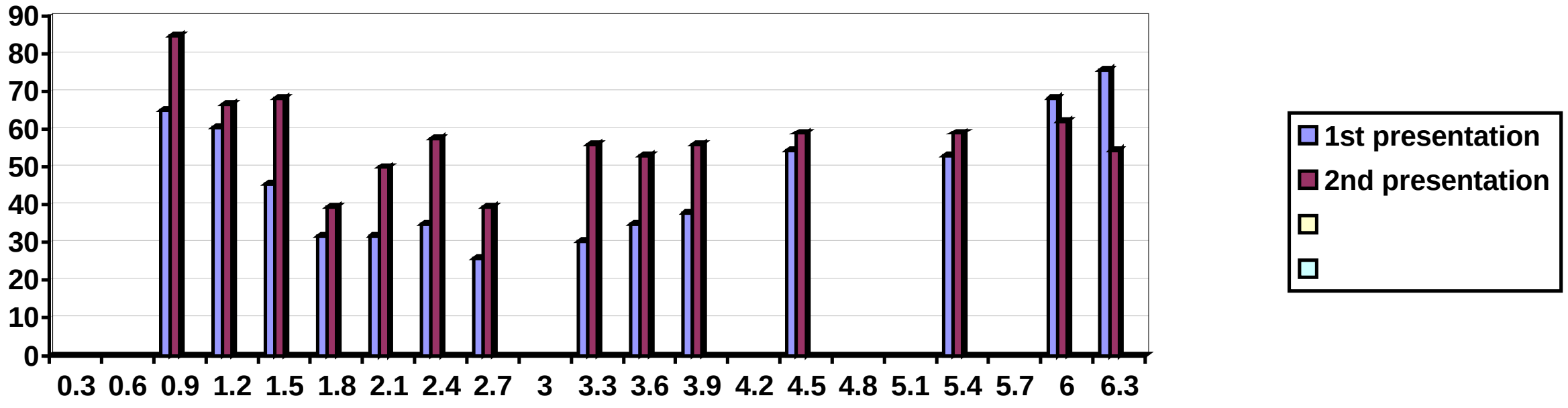

Figure 3. The 2nd A series . The distribution of correct responses along the scale of possible stimuli during the 1st and 2nd time signal presentations.

## Table and figure 4

## The 2nd B series

The strength of presented conjugate pairs along the partly complete scale

| | | Numbers of places on the complete scale | | | | | | | | | | | | |
|---|---|---|---|---|---|---|---|---|---|---|---|---|---|---|
| Stimulus categories | | 1 | 2 | 3 | 4 | 5 | 6 | 7 | 8 | 9 | 10 | 11 | 12 | 13 |
| Duration | Seconds | | | 0.9 | 1.2 | 1.5 | **1.8** | 2.1 | 2.4 | 2.7 | | 3.3 | 3.6 | 3.9 |
| Length | Centimeters | | | 1.5 | 2 | 2.5 | **3** | 3.5 | 4 | 4.5 | | 5.5 | 6 | 6.5 |
| **Numbers of places on the presented scale** | | | | 1 | 2 | 3 | 4 | 5 | 6 | 7 | | 8 | 9 | 10 |
| 1st presentation | | | | **WP** | **RSP** | **WtP** | **WtP** | **WtP** | **RSP** | **WtP** | | **WtP** | **WP** | **WP** |
| 2nd presentation | | | | **SP** | **SP** | **SP** | **WP** | **WP** | **SP** | **RSP** | | **RSP** | **RSP** | **RSP** |

## The 2nd B series

Continuation

| | | Numbers of places on the complete scale | | | | | | | | |
|---|---|---|---|---|---|---|---|---|---|---|
| Stimulus categories | | 14 | 15 | 16 | 17 | 18 | 19 | 20 | 21 | |
| Duration | Seconds | | 4.5 | | | 5.4 | | 6 | 6.3 | |
| Length | Centimeters | | 7.5 | | | 9 | | 10 | 10.5 | |
| **Numbers of places on the presented scale** | | | 11 | | | 12 | | 13 | 14 | |
| 1st presentation | | | **RSP** | | | **SP** | | **SP** | **SP** | |
| 2nd presentation | | | **SP** | | | **RSP** | | **SP** | **SP** | |

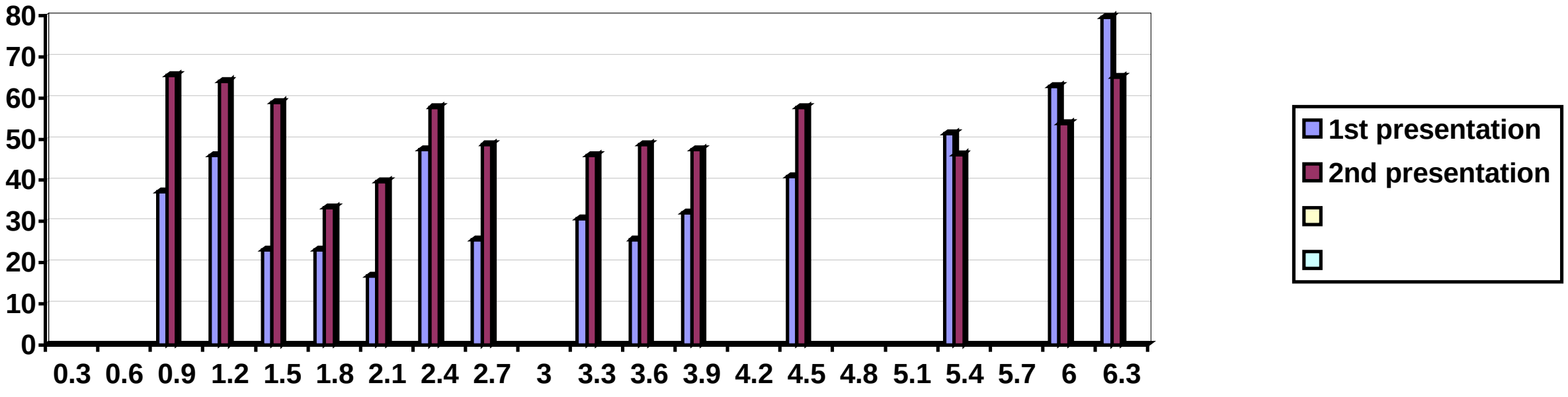

Figure 4. The 2nd B series . The distribution of correct responses along the scale of possible stimuli during the 1st and 2nd time signal presentations.

Appendix 4

The indirect confirmation of the hypothesis of the mental measuring scale of the correlation of possible stimuli which does not have an analog in the presented physical stimuli, can be found in the answers of observers to questions of form the questionnaire about the mental actions which they performed during the correlation of the presented to them intervals of space and time. Observers, which most in detail described these mental actions, indicated that during the correlation of the presented intervals they used as rules for measuring all remaining less long intervals longest of them. As an example one can adduce several answers to the following question

How did you manage to define and compare the duration of signal and the length of space interval how did you correlate them? Describe all this it in the manner that you seem, without fearing to prove to be ridiculous.

Observer N. At first I selected the **maximum length** of a pipe and the **maximum duration** of signal by a random search. Then I assigned to the peak signal as 8 intervals, established subjectively. I consider that my answers became more precisely, when I established such interval.

Observer K. After the first of 1/3 experiments **my calculation was arbitrary in the form of a specific frequency, accompanied by tracking of the signal on the longest pipe**. In this case I recalled about the work of computer at the master frequency and toward the end of the experiment **I beat the corresponding rhythm.**

Observer P. **I moved my look along largest ruler parallel with the light-sound signal duration.** After the cessation of signal I stopped the motion of my look thus determining the duration of signal. The necessary line was selected by me after this.

QUESTIONNAIRE

1. Did you count when the signal was produced? (The 1st variant is meant when the signal's duration was compared to the one of the pipe.)

a) Was the count free? What form did it take? What type of a count was it? Describe it in detail.

b) Was the attempt to count seconds?

c) Was the attempt to define lengths in centimeters done?

d) If there were such attempts, how did you manage to do it? Possible answer: made use of imaginary watch and ruler or some other objects? Or did not use them. Then what means are they?

Describe them in detail, please?

2. If you did not count how did you manage to define and compare the duration of signal and the length of the object, how did you correlate them? Please, describe all of this, the way you felt, without danger of seeming ridiculous?

3. As for you what does the correlate between sizes of a temporal signal and space object (the pipes or the lines) mean? How could you produce looking for the corresponding line after the signal was given?

4. How did you value time of a pipe (by pressing the button) in the second part of an experiment?

# 24

James H

1. Yes. Free count, no relation to seconds or centimeters or watch or ruler

a) Free-internal counting and comparison of signals to establish a norm

b) No

c) No

d) Does not apply

2. Does not apply

3. I compare in my mind relative length of pipes, length of signal using past experience of signal length

4. I pressed button in relation to my free count

For example, the subjects imagined the amodal mental scale of possible stimuli as a ruler, thermometer, and clock dial and so on. The process of measuring perceived intervals by means of this scale was accompanied by a count, tapping by finger or leg, by the rhythmical motion of their body and so on

As an example, one can adduce several answers to the following question:

How did you manage to define and compare the duration of the signal, and the length of the space interval, and how did you correlate them? Describe all this in the manner that it seemed to you, without fear of disapproval.

Observer N. "At first I selected the **maximum length** of a pipe and the **maximum duration** of the signal by a random search." "Then I assigned to the peak signal as 8 intervals, established subjectively". "I consider that my answers became more precise, when I established such an interval."

Observer K. "After the first 1/3 of the experiments **my calculation was arbitrary in the form of a specific frequency, accompanied by the tracking of the signal on the longest pipe**." "In this case I recalled about the work of a computer at the master frequency and toward the end of the experiment **I beat the corresponding rhythm."**

Observer P. "**I moved my look (visual scan) along largest (longest) ruler parallel with the light-sound signal duration."** "After the cessation of the signal I stopped the motion of my look (scan) thus determining the duration of the signal." "The necessary line was selected by me after this."